\documentclass[usenatbib]{mn2e}
\usepackage{amsmath}
\usepackage{natbib}
\usepackage{epsfig}
\usepackage{txfonts}

\newcommand{\xe}{x_{\rm e}}

\newcommand{\xc}{x_{\rm c}}

\newcommand{\id}{{\,\rm d}}

\newcommand{\beq}{\begin{equation}}   %

\newcommand{\eeq}{\end{equation}}   %

\newcommand{\beqa}{\begin{eqnarray}}   %

\newcommand{\eeqa}{\end{eqnarray}}   %

\newcommand{\beal}{\begin{align}}

\newcommand{\enal}{\end{align}}

\newcommand{\bspl}{\begin{split}}

\newcommand{\espl}{\end{split}}

\newcommand{\bsub}{\begin{subequations}}

\newcommand{\esub}{\end{subequations}}

\newcommand{\bmulti}{\begin{multline}}   %

\newcommand{\beqm}{\begin{mathletters}}   %

\newcommand{\eeqm}{\end{mathletters}}   %

\newcommand{\Abst}[1]{\,#1}

\newcommand{\me}{m_{\rm e}}

\newcommand{\Te}{T_{\rm e}}

\newcommand{\Tg}{T_{\gamma}}

\newcommand{\The}{\theta_{\rm e}}

\newcommand{\Thg}{\theta_{\gamma}}

\newcommand{\sigT}{\sigma_{\rm T}}

\newcommand{\nPl}{n_{\rm Pl}}

\newcommand{\nBE}{n_{\rm BE}}

\newcommand{\pd}{\partial}

\newcommand{\pAb}[2]{\frac{\displaystyle\pd #1}{\displaystyle\pd #2}}

\newcommand{\PAb}[3]{\frac{\displaystyle\pd^{#3} #1}{\displaystyle\pd {#2}^{#3}}}

\newcommand{\Abl}[2]{\frac{{\rm d} #1}{{\rm d} #2}}

\newcommand{\pot}[2]{#1 \times 10^{#2}}

\newcommand{\Thz}{\theta_{z}}
\newcommand{\Yp}{Y_{\rm p}}

\newcommand{\ion}[2]{{\text{{\sc #1}\,{\sc #2}}}}

\newcommand{\Tz}{{T_{z}}}
\newcommand{\Tgs}{{T_{\gamma}^\ast}}
\newcommand{\TCMB}{{T_{\rm CMB}}}
\newcommand{\tX}{{t_{\rm X}}}
\newcommand{\zX}{{z_{\rm X}}}
\newcommand{\zh}{{z_{\rm h}}}
\newcommand{\zmu}{{z_{\mu}}}
\newcommand{\ye}{{y_{\rm e}}}
\newcommand{\yff}{{y_{\rm ff}}}
\newcommand{\sigmah}{{\sigma_{\rm h}}}

\newcommand{\zs}{{z_{\rm s}}}
\newcommand{\ze}{{z_{\rm e}}}

\newcommand{\nS}{n_{\rm S}}

\newcommand{\Planck}{{\sc Planck}}

\newcommand{\zdc}{{{\rm few}\times 10^5}}

\usepackage{color}
\newcommand{\changeA}[1]{{#1}}
\usepackage{hyperref}
\usepackage{grffile}
\usepackage{graphics}
\usepackage{booktabs}

\title[CMB spectral distortions]
{The evolution of CMB spectral distortions in the early Universe}

\author[Chluba and Sunyaev]{J. Chluba$^{1}$\thanks{E-mail:
  jchluba@cita.utoronto.ca} and R.~A. Sunyaev$^{2,3}$ 
  \\
$^{1}$ Canadian Institute for Theoretical Astrophysics, 60 St. George Street,
Toronto, ON M5S 3H8, Canada
\\
$^{2}$ Max-Planck Institut f\"ur Astrophysik, Karl-Schwarzschild-Str. 1,
D-85740 Garching, Germany
\\
$^{3}$ Space Research Institute, Russian Academy of Sciences, Profsoyuznaya 84/32,
117997 Moscow, Russia
}

\begin{document}

\date{{Accepted 2011 September 8. Received 2011 September 3}}

\maketitle

\begin{abstract}
The energy spectrum of the cosmic microwave background (CMB) allows constraining episodes of energy release in the early Universe. 
In this paper we revisit and refine the computations of the cosmological thermalization problem. For this purpose {a new code, called {\sc CosmoTherm}, was developed} that allows solving the coupled photon-electron Boltzmann equation in the expanding, isotropic Universe for small spectral distortion in the CMB. 
We explicitly compute the shape of the spectral distortions caused by energy release due to (i) annihilating dark matter; (ii) decaying relict particles; (iii) dissipation of acoustic waves; and (iv) quasi-instantaneous heating. 
We also demonstrate that (v) the continuous interaction of CMB photons with adiabatically cooling non-relativistic electrons and baryons causes a {\it negative} $\mu$-type CMB spectral distortion of $\Delta I_\nu/I_\nu\sim 10^{-8}$ in the GHz spectral band. 
We solve the thermalization problem including improved approximations for the double Compton and Bremsstrahlung emissivities, as well as the latest treatment of the cosmological recombination process.
At redshifts $z\lesssim 10^3$ the matter starts to cool significantly below the temperature of the CMB so that at very low frequencies free-free absorption alters the shape of primordial distortions significantly. 
In addition, the cooling electrons down-scatter CMB photons introducing a small late {\it negative} $y$-type distortion at high frequencies.
We also discuss our results in the light of the recently proposed CMB experiment {\sc Pixie}, for which {\sc CosmoTherm} should allow detailed forecasting.  
Our current computations show that for energy injection because of (ii) and (iv) {\sc Pixie} {should} 
allow to improve existing limits, while the CMB distortions caused by the other processes seem to remain unobservable {with the currently proposed sensitivities and spectral bands of {\sc Pixie}}.
%
%
\end{abstract}

\begin{keywords}
Cosmology: cosmic microwave background -- theory -- observations
\end{keywords}

\section{Introduction}
\label{sec:Intro}

Investigations of the cosmic microwave background (CMB) temperature and polarization anisotropies without doubt have allowed modern cosmology to mature from an order of magnitude branch of physics to a precise scientific discipline, were theoretical predictions today are challenged by a vast amount of observational evidence.
{Since its discovery in the 60's \citep{Penzias1965}, the full CMB sky has been mapped by {\sc Cobe/Dmr} \citep{Smoot1992} and {\sc Wmap} \citep{WMAP_params, Page2006}, 
while at small angular scales many balloon-borne and ground-based CMB experiments like {\sc Maxima} \citep{Hanany2000}, {\sc Boomerang} \citep{Netterfield2002}, {\sc Dasi} \citep{DASI2002}, {\sc Archeops} \citep{Archeops2003}, {\sc Cbi} \citep{CBI2003} and {\sc Vsa} \citep{VSA2003} provided important additional leverage, significantly helping to tighten the joint constraints on cosmological parameters.}
Presently \changeA{{\sc Planck}}\footnote{\url{http://www.esa.int/Planck}} is producing CMB data with unprecedented precision, while both {\sc ACT} \citep[e.g., see][]{Hajian2010, Dunkley2010, Das2011} and {\sc SPT} \citep{Lueker2010, Vanderlinde2010} are pushing the frontier of $TT$ CMB power spectra at small angular scales.
In the near future 
{\sc SPTpol}\footnote{\url{http://pole.uchicago.edu/}} \citep{SPTpol} and
{\sc ACTPol}\footnote{\url{http://www.physics.princeton.edu/act/}}
\citep{ACTPol} will provide additional small scale $E$-mode
polarization data, complementing the polarization power spectra obtained with  {\sc Planck} 
and further increasing the significance of the $TT$ power spectra. 
Also {\sc Quiet}\footnote{\url{http://quiet.uchicago.edu/}} \citep{Bischoff2010} already now is observing the CMB in $Q$ and $U$ polarization with the aim to tighten the constraints on the potential level of $B$-modes at multipoles $l\lesssim 500$, 
{while {\sc Spider}\footnote{\url{http://www.astro.caltech.edu/~lgg/spider/spider_front.htm}} and {\sc Polarbear}\footnote{\url{http://sites.google.com/site/mcgillcosmology/Home/polarbear}} are getting ready to deliver precision CMB polarization data in the near future.}

With the new datasets, cosmologists will be able to
determine the key cosmological parameters with extremely high
precision, {also} making it possible to distinguish between {various models}
of {\it inflation} by measuring the
precise value of the spectral index of scalar perturbations, $n_{\rm
  S}$, and constraining its possible running, $n_{\rm run}$.
Furthermore, with the {\sc Planck} satellite {primordial} $B$-mode polarization may be discovered, 
if the tensor to scalar ratio, $r$, is larger than a few percent  \citep{Planckblue2006}. 
This would be the smoking gun of inflation and indicate the existence of gravitational waves, {one of the main scientific goals of {\sc Planck}, {\sc Quiet}, {\sc Spider}\changeA{, {\sc Polarbear} and other CMB experiments}}.

However, the anisotropies are not the only piece of information about the early Universe the CMB offers us. 
In addition, the {\it energy spectrum} of the CMB tells the tale about the thermal history of the Universe at very early times, well before any structures had formed and when the baryonic matter and radiation were tightly coupled via Compton scattering.
It is well known that any perturbation of the full thermodynamic equilibrium between photons and baryons in the early Universe inevitably leads to spectral distortions in the CMB \citep{Zeldovich1969, Sunyaev1970mu, Sunyaev1970SPEC, Zeldovich1972, Sunyaev1974, Illarionov1974}.
This for example could be caused by some release of energy by {\it decaying relic particles} \citep{Hu1993b} or {\it evaporating primordial black holes} \citep{Carr2010}, and depending on their lifetime\footnote{For black holes the evaporation timescale is $t_{\rm ev}\propto M^{3}$ \citep{Hawking1974}.} and the total amount of energy that was transferred to the medium, the residual CMB spectral distortion will be smaller or larger.

The process that erases possible spectral distortions of the CMB in the early Universe and thereby attempts to restores full thermodynamic equilibrium is called {\it thermalization}. In the past it has been studied extensively, both analytically and numerically \citep{Illarionov1975, Illarionov1975b, Danese1977, Danese1982, Burigana1991, Burigana1991b, Daly1991, Barrow1991, Burigana1993, Hu1993, Hu1993b, Hu1994, Burigana1995, Burigana2003, Lamon2006, Procopio2009}.
For very early energy release ($z\gtrsim \pot{2}{6}$), photon production at low frequencies by double Compton emission and the up-scattering of photons to high frequencies by Compton scattering are so efficient that the thermalization process is practically perfect for nearly arbitrary amounts of energy release, such that no distortion should remain today \citep[e.g., see][]{Burigana1991}.
However, below this redshift the CMB spectrum becomes vulnerable, and spectral distortions that could still be observable today can form.

In connection with early energy release two types of CMB distortions, called $\mu$- and $y$-type distortions, are usually distinguished \citep[e.g., see][]{Illarionov1975, Illarionov1975b}. 
The first type of distortion is characterized by a frequency-dependent chemical potential, $\mu(\nu)$, that is very close to constant at high frequencies, and vanishes at very low frequencies. This type of distortion forms at times when the Compton process is still able to achieve full kinetic equilibrium with the electrons (redshifts $z\gtrsim \pot{\text{few}}{5}$), and the CMB spectrum reaches a Bose-Einstein distribution with occupation number $n_\nu=1/[e^{x+\mu(\nu)}-1]$, where $x=h\nu/k\Tg$, with $\Tg$ denoting the CMB temperature.
However, photon production by double Compton and Bremsstrahlung emission stopped being very fast, so that only at low frequencies full equilibrium can be restored.

At low redshifts ($z\lesssim \pot{\text{few}}{3}-10^4$), on the other hand, $y$-type distortions form. These are also known in connection with SZ clusters \citep{Zeldovich1969} and are characterized by a constant temperature decrement at low and an increment at high frequencies.
This type of distortion is formed when the Compton process ceases to be extremely efficient, so that full kinetic equilibrium between electrons and photons can no longer be achieved.
Photons produced at low frequencies by Bremsstrahlung are therefore not up-scattered very much and remain close to the frequency band they were emitted.

Both types of distortions were tightly constrained by {\sc Cobe/Firas} \citep{Mather1994, Fixsen1996} only leaving room for $|y|\lesssim 10^{-5}$ and $|\mu|\lesssim 10^{-4}$ at cm and dm wavelength.
This imposes strong limits on possible energy release by decaying particles \citep{Hu1993b}, which for some particle lifetimes even today are comparable to those derived, for example, from big bang nucleosynthesis \citep{Kusakabe2006, Kohri2010}.
However, since the 90's detector technology has advanced very much. Already about ten years ago, a {\sc Firas} type-II experiment with roughly 100 times better sensitivity in principle would have been possible \citep{Fixsen2002}.
\changeA{Additional} new technology was used in several flights of the {\sc Arcade} balloon-borne instrument \citep{Kogut2004, Fixsen2011, Kogut2011}, \changeA{from which improved limits on the CMB temperature at $\nu \sim 10\,$GHz were derived \citep{Seiffert2011}. However, until now} no signatures of primordial distortions were found, and the limits on \changeA{overall} CMB spectral distortions derived from {\sc Cobe/Firas} until today remain most stringent.

However, very recently a new space-based mission, called {\sc Pixie} \citep{Kogut2011PIXIE}, was proposed. {\sc Pixie} will have exquisite spectral capabilities with 400 channels at frequencies ranging from $\nu=30\,$GHz to $6\,$THz. 
The claimed sensitivities should allow to tighten the constraints on $\mu$- and $y$-type distortions by about three orders of magnitude.
A detection of $y\sim 10^{-8}$ and $\mu\sim \pot{5}{-8}$ could therefore become possible at the $5\sigma$ level \citep[see Fig.~12 of][]{Kogut2011PIXIE}, {at least in terms of spectral sensitivity}.
\changeA{However, it is not yet totally clear what the foreground limitations to measuring primordial spectral distortions will be.}
Under simplifying assumptions analytic approximation in the two extreme cases of $\mu$ and $y$-type distortions can be found, however, for energy release in the redshift range $10^4\lesssim z\lesssim \pot{\text{few}}{5}$ it is clear that a {\it mixture} of both types of distortions will form. In that case analytic approximations are more difficult and numerical studies are required.
Furthermore, in the past, mainly the special cases of {\it instantaneous} energy release at some initial redshift were numerically studied in detail for this regime \citep[e.g. see][]{Burigana1995, Burigana2003}.
However, from the physical point of view cases of {\it continuous} energy release over extended periods are more natural.
In addition, so far only problems with rather large energy injection, close to the allowed upper bounds obtained by {\sc Cobe/Firas}, were investigated thoroughly.
Again from the observational point of view, significantly smaller distortions seem to be favoured, and in addition the sensitivities reached by {\sc Pixie} demand studies with very small energy release.
To allow forecasting the observational possibilities of {\sc Pixie} with respect to constraints on the thermal history of our Universe, significantly refined computations of the possible spectral distortions with particular physical processes in mind therefore become necessary.

In this work we revisit the cosmological thermalization problem in the light of {\sc Pixie}. We formulate the thermalization problem allowing to resolve very small spectral distortions at levels well below the current upper limits of {\sc Cobe/Firas}. 
We use improved approximations for the double Compton and Bremsstrahlung emissivities, as well as a detailed treatment of the cosmological recombination problem based on {\sc CosmoRec} \citep{Chluba2010b}.
We explicitly compute the expected CMB spectral distortions for different energy injection scenarios.
In particular, we discuss two scenarios that are even present in the standard cosmological model. One is connected with the interaction of CMB photon with {\it adiabatically cooling electrons} and baryons (see Sect.~\ref{sec:Q_cooling}), while the other is caused by the {\it dissipation of acoustic waves} in the early Universe (see Sect.~\ref{sec:Q_waves}).
We also demonstrate that the evolution of spectral distortions at very low frequencies $\nu\lesssim 1\,$GHz is significantly affected by free-free absorption at late times ($z\lesssim 10^3$), when the Compton process no longer is able to keep the electrons and baryons from cooling strongly below the CMB temperature.
For large energy release this modification is not very significant, however, for very small energy release, comparable to $\mu$ and $y$ at the level $\sim 10^{-8}$, it becomes important.
%

%

 One of {the outcomes} of this work is {\sc CosmoTherm}\footnote{{\sc CosmoTherm} will be made available at \url{www.Chluba.de/CosmoTherm}.}, an open source code that solves the cosmological thermalization problem for different energy injection scenarios in the limit of small distortions. 
 In the future {\sc CosmoTherm} should be useful for detailed forecasts of possibilities to constrain the thermal history of our Universe by measuring the CMB energy spectrum.

\section{Formulation of the thermalization problem}
\label{sec:thermalization}
A solution of the thermalization problem in the isotropic Universe generally can only be obtained numerically. 
Under simplifying assumptions several analytic approximations exist, however, these approximations have their limitations.
For example, for very large energy injection, non-linear aspects of the problem become important, while for very small energy release, the precise shape of the distortion depends strongly on the small difference in the electron and photon temperature, which is comparable to the distortion itself, and is more difficult to account for in analytic approximations.
Below we provide the required equations for the formulation of the problem that allows to include all these aspects consistently.

\subsection{General aspects of the thermalization problem} 
\label{sec:general}
To thermalize spectral distortions after some significant energy release in the early Universe two main ingredients are needed: (i) the number of photons has to be readjusted and (ii) photons have to be redistributed over frequency to restore the shape of a blackbody.

At redshifts $z\gtrsim 10^8-10^9$ electrons and positrons were very abundant
and electron-positron, electron-electron and positron-positron Bremsstrahlung
were orders of magnitude more efficient in producing soft photons than any other process.
Hence thermalization of CMB spectral distortions was extremely rapid.
Therefore we shall restrict ourselves to redshifts below $z\lesssim \pot{5}{7}$.
By then the number densities of electrons and positrons had frozen out\footnote{More precisely the freeze-out of positrons happens at temperatures $kT\sim \me c^2 /40$ or redshift $z\sim \pot{5.4}{7}$.}
and only insignificant amounts of positrons were left from the era of
electron-positron annihilation.
During the subsequent evolution of the (isotropic) Universe the main interactions between
the photons and baryonic matter are governed by Compton scattering (CS), 
double Compton scattering (DC) and  normal electron-ion Bremsstrahlung (BR). Furthermore due to the expansion of the Universe the photons suffer from redshifting and the non-relativistic electrons and baryons cool adiabatically.

The photons do not interact directly with the baryonic matter (hydrogen,
helium and the nuclei of other light elements), but only indirectly mediated by
electrons, which themselves are strongly coupled to the baryons via Coulomb
scattering. 
The timescale, on which the electrons and baryons adjust their energy
distributions is much shorter than any other timescale of importance here.
Therefore it can be assumed that the electrons and baryons always follow a
(relativistic) Maxwell Boltzmann distribution with one temperature $\Te$.  
In this case the problem can be formulated with the Boltzmann equations for the
evolution of the photon phase space distribution function and a coupled
equation describing the time evolution of the electron (and baryon)
temperature. 
In addition the expansion timescale, $t_{\rm exp}=1/H$, where $H$ is the Hubble factor,  
is affected by the energy injection process, however, as we argue in Sect.~\ref{sec:exp_hist} the corresponding correction is negligible.
Below we shall write down the photon Boltzmann equation in
the expanding isotropic Universe and the evolution equation for the electron temperature
with inclusion of heating and cooling.

\subsection{Evolution of the photons in the expanding Universe} 
\label{sec:PhotonEvo}
In the early Universe the photons undergo many interactions with the electrons 
in the Universe. As mentioned earlier, the most
important processes are Compton scattering, double Compton scattering,
Bremsstrahlung and the adiabatic expansion of the Universe. Among these
processes for most times Compton scattering is the fastest. Therefore
it is convenient to express all the involved timescales in units of
$t_{\rm C}=1/\sigT\,N_{\rm e}\,c$, the Thomson scattering time.

The Boltzmann equation for the evolution of the photon occupation number, $n_\nu$, then can be expressed as:
\beq\label{eq:BoltzEq_Photons}
\pAb{n_{\nu}}{\tau}-H\,t_{\rm C}\,\nu\,\pAb{n_{\nu}}{\nu}
=\left.\Abl{n_{\nu}}{\tau}\right|_{\rm C}
+\left.\Abl{n_{\nu}}{\tau}\right|_{\rm DC}
+\left.\Abl{n_{\nu}}{\tau}\right|_{\rm BR}
\Abst{,}
\eeq
where we introduced the optical depth $\id\tau=\id t/t_{\rm C}$ to electron
scattering as the dimensionless time variable. 
The second term on the left hand side
is due to the expansion of the Universe and the right hand side terms
correspond to the physical processes quoted above. 

For computational purposes it is also convenient to transform to dimensionless frequency. 
Introducing $x=h\nu/k\Tz$, where $\Tz=T_{z,0}[1+z]$, the expansion term in Eq.~\eqref{eq:BoltzEq_Photons} 
can be absorbed so that photons no longer redshift out of the chosen computational domain.
Here one point is very important: the temperature $T_{z,0}$ can really be {\it any} temperature, as long as it remains constant in time. One obvious choice is the observed CMB temperature today, $T_0=2.726\,$K \citep{Fixsen2009}, but we will use this freedom to fix our initial conditions conveniently (see Sect.~\ref{sec:initial_cond}).

For $T_{z,0}\equiv T_0$, we have the standard redshift dependent CMB temperature, 
$\Tz\equiv \TCMB=T_0[1+z]$.
However, it is important to mention that with this definition $\Tz$ is not necessarily identical to the effective (thermodynamic) temperature of the photon field, 
\beal\label{eq:T_eff}
T^\ast_\gamma
&=\Tz \left[\frac{\mathcal{G}_{3}}{\mathcal{G}^{\rm pl}_{3}}\right]^{1/4}
=\Tz \left[\frac{15\,\mathcal{G}_{3}}{\pi^4}\right]^{1/4}
\end{align}
which is found by comparing the total energy density of the photon distribution 
with the one of a pure blackbody. Here $\mathcal{G}_{3}$ is defined by the integral
\beal
\label{eq:I_G_Int}
\mathcal{G}_{3}&=\int x^3 n_x  \id x=\mathcal{G}^{\rm pl}_{3}+\int x^3 \Delta n_x  \id x,
\end{align}
over the photon occupation number, $n_x$. For a blackbody with thermodynamic temperature $\Tz$ one has $\nPl(x)=1/[e^x-1]$, and hence $\mathcal{G}^{\rm pl}_{3}=\pi^4/15\approx 6.4939$, trivially implying that $T^\ast_\gamma \equiv \Tz$.
However, in the presence of spectral distortions after release of energy generally $T^\ast_\gamma \neq \Tz$. 
This for example means that at higher redshifts the effective temperature of the CMB could have been slightly lower than today, and only because of energy injection and subsequent thermalization it has reached its present value. Furthermore, it is important to mention that $T_0$ is known with precision $\pm 1\,$mK \citep{Fixsen2002, Fixsen2009}, so that deviations of the effective temperature from $T_0$ which are {much} smaller than this are presently indistinguishable. We will return to this aspect of the problem again in Sect.~\ref{sec:initial_cond}.

In Eq.~\eqref{eq:I_G_Int} we also introduced the distortion, $\Delta n_x=n_{x}-\nPl(x)$, from a blackbody with temperature $\Tz$. 
For numerical computations it in general is useful to substitute $n_x$ in this way, as this allows one to cancel the dominant terms and thereby linearize the problem, by only solving for the (small) correction $\Delta n_x$.
%
%
 
\subsubsection{Compton scattering}
The contribution of Compton scattering to the right hand side of the photon
Boltzmann equation \eqref{eq:BoltzEq_Photons} 
can be treated using the Kompaneets equation \citep{Kompa56}:
\beal
\label{eq:Komp_equ}
\left.\Abl{n_x}{\tau}\right|_{\rm C}
&=
\frac{\theta_{\rm e}}{x^2}\,
\pAb{}{x}\, x^4\!\left[\pAb{n_x}{x}+\phi\,n_x(n_x+1)\right]\
\Abst{.}
\end{align}
Here we introduced the abbreviations $\phi=\Tz/\Te$ and $\theta_{\rm e}=\frac{k\Te}{\me c^2}$. 
This equation describes the redistribution of photons over frequency, were the effect of Doppler diffusion and boosting, electron recoil, and stimulated electron scattering are accounted for.

The Kompaneets equation was found as lowest order
Fokker-Planck expansion of the collision term for Compton
scattering. 
We neglect higher order relativistic corrections \changeA{\citep[e.g., see][]{Rephaeli1995, Itoh98, Challinor1998, Sazonov1998, Nozawa2006}}. These are expected to affect the results at the level of a few percent at $z\gtrsim 10^6-10^7$ \citep{Chluba2005}, so that for the purpose of this paper Eq.~\eqref{eq:Komp_equ} will be sufficient.

One can easily verify, that the number of photons is conserved under Compton
scattering by multiplying Eq.~\eqref{eq:Komp_equ} with $x^2$ and integrating over frequency using
integration by parts and the fact that the photon distribution vanishes
sufficiently fast for $x\rightarrow0$ and $x~\rightarrow~\infty$. Furthermore,
one can easily check that for photons, which follow a Bose-Einstein
distribution, $\nBE(x)=1/[e^{x+\mu}-1]$, with constant chemical potential
$\mu=\mu_0$, Eq.~\eqref{eq:Komp_equ} vanishes identically if $\phi=1$.
For $\phi\neq 1$ in the chosen coordinates the photon distribution in kinetic equilibrium with respect to Compton scattering is $\nBE(x)=1/[e^{x+\mu(x)}-1]$ with frequency-dependent chemical potential $\mu(x)=\mu_0+\xe-x$, where we defined $\xe=\phi x$.

If we now insert $n_{x}=\nPl(x)+\Delta n_x$ into Eq.~\eqref{eq:Komp_equ} and neglect terms of $\mathcal{O}(\Delta n_x^2)$, we obtain the linear equation
\beal
\label{eq:Komp_equ_line}
\left.\Abl{\Delta n_x}{\tau}\right|_{\rm C}
&=
\frac{\Delta \theta_{\rm e}}{x^2}\,
\pAb{}{x}\, x^4\, \zeta
+\frac{\theta_{\rm e}}{x^2}\,
\pAb{}{x}\, x^4\!\left[\pAb{\Delta n_x}{x}+\phi\,\xi \,\Delta n_x\right],
\end{align}
where for convenience we have introduced
\bsub
\label{eq:occupation_abr}
\beal
\zeta(x)&=-\nPl(\nPl+1)=-\frac{e^{-x}}{[1-e^{-x}]^2}\equiv \frac{1}{2}\,\pAb{\xi(x)}{x}
\\
\xi(x)&=2 \nPl+1=\frac{1+e^{-x}}{1-e^{-x}}
\end{align}
\esub
and $\Delta \theta_{\rm e}=\theta_{\rm e}- \Thz$, with $ \Thz=\frac{k\Tz}{\me c^2}\approx \pot{4.60}{-10}[1+z]$.
One can see from Eq.~\eqref{eq:Komp_equ_line} that the remaining part arising from the CMB background spectrum appears as a source term in the evolution equation, which vanishes identically unless $\Tz\neq \Te$.

Explicitly carrying out part of the derivatives in Eq.~\eqref{eq:Komp_equ_line} and rearranging terms we find
\beal
\label{eq:Komp_equ_line_rearr}
\left.\Abl{\Delta n_x}{\tau}\right|_{\rm C}
&=
D_{\rm e}\PAb{\Delta n_x}{x}{2}
+D_{\rm e} \left[\frac{4}{x}+\phi\,\xi \right] \pAb{\Delta n_x}{x}
\nonumber\\
&\qquad
+D_{\rm e} \,\phi\,\xi \left[\frac{4}{x}+\pAb{\ln \xi}{x}\right] \,\Delta n_x
-D_{\rm e}\,\Delta \phi\, \zeta  \left[\frac{4}{x}-\xi \right]  
\end{align}
with $D_{\rm e}=\theta_{\rm e} \, x^2$, $\Delta \phi=\phi-1$, and $\partial_x \ln \xi=-2e^{-x}/[1-e^{-2x}]$.

For numerical computations it is convenient to precompute the functions $\zeta $, $\xi$, $\partial_x \ln \xi$, $\nPl$ and $e^{-x}$ once the frequency grid is chosen. This accelerates the computation significantly.
Furthermore, it is important to use series expansion of the expressions for $x\ll1$, in particular when differences $1-e^{-x}$ are encountered.
This is very important to achieve numerical stability and the correct limiting behaviour, and  will also be crucial for the formulation of the emission and absorption term in the next section.

\subsubsection{Double Compton scattering and Bremsstrahlung}
\label{sec:Emission}
In the cosmological thermalization problem, the double Compton process 
and Bremsstrahlung provide the source and sink of photons.
Their contribution to photon Boltzmann equation, Eq.~\eqref{eq:BoltzEq_Photons}, can be
cast into the form \citep{Rybicki1979, Lightman1981, Danese1982, Chluba2005}:
\beal
\label{eq:DC_term}
\left.\Abl{n_{x}}{\tau}\right|_{\rm DC+BR} 
&= \frac{1}{x^3} \Big[1-n_{x}\,(e^{\xe}-1)\Big]\times K(x, \Thz, \The) 
\end{align}
where the emission coefficient $K$ is given by the sum of the contribution due
to double Compton scattering and Bremsstrahlung,
$K=K_{\rm DC}+K_{\rm BR}$.
Since $K$ drops off exponentially for $x\rightarrow \infty$ (see below), the main emission and absorption of photons is occurring at low frequencies.
At small $x$ the photon distribution after a very short time is pushed into equilibrium with $n^{\rm eq}_x=1/[e^{\xe}-1]$, i.e., a blackbody of temperature $\Tz\equiv \Te$.
%
%
If once again we insert $n_{x}=\nPl(x)+\Delta n_x$ and rearrange terms, we find
\beal
\label{eq:DC_term_rewrite}
\left.\Abl{\Delta  n_{x}}{\tau}\right|_{\rm DC+BR} 
&= \frac{1-e^{\Delta \xe}}{1-e^{-x}}\times \frac{K(x, \Thz, \The)}{x^3} 
\nonumber \\ 
&\qquad +\Delta n_{x} (1-e^{\xe})\times  \frac{K(x, \Thz, \The)}{x^3},
\end{align}
with $\Delta \xe=\xe-x$. 
Note that the first term vanishes if $\Te\equiv \Tz$, however, the second is pushed to equilibrium only because of interplay with the first term: if a distortion is present then consequently $\Te\neq \Tz$ and hence the first term either leads to net absorption ($\Te <  \Tz$) or emission ($\Te >  \Tz$) until  (at particular frequencies) the distortion term is balanced.

\subsubsection*{Double Compton scattering}
Due to the large entropy of the Universe at sufficiently high redshifts ($z\gtrsim \zdc$)
DC emission dominates over BR \citep{Danese1982}. The DC
emission coefficient can be given as \citep{Lightman1981, Thorne1981, Chluba2007a}
\beal
\label{eq:K_DC}
K_{\rm DC}(x, \Thz, \theta_{\rm e})
&=\frac{4\alpha}{3\pi}\,\Thz^2\,\times g_{\rm dc}(x,\Thz,  \The)
\Abst{,}
\end{align}
where $\alpha$ is the fine structure constant and $g_{\rm dc}(x, \Thz,  \The)$
is the effective DC Gaunt factor.  
In lowest order of the photon and electron energies the DC Gaunt factor factorizes \citep[see][for more details]{Chluba2005}. Furthermore, if the photon distribution is not too far from full equilibrium one can approximate $g_{\rm dc}(x, \Thz, \The)$ using a blackbody ambient radiation field and assuming that $\Te\sim \Tz$.
In this case one has \citep[e.g., see][]{Chluba2005, Chluba2007a}
\beal
\label{eq:g_DC}
g_{\rm dc}(x, \Thz, \The)
&\approx \frac{\mathcal{I}^{\rm pl}_4}{1+14.16\,\Thz}\times H_{\rm dc}(x)
\Abst{,}
\end{align}
where $\mathcal{I}^{\rm pl}_4=\int x^4 \nPl(\nPl+1)\id x= 4\pi^4/15\approx 25.976$.
Here we have included the first order relativistic correction in the photon temperature, however, this term only becomes significant at $z\gtrsim {\rm few}\times 10^6$.
%

\begin{figure}
\centering
\includegraphics[width=0.99\columnwidth]{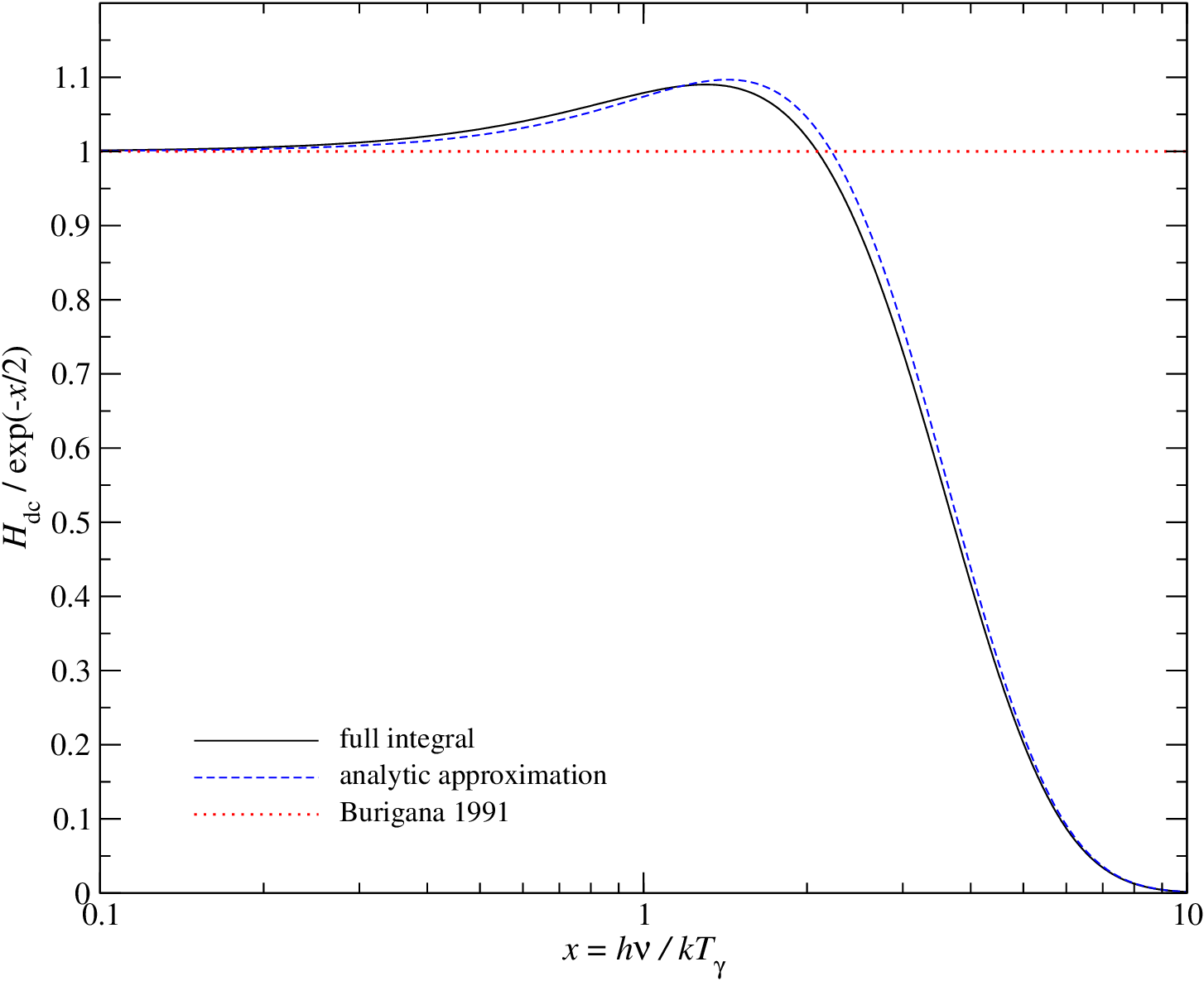}
\caption{Effective double Compton correction factor $H_{\rm dc}(x)$. We compare the result from a full integration of a blackbody spectrum with the approximation given by Eq.~\eqref{eq:H_DC_appr}. 
{
For comparison also the approximation of \citet{Burigana1991} is shown. Close to the maximum of the CMB blackbody spectrum the differences are $\sim 20\%-40\%$.}}
\label{fig:Gaunt_DC}
\end{figure}
%
The second factor in Eq.~\eqref{eq:g_DC} allows to go beyond the soft photon limit, for which $x\ll 1$ was assumed.
In lowest order $H_{\rm dc}(x)$ only depends on the ambient photon distribution, but is independent of the electron temperature.
It can be computed using \citep[see][for more details]{Chluba2005}
\beal
\label{eq:H_DC}
H_{\rm dc}(x)
&\approx \frac{1}{\mathcal{I}^{\rm pl}_4}\!
\int_{2x}^\infty x'^4 \nPl(x')[1+\nPl(x'-x)]\left[ \frac{x}{x'} H_{\rm G}\left(\frac{x}{x'} \right) \right] \id x'
\end{align}
where $H_{\rm G}(w)=[1-3y+3y^2/2-y^3]/y$ with $y=w[1-w]$. The factor $H_{\rm G}(w)$ was first obtained by \citet{Gould1984} to describe the corrections to the DC emissivity when going beyond the soft photon limit but assuming resting electrons\footnote{Note that $H_{\rm G}(w)$ is $1/2$ of $F(w)$ given by Eq.~(27) of \citet{Gould1984}. The factor of 2 is to avoid double counting of photons.}.
In the limit $x\rightarrow 0$ one finds $w\,H_{\rm G}(w) \rightarrow 1$, so that $H_{\rm dc}(x)\rightarrow 1$.

{%
Expression~\eqref{eq:H_DC} was also used in the work of \citet{Burigana1991}. There the approximation $H_{\rm dc}(x)\approx e^{-x\,\phi/2}$ was given.
However, as mentioned above with the assumptions leading to Eq.~\eqref{eq:H_DC} the electron temperature is irrelevant, and hence one should set $\phi\rightarrow 1$.
Furthermore, we reexamined the integral and found that for background photons that follow a blackbody spectrum
\beal
\label{eq:H_DC_appr}
H^{\rm pl}_{\rm dc}(x)
&\approx e^{-2x } \left[ 1+\frac{3}{2}x+\frac{29}{24} x^2+\frac{11}{16} x^3+\frac{5}{12} x^4\right]
\end{align}
provides a much better approximation to the full numerical result for $H_{\rm dc}$ (cf. Fig~\ref{fig:Gaunt_DC}).
This approximation was obtained by replacing $\nPl(x)\approx e^{-x}$ and neglecting the induced term in Eq.~\eqref{eq:H_DC}. Furthermore, the resulting expression was rescaled to have the correct limit for $x\rightarrow 0$.
In particular, for $x\gg 1$ Eq.~\eqref{eq:H_DC_appr} captures the correct scaling {$H_{\rm dc}(x)\sim x^4\,e^{-2x}$}.
However, since most of the photons are produced at low frequencies $x\ll1$ we do not expect any big difference because of this improved approximation.
Nevertheless, when using the old approximation we found that at early times the spectrum is erroneously brought into full equilibrium at very high frequencies, just by DC emission and absorption. 
}

We note here that if the distortions are not small, then in lowest order the correction to the DC emission can be accounted for by replacing $\nPl$ with the solution $n_x$ in the expression for $\mathcal{I}^{\rm pl}_4$.
{However, from the observational point of view it seems unlikely that distortions of interest ever exceeded the level $\Delta n_\nu/n_\nu\sim 10^{-3}$, even at $z\sim 10^7$. Therefore, the above approximation should be sufficient.}
{Of course this does not include DC emission from very high energy photons that are directly related to the energy injection process. However, in that case the simple approximation used above will anyhow need revision, {although} the total contribution to the photon production is still expected to be small.
}

\subsubsection*{Bremsstrahlung}
At lower redshifts ($z\lesssim \zdc$) Bremsstrahlung starts to become the main
source of soft photons. One can define the Bremsstrahlung emission coefficient
by \citep[cf.][]{Burigana1991, Hu1993}
\beal
\label{eq:K_BR}
K_{\rm BR}(x, \theta_{\rm e})
&=\frac{\alpha\,\lambda_{\rm e}^3}{2\pi\sqrt{6\pi}}\;\frac{\theta_{\rm e}^{-7/2}\,e^{-x\,\phi}}{\phi^3}
\sum_{\rm i} Z^2_{\rm i}\,N_{\rm i}\, g_{\rm ff}(Z_{\rm i},x,\theta_{\rm e})
\Abst{.}
\end{align}
Here $\lambda_{\rm e}=h/\me\,c$ is the Compton wavelength of the electron,
$Z_{\rm i},\,N_{\rm i}$ and $g_{\rm ff}(Z_{\rm i},x,\theta_{\rm e})$ are the
charge, the number density and the BR Gaunt factor for a nucleus of the atomic species i,
respectively. Various simple analytical approximations exists
\citep{Rybicki1979}, but nowadays more accurate fitting formulae, valid over a wide range of 
temperatures and frequencies, may be found in \citet{Nozawa1998} and \citet{Itoh2000}.
{
In comparison with the expressions summarized in \citet{Burigana1991} we find differences 
at the level of $10\%-20\%$ for small $x$.
}

In the early Universe only hydrogen and helium contribute to the BR Gaunt
factor, while the other light elements can be neglected. 
In the non-relativistic case the hydrogen and helium Gaunt factors are
approximately equal, i.e., $g_{\rm H, ff}\approx g_{\rm He, ff}$ to within a few
percent. Therefore assuming that the plasma is still fully ionized 
the sum in Eq.~\eqref{eq:K_BR} may be simplified to
$\sum \approx g_{\rm H, ff}\,N_{\rm b}$, 
where $N_{\rm b}$ is the baryon number density.
However, for percent accuracy one should take the full expressions for
$g_{\rm H, ff}$ and $g_{\rm He, ff}$ into account, which does not lead to any 
significant computational burden using the expressions of  \citet{Itoh2000}.

Furthermore, at redshifts $z\lesssim 7000-8000$ the plasma enters the 
different epochs of recombination. Therefore, the mixture of the different species ($N_{\rm e}$, \ion{H}{i}, \ion{H}{ii}, \ion{He}{i}, \ion{He}{ii}, \ion{He}{iii}) in the primordial medium has to be followed. 
We use the most recent computations of the recombination process including previously neglected physical corrections to the recombination dynamics according to \citet{Chluba2010b}.

\subsection{Evolution of the {ordinary matter} temperature}
\label{sec:electrons}
As mentioned above, the baryons in the Universe will all follow a Maxwell-Boltzmann distribution with a temperature that is equal to the electron temperature.
Caused by the Hubble expansion alone, the matter (electrons plus baryons) temperature would scale as $\Te~\propto~[1+z]^2$ {\citep{Zeldovich68}}, {implying $\Te<\Tg$}.
However, the electrons Compton scatter many times off background photons and therefore are pushed very close to the Compton equilibrium temperature
\beal
\label{eq:T_eq}
T_{\rm e}^{\rm eq}&=\frac{\mathcal{I}_4}{4\,\mathcal{G}_{3}}\,\Tz
=\Tz + \left[\frac{\Delta \mathcal{I}_4}{4\mathcal{G}_3}-\frac{\Delta \mathcal{G}_3}{\mathcal{G}_3}\right]\Tz,
\end{align}
within the (distorted) photon field, where $\mathcal{I}_4=\int x^4 n_x(n_x+1)\id x$.
If the CMB is undistorted, $\mathcal{G}^{\rm pl}_{3}=\mathcal{I}^{\rm pl}_4/4$ and $T_{\rm e}^{\rm eq} \equiv \Tz$, however, in the presence of spectral distortions generally $T_{\rm e}^{\rm eq} \neq \Tz$.
Here we also defined $\Delta \mathcal{I}_4= \mathcal{I}_4-\mathcal{I}^{\rm pl}_4$ and $\Delta \mathcal{G}_3= \mathcal{G}_3-\mathcal{G}^{\rm pl}_3$, which for numerical purposes is better, as the main terms can be cancelled out.

In addition, when spectral distortions are present, the matter cools/heats via DC and BR, and if some significant energy is released, this will in addition heat the medium.
Putting all this together the evolution equation for the electron temperature reads \citep[e.g., see][for a detailed derivation]{Chluba2005}
\beal
\label{eq:Te_equation}
\Abl{\rho_{\rm e}}{\tau}&=
%
%
%
\frac{t_{\rm C}\dot{Q}}{\alpha_{\rm h}\theta_{\gamma}}
+\frac{4\tilde{\rho}_\gamma}{\alpha_{\rm h}}[\rho^{\rm eq}_{\rm e}-\rho_{\rm e}] 
- \frac{4\tilde{\rho}_\gamma}{\alpha_{\rm h}}\mathcal{H}_{\rm DC, BR} (\rho_{\rm e})
-H\,t_{\rm C}\,\rho_{\rm e}.
\end{align}
Here we introduced $\rho_{\rm e}=1/\phi=\Te/\Tz$; the heat capacity of the medium\footnote{We neglected relativistic corrections to the heat capacity of the electrons, which at $z\sim 10^7$ would be of order percent {\citep{Chluba2005}}.}, $k\alpha_{\rm h}=\frac{3}{2}k[N_{\rm e}+N_{\rm H}+N_{\rm He}]=\frac{3}{2}k\,N_{\rm H}[1+f_{\rm He}+X_{\rm e}]$; the energy injection rate, $\dot{Q}$, which for example could be caused by some decaying particles (see Sect.~\ref{sec:injection_rates} for more details); 
and the energy density of the photon field in units of electron rest mass, $\tilde{\rho}_\gamma=\kappa_\gamma\theta_{\gamma}^4\,\mathcal{G}_{3}$, with $\kappa_\gamma=8\pi \lambda_{\rm e}^{-3}\approx \pot{1.760 }{30}\,{\rm cm}^{-3}$.
Furthermore, we defined $\rho^{\rm eq}_{\rm e}=T^{\rm eq}_{\rm e}/\Tz=1+\frac{\Delta \mathcal{I}_4}{4\mathcal{G}_3}-\frac{\Delta \mathcal{G}_3}{\mathcal{G}_3}$.
For numerical purposes it is better to rewrite the above equation in $\Delta \rho _{\rm e} =\rho _{\rm e}-1$, so that this formulation for $\rho^{\rm eq}_{\rm e}$ is very convenient.

The integral $\mathcal{H}_{\rm DC, BR}$ arises from the cooling/heating by DC and BR, and reads
\beal
\label{eq:DC_BR_cool}
\mathcal{H}_{\rm DC, BR}
&= \frac{1}{4\mathcal{G}_{3} \Thz}\int \Big[1-n_{x}\,(e^{\xe}-1)\Big]\times K(x, \Thz, \The) \id x.
\end{align}
For numerical purposes it again is useful to group terms of similar order, when calculating this integral numerically.
Usually, $\mathcal{H}_{\rm DC, BR}$ does not contribute very significantly to the total energy balance, but for energy conservation over the whole history it is crucial to obtain precise values for it.

\subsubsection{Approximate solution for $\Te$}
Because the timescale for Compton scattering is extremely short until $z\lesssim 800$, at high redshifts one can solve Eq.~\eqref{eq:Te_equation} assuming {\it quasi-stationarity}. Even at low redshifts this provides a very good first approximation for the correct temperature. Our full numerical computations confirm this statement.
Since the cooling from DC and BR is very small, it is furthermore possible to write $\mathcal{H}_{\rm DC, BR}(\rho^{i+1}_{\rm e})\approx \mathcal{H}_{\rm DC, BR}(\rho^{i}_{\rm e})+\partial_{\rho^{i}_{\rm e}} \mathcal{H}_{\rm DC, BR}(\rho^{i}_{\rm e})[\rho^{i+1}_{\rm e}-\rho^{i}_{\rm e}]$, so that
\bsub
\label{eq:Te_sol}
\beal
\rho^{i+1}_{\rm e}&=
\frac{\rho^{\rm eq,\ast}_{\rm e}}{1+\partial_{\rho^{i}_{\rm e}}\mathcal{H}_{\rm DC, BR}(\rho^{i}_{\rm e})+\Lambda}
\\
\rho^{\rm eq,\ast}_{\rm e}&=\rho^{\rm eq}_{\rm e}
+\frac{t_{\rm C}\dot{Q}}{4\tilde{\rho}_\gamma\theta_{\gamma}}
- \mathcal{H}_{\rm DC, BR} (\rho^i_{\rm e})
+\partial_{\rho^{i}_{\rm e}} \mathcal{H}_{\rm DC, BR}(\rho^{i}_{\rm e})\,\rho^{i}_{\rm e}
\\
\Lambda&=\frac{H\,t_{\rm C}\alpha_{\rm h}}{4\tilde{\rho}_\gamma} 
= \frac{3}{8}\,\frac{1+f_{\rm He} + X_{\rm e}} {\sigma_{\rm T}\,X_{\rm e}\,c\, \tilde{\rho}_\gamma} \, H
\end{align}
\esub
where $\rho^{j}_{\rm e}$ is the solution for $\rho_{\rm e}$ at time-step $\tau_j$. Also note that here we assumed that the spectral distortion from the time-step $\tau_i<\tau_{i+1}$ is used in all the integrals that have to be taken over the photon distribution. This will lead to some small error, which can be controlled by choosing an appropriate step-size.
Furthermore, as we will explain below, one can iterate the system at fixed time until the solution converges.

The derivative of $\mathcal{H}_{\rm DC, BR}$ with respect to $\rho^{i}_{\rm e}$ can be computed numerically, however, it is strongly dominated by the term connected with the exponential factor $e^{\xe}$ in the definition Eq.~\eqref{eq:DC_BR_cool}.
For numerical purposes we in general computed all integrals over the photon distribution by splitting the spectrum up into the CMB blackbody part and the distortion. This allows to achieve very high accuracy for the deviations form the equilibrium case, a fact that is very important when tiny distortions are being considered.
At low frequencies we again used appropriate series expansions of the frequency dependent functions.

\subsection{Changes in the expansion history and global energetics}
\label{sec:exp_hist}
Changes in the thermal history of the Universe generally also imply changes in the expansion rate. 
For example, considering a massive decaying relic particle, its contribution, $\rho_{\rm X}$, to the total energy density, $\rho_{\rm tot}$,  will decrease as time goes by. The released energy is then transferred to the medium and because of Compton scattering bulk\footnote{Since the temperature dependent contribution to the total matter energy density is negligibly small, well before recombination virtually all the injected energy will be stored inside the photon distribution.} of it ends up in the CMB background, potentially leaving a spectral distortion, and the matter with an increased temperature.
Because photons redshift, this implies that the expansion factor changes its redshift dependence, and hence the relation between proper time and redshift is altered.

To include this effect, one has to consider the changes in the energy density of the photons (or more generally all relativistic species) and the relic particles (or more generally the source of the released energy).
For simplicity we shall assume that the energy is released by some non-relativistic massive particle, but this can be easily relaxed. We then have
\bsub
\label{eq:energy_densities_dt}
\beal
\frac{1}{a^4}\Abl{a^4 \rho_\gamma}{t}&\approx \dot{Q}
\\
\frac{1}{a^3}\Abl{a^3 \rho_{\rm X}}{t}  & =-\dot{Q},
\end{align}
\esub
where $\rho_\gamma$ is the energy density of the photons, and $a \propto [1+z]^{-1}$ is the scale factor.

We emphasize that here it was assumed that the matter temperature re-adjusts, but that the corresponding change in the total energy density of the normal baryonic matter is negligible.
Furthermore, Eq.~\eqref{eq:energy_densities_dt} implies that all the injected energy ends up in the photon distribution. For energy injection at very late times, close to the end of  recombination, this is certainly not true anymore.
In that case, it will be important to add the electron temperature equation to the system, and then explicitly use the Compton heating term for the photons. In this way one has a more {accurate} description of the heat flow between electrons and photons, and only the spectral distortion is neglected in the problem.
However, in terms of the global energetics this should not make any large difference.
As we will see below, in the case of adiabatically cooling electrons it turns out that such treatment is necessary in order to define the correct initial condition (see Sect.~\ref{sec:cooling}).

Equation~\eqref{eq:energy_densities_dt} has the solutions $\rho_\gamma(z)= \rho^{(0)}_{\gamma}(z)+\Delta \rho_{\gamma}(z)$ and $\rho_{\rm X}(z)= \rho^{(0)}_{\rm X}(z)+\Delta \rho_{\rm X}(z)$ with 
\bsub
\label{eq:energy_densities_sol}
\beal
\label{eq:energy_densities_sol_a}
\Delta \rho_{\gamma}(z)&=-\frac{1}{a^4(z)} \int_0^z \frac{a^4(z')\,\dot{Q}\id z'}{H(z')[1+z']}
\nonumber\\
&\equiv -\rho^{(0)}_{\gamma}(z) \int_0^z \frac{\dot{Q}}{\rho^{(0)}_{\gamma}(z')}
\frac{\id z'}{H(z')[1+z']}
\\[1mm]
\Delta \rho_{\rm X}(z)&=\frac{1}{a^3(z)} \int_0^z \frac{a^3(z')\,\dot{Q}\id z'}{H(z')[1+z']}
\end{align}
\esub
where $\rho^{(0)}_{\gamma} =\rho_{\gamma, 0} [1+z]^4\approx 0.26\,[1+z]^4\,{\rm eV\,cm^{-3}}$ is the unperturbed photon energy density, and similarly $\rho^{(0)}_{\rm X} =\rho_{\rm X, 0} [1+z]^3$.
Here $\rho_{\rm i, 0}$, with $\rm i\in \{\gamma, \rm X\}$, denotes the present day values of the corresponding energy densities.
For decaying relic particles with lifetimes shorter than the Hubble time this consequently means $\rho_{\rm X, 0}=0$.
We will use Eq.~\eqref{eq:energy_densities_sol_a} later to compute the appropriate initial condition for the energy density of the photon field (see Sect.~\ref{sec:initial_cond}).

The integrals in Eq.~\eqref{eq:energy_densities_sol} themselves depend on $H(z)$, however, assuming that the total change in the energy density is small, one can use  $H(z)\approx H^{(0)}(z)$, where $H^{(0)}(z)$ denotes the unperturbed expansion factor with $\dot{Q}=0$.
Then the modified expansion factor is simply given by $H(z)=H^{(0)}(z)+\Delta H(z)$ with
\beal
\label{eq:H_corr}
\frac{\Delta H(z)}{H^{(0)}(z)}&\approx
\,\frac{\Delta \rho_{\gamma}(z)+\Delta \rho_{\rm X}(z)}{2\,\rho^{(0)}_{\rm tot}(z)}
\nonumber
\\
&=\frac{\rho^{(0)}_{\gamma}(z)}{2\,\rho^{(0)}_{\rm tot}(z)}
 \int_0^z \frac{z'-z}{1+z}\,\frac{\dot{Q}}{\rho^{(0)}_{\gamma}(z')}\frac{\id z'}{H^{(0)}(z')[1+z']}
\end{align}
where $\rho^{(0)}_{\rm tot}(z)=3\,[H^{(0)}(z)]^2 / 8\pi G$ is the total energy density of the unperturbed Universe.
This equation shows clearly that the change in the total energy density is arising from the fact that energy released by non-relativistic particles is transformed into contributions to the photon energy density: if photons had the same adiabatic index as matter this would not change anything, a conclusion that can be easily reached when thinking about energy conservation. 
However, the additional redshifting of photons implies a small net change in the expansion rate.

In terms of the thermalization problem, the Hubble factor explicitly only appears in the evolution equation for the matter temperature, Eq.~\eqref{eq:Te_equation}, while otherwise it is only present implicitly, via $\tau_{\rm C}$.
For $\tau_{\rm C}$ it can be neglected, since this will only lead to an extremely small shift in the relation between proper time and redshift that corrects the small spectral distortion, $\Delta n_x$. This conclusion can be reached as well when transforming to redshift as time variable, since then $H$ appears in all terms for the photon Boltzmann equation, where $\Delta H$ is a correction-to-correction.
A similar reasoning holds for the matter temperature equation. There one could in principle replace $H$ with $H^{(0)}+\Delta H$, however, looking at the quasi-stationary solution for $\rho_{\rm e}$, Eq.~\eqref{eq:Te_sol}, it is evident that this will only lead to a minor {difference}. We checked this statement numerically and found the addition of $\Delta H$ to be negligible.
We conclude that the correction to the Hubble factor enters the problem in second order.

We also comment that above we made several simplifying assumptions about the form of the energy release.
For example, if all the energy is released in form of neutrinos, then both the CMB spectrum and the matter temperature remain unaffected, however, the Hubble factor will again change, assuming that the energy density of the source of the energy release itself does not scale like $[1+z]^4$.
However, considering this problem in more detail is beyond the scope of this paper and we do not expect major changes in our conclusions for this case.

\subsection{Energy injection rates for different processes}
\label{sec:injection_rates} 
Having a particular physical mechanism for the energy release in mind, one can specify the energy injection rate associated with this process.
Here we assume that most of the energy is going directly into heating of the medium, but only very few photons are produced around the maximum of the CMB energy spectrum, or reach this domain during the evolution of the photon distribution.
Also, here we do not treat the possible up-scattering of CMB photons by the decay products, nor do we allow for Bremsstrahlung emission from these particles directly in the regime of the CMB. 
For a consistent treatment of the thermalization problem it will be very interesting to consider these aspects in more detail, however, this is far beyond the scope of this paper.

With these simplifications it is sufficient to provide $\dot{Q}$ in Eq.~\eqref{eq:Te_equation} for each of the processes and then solve the thermalization problem assuming full equilibrium initial conditions.
{Below we first discuss the effect of adiabatically cooling electrons and the energy injection caused by dissipation of acoustic waves in the early Universe.
Furthermore,  we give $\dot{Q}$ for annihilating and decaying particles, as well as for short bursts of energy injection.
The former two processes {\it are present} even for the standard cosmological model, while the latter three are subject to strong uncertainties and require non-standard extensions of the physical model.
}

\subsubsection{Adiabatically cooling electrons {and baryons}}
\label{sec:Q_cooling} 
As mentioned above, if the interactions with the CMB are neglected, the temperature of the matter scales as $\Te\propto (1+z)^2$ once the baryons and electrons become non-relativistic.
{At very low redshifts ($z\lesssim 200$) it is well known that this difference in the adiabatic index of baryonic matter and radiation leads to a large difference in the CMB and electron temperature \citep{Zeldovich68} {with $\Te<\Tg$}, which is also very important for the formation of the global 21cm signal from the epoch of reionization \citep[e.g., see][and references therein]{Pritchard2008}.}
However, at higher redshifts Compton scattering couples the electrons tightly to the CMB photons, implying that the baryonic matter in the Universe must continuously {\it extract} energy from the CMB in order to establish $\Te\sim \Tg$ until Compton scattering eventually becomes very inefficient at $z\lesssim 200$.
Consequently, this should lead to a tiny spectral distortion of the CMB, tiny because the heat capacity of the CMB is extremely large in comparison to the one of matter.
A simple estimate for this case can be given by equating the Compton heating term with the term from the adiabatic Hubble expansion: 
%
$\id \tilde{\rho}_\gamma/\id t = - H\alpha_{\rm h}\Thz\equiv [\me c^2]^{-1}\id E/\id t $. 
This term is already included in Eq.~\eqref{eq:Te_equation}, so that no extra $\dot{Q}$ has to be given.
With this the total extraction of energy from the CMB is roughly given by 
\beal
\label{eq:Drho_tot_cool}
\left.\frac{\Delta \rho_\gamma}{\rho_\gamma} \right|_{\rm cool}\!\!
&\approx
-\frac{\alpha_{\rm h}\Thg}{\tilde{\rho}_{\gamma, 0}}\,\ln\left[\frac{1+\zs}{1+\ze}\right]
\approx 
-\pot{5.6}{-10}\ln\left[\frac{1+\zs}{1+\ze}\right].
\end{align}
For the evolution from $z=2\times 10^7$ until $z\sim 10^3$ this implies a relative energy extraction of $\left.\frac{\Delta \rho_\gamma}{\rho_\gamma} \right|_{\rm cool}\sim -\pot{5.6}{-9}$, {which is only a few times below the claimed sensitivity of {\sc Pixie} \citep{Kogut2011}}.

Here it is interesting to observe that well before the recombination epoch the effective cooling term caused by the electrons has the same redshift dependence as annihilating matter (see below), i.e. $\dot{Q}\propto (1+z)^6$.
This implies that the type of distortion will be similar in the two cases, however, the distortion caused by the adiabatic cooling of {ordinary matter} has opposite sign (see Fig.~\ref{fig:Te_x_cool}), {and tend to decrease each other}.

Furthermore, with Eq.~\eqref{eq:Drho_tot_ANNl} one can estimate the annihilation efficiency that is equivalent to the {matter} cooling term. It turns out that if some annihilating relic particle is injecting energy with an efficiency $f_{\rm ann}\sim \pot{1.6}{-23}\,\rm eV \, s^{-1}$ then the net distortion introduced by both processes should practically vanish.
As we will show below, our computations confirm this estimate (see Fig.~\ref{fig:Te_x_ann}), demonstrating the precision of our numerical treatment.

\subsubsection{Dissipation of acoustic waves}
\label{sec:Q_waves} 
Acoustic waves in the photon-baryon fluid dissipate part of their energy because of diffusion damping \citep{Silk1968}.
As first {estimated by \cite{Sunyaev1970diss}, and later refined by} \citet{Daly1991} and \citet{Barrow1991}, this effect should lead to a spectral distortion in the CMB which depends on the spectral index of scalar perturbation, $n_{\rm S}$, and therefore could {potentially} allow constraining inflationary models {\citep{Mukhanov1981} using future CMB spectral measurements} \citep{Kogut2011PIXIE}. 
Following \citet{Hu1994}, the heating rate in a radiation dominated Universe can be cast into the form
\beal
\label{eq:DE_Dt_waves}
\left. \dot{Q} \right|_{\rm ac}
&= \mathcal{F}(n_{\rm S})\, \tilde{\rho}_\gamma\,H(z) [1+z]^{3(n_{\rm S}-1)/2},
\end{align}
with\footnote{We assumed that all cosmological parameters aside from $\nS$ have their standard values. Furthermore, as in \citet{Hu1994} we simply used the {\sc Cobe} normalization for our estimate.} $\mathcal{F}(x)\approx \pot{5.2}{-9}\,x^{-2.1}\,[0.045]^{x-1}$ for $x\sim 1$.
For $n_{\rm S}=1$ this implies a total energy dissipation of $\left.\frac{\Delta \rho_\gamma}{\rho_\gamma} \right|_{\rm ac}\sim \pot{3.9}{-8}$ between $z\sim \pot{2}{6}$ and $10^3$.
This is about 10 times larger than the energy {extracted} in the case of  adiabatically cooling {matter} (Sect.~\ref{sec:Q_cooling}) over the same period.
One therefore expects the associated distortion to be about an order of magnitude larger.

Also, like in the case of cooling {matter} and annihilating particles one has $\dot{Q}\propto (1+z)^6$ for $\nS\sim 1$.
The equivalent effective annihilation rate in this case is $f_{\rm ann}\sim \pot{1.5}{-22}\,{\rm eV\,s^{-1}}$.
This suggests that the bounds on $f_{\rm ann}$ derived from the CMB anisotropies \citep{Slatyer2009, Galli2009, Huetsi2009, Huetsi2011} already now might be in tension with this heating mechanism, however, the energy release from acoustic waves only goes into heating of the medium, while for annihilating particles extra ionizing and exciting photons are produced, implying that the effect on the ionization history could be much smaller here.

We therefore computed the modification to the recombination history caused by this process using {\sc CosmoRec}, and found a small change of $\Delta N_{\rm e}/N_{\rm e}\lesssim 0.06\%$ around the maximum of the Thomson visibility function \citep{Sunyaev1970} close to $z\sim 1100$, while the electron fraction in the freeze-out tail indeed was affected by several percent.
The changes in the freeze-out tail of recombination could have important consequences for the CMB power spectra, however, at late times, close to recombination, the simple formula Eq.~\eqref{eq:DE_Dt_waves} should be modified to take into account deviations from tight-coupling and also the assumption of radiation domination is no longer valid, implying that the rate of energy release is overestimated at those epochs.
{Furthermore, in this case one should more carefully treat the fraction of energy that directly goes into heating of the matter as opposed to terms appearing directly in the photon field.}
We believe that the overall effect on the recombination dynamics arising from this process is negligible.
Nevertheless, refined computations of the energy release rate might become important for precise computations of this problem, but for the purpose of this paper the above simplification will be sufficient.

\subsubsection{Annihilating particles}
\label{sec:injection_rates_annihil} 
The rate at which annihilating particles inject energy scales like $\dot{E} \propto N_{\rm X}^2$ with the number density of the annihilating particles.
To parametrize the rate of energy injection we use \citep[compare][]{Padmanabhan2005, Chluba2010a}
\beal
\label{eq:DE_Dt_ANNl}
\left. \Abl{E}{t}\right|_{\rm ann}
&= [1-f_{\nu}] f_{\rm ann} N_{\rm H} (1+z)^3,
\end{align}
where $f_{\rm ann}$ has dimensions ${\rm eV \, s^{-1}}$, and $f_{\nu}$ is the fraction of the total energy carried away by neutrinos. 
$f_{\rm ann}$  is related to the abundance of the annihilating particle, its mass, and the thermally averaged annihilation cross section, however, here we treat all these dependencies with one parameter.

Current measurements of the CMB anisotropies already tightly constrain $f^\ast_{\rm ann}=[1-f_{\nu}] f_{\rm ann}\lesssim \pot{2}{-23}{\rm eV \, s^{-1}}$ \citep{Slatyer2009, Galli2009, Huetsi2009, Huetsi2011}.
This limit is orders of magnitude stronger than the one obtained from {\sc Cobe/Firas} \citep[e.g., see][]{McDonald2001, Chluba2010a}, implying that there is only little room for CMB spectral distortions from annihilating matter, unless a large fraction of the energy is actually released in form of neutrinos, i.e. $f_\nu\sim 1$.

With Eq~\eqref{eq:DE_Dt_ANNl} it then follows
\beal
\label{eq:Drho_Dt_ANNl}
\left. \dot{Q} \right|_{\rm ann}
&=\frac{ g_{\rm h}(z)}{\me c^2} \left. \Abl{E}{t}\right|_{\rm ann},
\end{align}
where we parametrized the fraction of the total energy going into heating using $g_{\rm h}(z)$. 
Similar to \citet{Padmanabhan2005} we will adopt $g_{\rm h}(z)\approx [1+f_{\rm He}+ 2(X_{\rm p}+f_{\rm He}(Z_{\ion{He}{ii}}+Z_{\ion{He}{iii}})]/3[1+f_{\rm He}]$.
Here, respectively $X_{\rm p}$, $f_{\rm He}Z_{\ion{He}{ii}}$ and $f_{\rm He}Z_{\ion{He}{iii}}$ denote the free proton, singly and doubly ionized helium ions relative to the total number of hydrogen nuclei, $N_{\rm H}$, in the Universe.
At high redshifts, well before recombination starts ($z\gtrsim 8000$), one has $X_{\rm p}\sim Z_{\ion{He}{ii}}\sim Z_{\ion{He}{iii}}\sim~1$, such that $g_{\rm h}(z) \sim 1$, while during hydrogen recombination $g_{\rm h}(z)$ drops rapidly towards $g_{\rm h}(z)\sim 1/3$.
At low redshifts, a significant part of the released energy is going into excitations and ionizations of ions, rather than pure heating, explaining why $g_{\rm h}(z)<1$ \citep{Shull1985, Chen2004}.

To estimate the total amount of energy that is injected relative to the energy density of the CMB one can simply assume radiation domination ($1/H\sim \pot{4.79}{19}[1+z]^{-2}\,\rm s$) and compute the integral $\Delta \rho_\gamma/\rho_\gamma= \int \Abl{E}{t}/\rho_\gamma\id t$.
In the case of annihilating matter this yields 
\beal
\label{eq:Drho_tot_ANNl}
\left.\frac{\Delta \rho_\gamma}{\rho_\gamma} \right|_{\rm ann}
%
%
&\approx
\pot{5.2}{-9}\left[\frac{[1-f_{\nu}] \, f_{\rm ann}}{\pot{2}{-23}\,{\rm eV s^{-1}}}\right]
\left[\frac{1-\Yp}{0.75}\right]
\left[\frac{\Omega_{\rm b}h^2}{0.022}\right].
\end{align}
Here we have assumed that $f_{\rm ann}^\ast$ is independent of redshift. Furthermore, we assumed that before $z\sim \pot{2}{6}$ everything thermalizes \citep{Burigana1991}, while effectively no energy is transferred to the CMB once recombination ends ($z\sim 10^3$).

This estimate shows\footnote{Here we neglect corrections because of partial thermalization of distortions at $z\lesssim \pot{2}{6}$} that for $f^\ast_{\rm ann}\sim \pot{2}{-23}{\rm eV \, s^{-1}}$ one can expect distortions at the level of $\Delta I_\nu/I_\nu \sim 10^{-9}$.
Interestingly, part of the energy is released at very high $z$, where one expects a $\mu$-type distortion \citep{Illarionov1975, Illarionov1975b}, however, also at low redshifts some energy is released.
This in contrast should lead to a $y$-type distortion \citep{Illarionov1975, Illarionov1975b, Hu1993}, so that in total one expects a mixture of both. We will compute the precise shape of the distortion for some cases below (see Sect.~\ref{sec:annihil}), confirming these statements.

\subsubsection{Decaying relic particles}
\label{sec:injection_rates_decay} 
For decaying particles the energy injection rate is only proportional to the density of the particle.
Motivated by  the parametrization of \citet{Chen2004} we therefore write
\beal
\label{eq:DE_Dt_dec}
\left. \Abl{E}{t}\right|_{\rm dec}
&= f^\ast_{\rm X} \, \Gamma_{\rm X}\, N_{\rm H} \, e^{-\Gamma_{\rm X} t}.
\end{align}
Again all the details connected with the particle, its mass and abundance (here relative to the number of hydrogen atoms in the Universe) are parametrized by $f^\ast_{\rm X}$.
The exponential factor arises because the relic particles rapidly disappear in the decay process once the age of the Universe is comparable to its lifetime $t_{\rm X}\sim 1/\Gamma_{\rm X}$.

For particles with very short lifetime a large amount of energy could be injected, without violating any of the CMB constraints, since at early times the thermalization process is very efficient \citep[e.g. see][]{Sunyaev1970mu, Hu1993b}.
This implies that in particular early energy release (at $z\gtrsim 10^4$) could still lead to interesting features in the CMB spectrum.
Again it will be important to see how the characteristics of the distortion change from $\mu$-type to $y$-type as the lifetime of the particle increases.

Under the above circumstances the constraints derived from {\sc Cobe/Firas} to date are still comparable to those from other probes \citep{Kusakabe2006, Kohri2010, Kogut2011PIXIE}, at least for particles with intermediate lifetimes close to $\tX\sim 10^{9}-10^{10}\,{\rm sec}$.
{However, these constraints could be significantly improved with future spectral measurement of the CMB \citep{Kogut2011}.}
In addition, constraints obtained with the CMB temperature and polarization anisotropies strongly limit the possible amount of energy injection by decaying particles with long lifetimes \citep{Chen2004, Zhang2007}.

To estimate the total amount of energy injected by the particle we again take the integral $\Delta \rho_\gamma/\rho_\gamma= \int \Abl{E}{t}/\rho_\gamma\id t$. This then leads to 
\beal
\label{eq:Drho_tot_dec}
\left.\frac{\Delta \rho_\gamma}{\rho_\gamma} \right|_{\rm dec}\!\!
&\!\approx\!
10^{-5}\!\left[\frac{f^\ast_{\rm X}}{\pot{8}{5}\,{\rm eV}}\right]
\left[\frac{1-\Yp}{0.75}\right]
\left[\frac{\Omega_{\rm b}h^2}{0.022}\right]
\left[\frac{1+z_{\rm X}}{\pot{5}{4}}\right]^{-1} \!\!\mathcal{J},
\end{align}
where for simplicity we once again assumed radiation domination (implying $t=1/2H\approx \pot{2.40}{19}[1+z]^{-2}\,\rm s$).
Furthermore, we introduced the redshift corresponding to the lifetime of the particle, $z_{\rm X}\sim \sqrt{\Gamma_{\rm X}\,t_0}$, with $t_0\approx \pot{2.40}{19}\,\rm s$.
In addition, we defined the integral $\mathcal{J}=\frac{2}{\sqrt{\pi}}\int^{z^2_{\rm X}}_0\!\! \id \xi\sqrt{\xi}\,e^{-\xi}$, which for $z_{\rm X}\gg 0$ is very close to unity.
Ignoring corrections because of the thermalization process\footnote{We will include these in our estimates of Sect.~\ref{sec:decay_estimate}.} \citep[e.g., see][]{Hu1993b, Chluba2005}, Eq.~\eqref{eq:Drho_tot_dec} implies that for particles with lifetimes of $t_{\rm X}\sim 10^9 \rm s$ the limits from {\sc Cobe/Firas}  are $f^\ast_{\rm X}\lesssim 10^{6}\rm eV$ for our parametrization.
Note that $\left. \dot{Q} \right|_{\rm dec}$ can be found by replacing $\,\left. \Abl{E}{t}\right|_{\rm ann}$ in Eq.~\eqref{eq:Drho_Dt_ANNl} with the expression Eq.~\eqref{eq:DE_Dt_dec}.
%

\subsubsection{Quasi-instantaneous energy release}
\label{sec:injection_rates_delta} 
As next case we discuss the possibility of short bursts of energy release. 
%
We model this case using a narrow Gaussian in cosmological time around the heating redshift $z_{\rm h}$ with some width $\sigmah$:
\beal
\label{eq:DE_Dt_delta}
\left. \Abl{E}{t}\right|_{\delta}
&= f^\ast_{\delta} [1+z_{\rm h}]^4\, 
\frac{e^{- [t-t_{\rm h}]^2/2\sigmah^2}}{\sqrt{2\pi \sigmah^2}}.
\end{align}
%
With this  parametrization, assuming sufficiently small $\sigmah$, one has 
\beal
\label{eq:Drho_tot_delta}
\left.\frac{\Delta \rho_\gamma}{\rho_\gamma} \right|_{\delta}
&\approx 10^{-5} \,\!\left[\frac{f^\ast_{\delta}}{\pot{2.6}{-6}\,{\rm eV}}\right].
\end{align}
Computationally it is demanding to follow very short bursts of energy injection, however, our code allows to treat such cases by setting the step-size appropriately.
Again $\left. \dot{Q} \right|_{\delta}$ can be found by replacing $\,\left. \Abl{E}{t}\right|_{\rm ann}$ in Eq.~\eqref{eq:Drho_Dt_ANNl} with the expression Eq.~\eqref{eq:DE_Dt_delta}.

With the above formula we can also study the effect of going from quasi-instantaneous to more extended energy release by changing the value of $\sigmah$. For short bursts we chose $\sigmah\sim 0.05\,t_{\rm h}$, but we also ran cases with $\sigmah\sim 0.25\,t_{\rm h}$ to demonstrate the transition to {more} extended energy release (see Fig.~\ref{fig:Te_x_delta_PIXIE}).

\section{Results for different thermal histories}
\label{sec:en_inj}
In this section we discuss some numerical results obtained by solving the coupled system of equations described above. 
We first give a few details about the new thermalization code, {\sc CosmoTherm}, which we developed for this purpose.
It should be possible to skip this section, if one is not interested in computational details.
In Sect.~\ref{sec:cooling}-\ref{sec:end_senarios} we then discuss several physical scenarios and the corresponding potential CMB spectral distortion.

\subsection{Numerical aspects}
\label{sec:numerics}
To solve the cosmological thermalization problem we tried two approaches.
In the first we simplified the computation with respect to the dependence on the electron temperature. At each time-step the solution for the photon distribution is obtained using the partial differential equation (PDE)  solver developed in connection with the cosmological recombination problem \citep{Chluba2010b}, which allows us to setup a non-uniform grid spacing using a second order semi-implicit scheme in both time and spatial coordinates.
However, we assume that the differential equation for the electron temperature can be replaced with the quasi-stationary approximation, Eq.~\eqref{eq:Te_sol}, and solved once the spectrum at $\tau_{i+1}$ is obtained.
Since both the evolution of the electron temperature and the spectral distortions is usually rather slow one can iterate the system of equations for fixed $\tau_{i+1}$ in a predictor-corrector fashion, until convergence is reached, and then proceed to the next time.

In the second approach, we modified the PDE solver of \citet{Chluba2010b} to include the additional integro-differential equation for the electron temperature. 
This results in a banded matrix for the Jacobian of the system, which has one additional off-diagonal row and column because of the integrals over the photon field needed for the computation of the electrons temperature.
Such system can be solved in $\mathcal{O}(N\log N)$ operations, and hence only leads to a small additional computational burden in comparison to the normal PDE problem. 
We typically chose $N\sim 4000-6000$ grid-points logarithmically spaced in frequency over the range $x\sim 10^{-5}$ to $2\times10^2$.
However, for precise conservation of energy we implemented a linear grid at frequencies above $x=0.1$.
Furthermore, we distributed more points in the range $x=0.1$ to $15$, where most of the total energy in the photon field is stored. 
Even for about $4000$ points we achieved very good conservation of energy and photon number.
We tested the convergence of the results increasing $N$ up to $4-5$ times, as well as widening/narrowing the frequency range.

We then included the coupled PDE/ODE system into the ODE stepper of {\sc CosmoRec} \citep{Chluba2010}. This stepper is based on an implicit Gear's method up to six order and allows very stable solution of the problem in time, with no need to iterate extensively. Furthermore, this approach in principle enables us to include the full non-linearity of the problem in $\Delta n_x$, however, for the formulation given above this did not make any difference.

We found that the two approaches described above give basically identical results however the second is significantly faster and more reliable for short episodes of energy injection.
In the current version of {\sc CosmoTherm} this approach was adopted.
For typical settings one execution takes a few minutes on a single core, however, for convergence tests one execution with $N\sim 20000$ and small step-size ($\Delta z/z\sim 10^{-4}$) took several hour.
Also for cases with quasi-instantaneous energy release a small step-size was necessary during the heating phase, such that the runtime was notably longer.

We also would like to mention that in the current implementation of {\sc CosmoTherm} we assume that the recombination history is not affected by the energy injection process and therefore can be given by the standard computation carried out with {\sc CosmoRec} \footnote{{\sc CosmoRec} is available at \url{www.Chluba.de/CosmoRec}.}. 
This assumption is incorrect at low redshifts for cases with late energy release, as small amounts of energy can have a significant impact on the ionization history.
We plan to consider this aspect of the problem in some future work.

\subsubsection{Initial condition}
\label{sec:initial_cond}
We performed the computations assuming full equilibrium at the starting redshift, $z_{\rm s}$. Usually we ran our code starting at $z_{\rm s}=\pot{4}{7}$.
However, in order to end up with a blackbody that has an energy density corresponding to the measured CMB temperature, $T_0$, today, we had to modify the temperature of the photons and baryons at $z_{\rm s}$.
No matter if the injected energy is fully thermalized or not, the energy density of the photon field before the energy  injection starts should be very close to $\rho_\gamma(z_{\rm s})\sim\rho^{(0)}_\gamma(z_{\rm s})-\Delta \rho(z_{\rm s})$, where $\Delta \rho(z_{\rm s})$ is approximately given by Eq.~\eqref{eq:energy_densities_sol_a}.
For more precise initial conditions, one has to numerically solve the problem for the global energetics (see Sect.~\ref{sec:exp_hist}) prior to running the thermalization code.

Although for extremely small energy injections neglecting $\Delta \rho(z_{\rm s})\neq 0$ makes a difference in the final effective temperature at the ending redshift, $\ze$, that is well below the current limit of $1\,$mK for $T_0$ \citep{Fixsen2011}, for consistent inclusion of the energy injection one should compute the initial effective temperature of the CMB by $T^\ast_{\gamma}=\Tz\left[1-\Delta \rho(z_{\rm s})/\rho^{(0)}_\gamma(z_{\rm s})\right]^{1/4}$. 
We then scaled $\Tz$ such that at $\zs$ it is identical to $T^\ast_{\gamma}$. This allows setting $\Delta n_x=0$ which avoids spurious photon production.
At the end of the computation one should therefore expect $T^\ast_{\gamma}(\ze)\sim T_0 [1+z_{\rm e}]>\Tz(\ze)$, if $\dot{Q}>0$.
This furthermore provides a good test for the conservation of photons and energy in the code.

\subsubsection{Boundary conditions}
\label{sec:bound_cond}
Finally, to close the system, we have to define the boundary conditions for the photon spectrum at the upper and lower frequencies. One reason for us choosing such a wide range over $x$ is that because of efficient BR emission and absorption even down to $\ze=200$ the photon distribution is always in full equilibrium with the electrons at the lower boundary. This allows us to explicitly set the spectrum to a blackbody with temperature $\Te$.
At at the upper boundary we also used this condition, which for most of the evolution is fulfilled identically, just because of Compton scattering pushing the distribution into kinetic equilibrium in the very far Wien tail.
However, at very low redshifts this Dirichlet boundary condition is not fully correct, as the timescale on which kinetic equilibrium is reached drops. Fortunately, it turns out that this is not leading to any important difference in the solution.

We confirmed this statement by simply forcing $\Delta n_x=0$ at both the upper and lower boundary and found that this did not alter the solution in the frequency domain of interest to us.
Furthermore, we tried von Neumann boundary conditions of the type $\partial_x n_x + \phi n_x(1+n_x)=0$. 
This condition is equivalent to the chosen Dirichlet boundary condition, however, the amplitude of the distortions is left free and only the shape is assumed to be given by a blackbody with temperature $\Te$ (i.e., $n_x=A(t)/\xe$ for $x\rightarrow 0$ and $n_x=A(t)\,e^{-\xe}$ for $x\rightarrow \infty$). Again this choice for the boundary condition did not affect the results well inside the computational domain significantly.
The current version of the code has both options available.

\subsection{Simple analytic description for $\mu$- and $y$-type distortion}
\label{sec:analytic_mu_y}
Before looking in more detail at the numerical results, here we give a very brief (and even rather crude) analytical description for $\mu$- and $y$-type distortions.
The formulae presented here are motivated by the early works on this problem \citep{Zeldovich1969, Sunyaev1970mu, Zeldovich1972, Sunyaev1974, Illarionov1974, Illarionov1975, Illarionov1975b}, where here we use the shapes of the different components as {\it templates} to approximate the obtained solutions.
Some more advanced approximations can be found in \citep{Burigana1995}.

%
As mentioned in the introduction, for $\mu$-type distortions the CMB spectrum is described by a Bose-Einstein distribution with frequency dependent chemical potential, $n_x=1/[e^{x+\mu(x)}-1]$.
Photon production at very low frequencies always pushed the spectrum to a blackbody with temperature of the matter, and at high frequencies, because of efficient Compton scattering, a constant chemical potential is found.
These limiting cases can be described by \citep[e.g., see][]{Sunyaev1970mu, Illarionov1975}
\beal
\label{eq:mu_x}
\mu(x)&=\mu_\infty\,e^{-\xc/x} + x[\phi_{\rm f} - 1],
\end{align}
with the modification that for $x\ll 1$ and $x\gg1$ we have the additional freedom $x+\mu(x)\equiv x\,\phi_{\rm f}$, which allows us to renormalize to a blackbody with $T\neq \Tz$.
Here $\mu_\infty$ is the constant chemical potential at very large $x$, and $\xc$ is the critical frequency which depends on the efficiency of photon production versus Compton scattering.
%
%
In terms of the brightness  temperature this translates into
%
\beal
\label{eq:T_mu_x}
T(x)&=\frac{\Tz}{\phi_{\rm f}+\frac{\mu_\infty}{x}\,e^{-\xc/x}},
\end{align}
which both at very large and small $x$ implies $T(x)=\Tz/\phi_{\rm f}$.

%
The case of $y$-type spectral distortions is also known in connection with SZ clusters \citep{Zeldovich1969}. The spectral distortion is given by
\beal
\label{eq:y_x}
\Delta n_x ^{\ye}&=\ye\,\frac{x\,e^x}{[e^x-1]^2}\left[x\,\frac{e^x+1}{e^x-1}-4\right],
\end{align}
where $\ye$ is the Compton $y$-parameter, which directly depends in the Thomson optical depth and the difference between the electrons and photon temperature.
Again for us only the shape of the $y$-distortion really matters, and we will determine the value of $\ye$ from case to case.
Most importantly, at small values of $x$ one has $\Delta n ^{\ye}_x/\nPl \equiv \Delta T/T \approx -2\ye$, while at high $x$ it follows that $\Delta n ^{\ye}_x/\nPl\approx \ye\,x^2$ and hence $\Delta T/T \approx \ye\,x$.
In the case of SZ clusters one always has $\ye>0$, since the electron temperature is many orders of magnitude larger than the CMB temperature. 
However, as we will see below (Sect.~\ref{sec:cooling}), in the case of the expanding Universe one can also encounter negative values for $\ye$, even in the standard cosmological picture.

%
As last component we add a distortion that could be caused by the free-free process at very low frequencies.
This can be simply approximated by \citep{Zeldovich1972}
\beal
\label{eq:Dn_ff_x}
\Delta n_x^{\rm ff}&=\yff\,\frac{e^{-x}}{x^3},
\end{align}
where $\yff$ parametrizes the amplitude of the free-free term.

{In the next few sections} we present the distortions as a frequency dependent effective temperature, or brightness temperature, which is found by comparing the obtained spectrum at each frequency with the one of a blackbody: 
\beal
\label{eq:T_x_def}
T(x)=\frac{\Tz\,x}{\ln\left(1+n^{-1}_x\right)}.
\end{align}
For approximations of the final distortion we will simply combine the three components described above:
\beal
\label{eq:n_x_app}
n^{\rm app}_x&=
\frac{1}{e^{x+\mu(x)}-1}
+ \Delta n^{\ye}_x + \Delta n^{\rm ff}_x,
\end{align}
with the parameters $\mu_{\infty}, \xc, \phi_{\rm f}$, $\ye$ and $\yff$ chosen accordingly. 
If not stated otherwise, parameters that are not listed were set to zero.

\begin{figure}
\centering
\includegraphics[width=0.99\columnwidth]{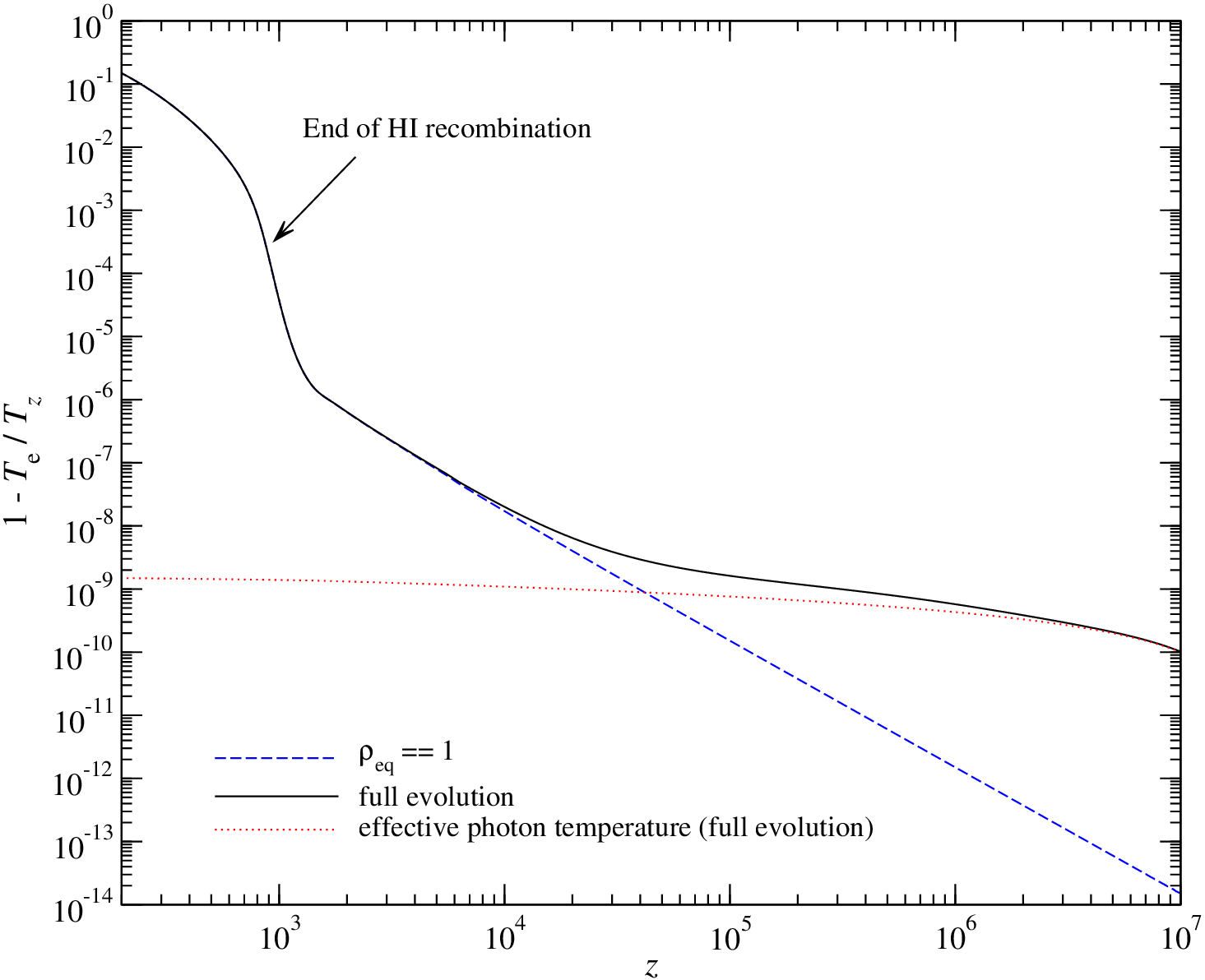}
\caption{Evolution of the electron temperature for the standard thermal history. The electrons are always slightly cooler than $\Tz$. For comparison we also show the evolution of the effective temperature of the photon field, $1-\rho^\ast=[\Tz-\Tgs]/\Tz$, which also implies $\Tgs<\Tz$.  }
\label{fig:Te_cooling}
\end{figure}
\subsection{Cooling of CMB photons by electrons in the absence of additional energy injection}
\label{sec:cooling}
As first case we studied the problem of adiabatically cooling electrons and what kind of distortion this causes in the CMB spectrum. As our estimate shows (see Sec.~\ref{sec:Q_cooling}) the distortion is expected to be very small. Therefore, this case provides a very good test problem for our numerical scheme and we shall see that {\sc CosmoTherm} indeed is capable of dealing with it.

We started the evolution at $\zs=4\times 10^7$ assuming an initial blackbody spectrum ($\Delta n_x=0$) with temperature $\Te(\zs)=\Tz(\zs)=\Tgs(\zs)$, where $\Tgs(\zs)$ was computed as explained in Sect.~\ref{sec:initial_cond} using the approximation Eq.~\eqref{eq:energy_densities_sol_a} with $\dot{Q}=- H\alpha_{\rm h}k\,\Tg$ for $\Delta \rho(z_{\rm s})$. 
We solved the problem down to $\ze=200$, i.e. well after recombination ends, however, the distortion froze in before that redshift, so that for our purposes $\ze=200$ is equivalent to $\ze = 0$.
{This statement of course ignores other modifications to the CMB spectrum that could be introduced at lower redshifts, for example by heating because of supernovae \citep{Oh2003}, or shocks caused by large scale structure formation \citep{Sunyaev1972b, Cen1999, Miniati2000}.
}

\subsubsection{Evolution of the electron temperature}
Figure~\ref{fig:Te_cooling} illustrates the evolution of the electron temperature for different settings.
The solid line is the result of the full integration including all heating and cooling terms, while for the dashed/blue line we neglected the cooling of electrons by DC and BR, and set $\rho_{\rm e}^{\rm eq}=1$, meaning that for the temperature equation we ignored the spectral distortions that are introduced by this process.
As one can see, at early times $1-\rho_{\rm e}$ is about four orders of magnitude larger when including all terms than in the case that enforces $\rho_{\rm e}^{\rm eq}=1$.
This can be explained by the fact that the heating of the electrons actually introduces a small distortion into the CMB, which results in a Compton equilibrium temperature $T_{\rm e}^{\rm eq} <\Tz$.
At lower redshifts the temperatures in the two discussed cases again coincide. This is because at those times the cooling by the expansion of the Universe starts to dominate over the Compton cooling related to the generated spectral distortion.

\begin{figure}
\centering
\includegraphics[width=0.99\columnwidth]{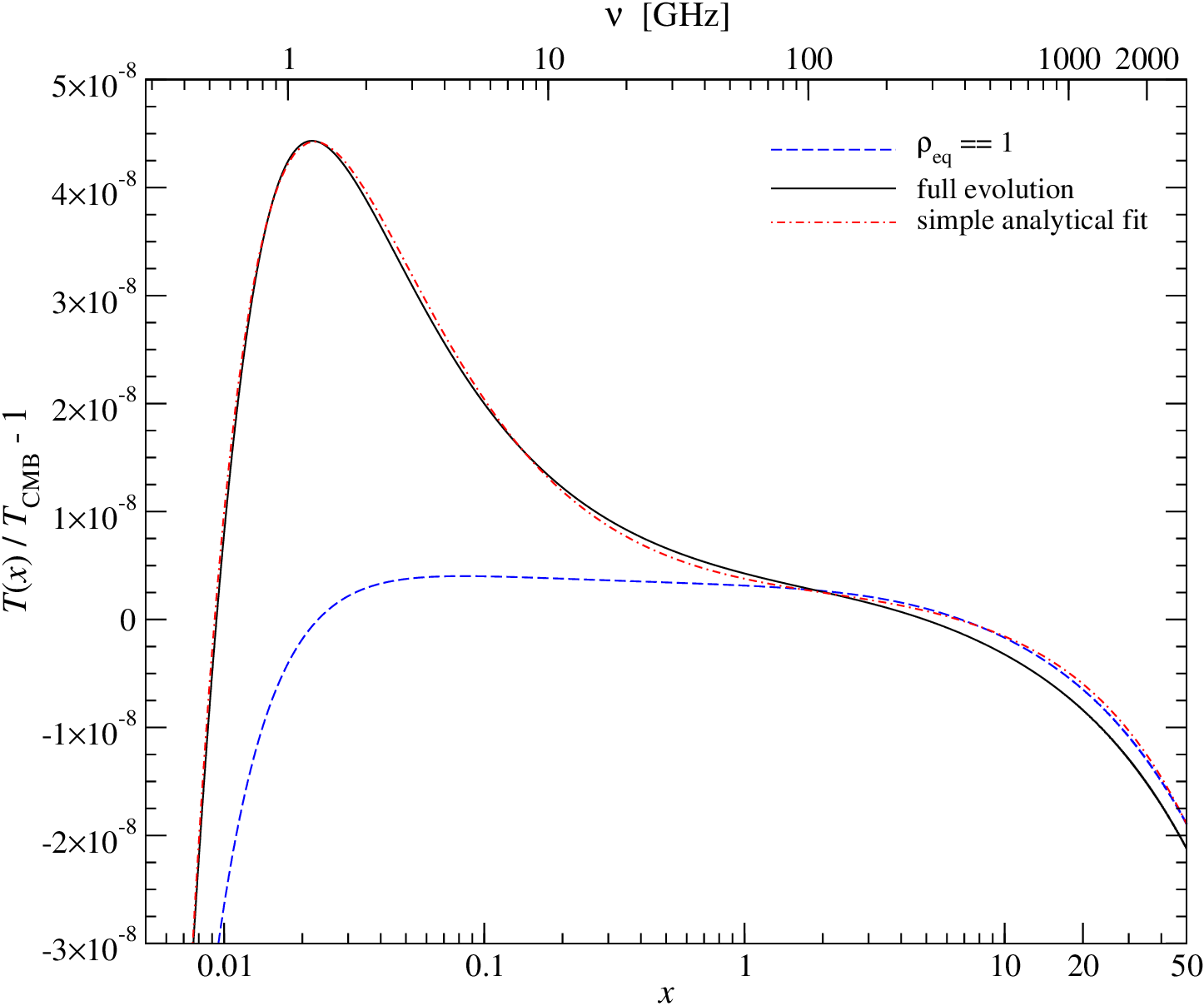}
\caption{CMB spectral distortion at $z=200$ caused by the continuous cooling from electrons. Neglecting the spectral distortion in the computation of the electron temperature leads to an underestimation of the final distortion at low frequencies.
We also show a simple analytical fit according to Eq.~\eqref{eq:n_x_app} with parameters $\mu_\infty=-\pot{2.22}{-9}$, $\xc=\pot{1.5}{-2}$, $\phi_{\rm f}-1=-\pot{8.0}{-10}$, $\ye=-\pot{4.3}{-10}$ and $\yff=-\pot{4.17}{-12}$. }
\label{fig:Te_x_cool}
\end{figure}

For comparison we also show the evolution of the effective temperature of the photon field, $1-\rho^\ast=[\Tz-\Tgs]/\Tz$, which also implies $\Tgs<\Tz$. We started the computation with $\Tz$ slightly higher than $\TCMB$, such that at $\ze$ we correctly have $\Tgs\sim \TCMB$.
Numerically, we obtain this value to within $\sim 10\%$, however, when computing the initial condition we assumed that the photons were cooled all the time. At low redshifts, during the recombination epoch this is no longer correct, such that one expects a slightly smaller coupling. Furthermore, our computation of the initial condition assumed no distortion, which again changes the balance toward slightly lower initial temperature. 
In particular, the cooling electrons start to significantly alter the high frequency tail of the CMB distortion (cf. Fig.~\ref{fig:Te_x_cool}), implying a smaller effective temperature.
These aspects of the problem are difficult to included before the computation is done. 
When considering cases in which the heating ends well before recombination and is much larger than the cooling by adiabatic expansion of the medium, this no longer is a problem, since truly bulk of the heat ends up in the photon field.

{
We also confirmed this statement by first computing the global energy balance problem (see Sect.~\ref{sec:exp_hist}), only neglecting the introduced distortions.
This allowed us to define the initial temperature for the run of {\sc CosmoTerm} more precisely, such that we obtained $\Tgs\sim \TCMB$ to within $0.1\%$ at $\ze=200$.
We conclude that {\sc CosmoTherm} conserves energy at a level well below $1\%$.
}


\subsubsection{Associated spectral distortion}
\begin{figure}
\centering
\includegraphics[width=0.99\columnwidth]{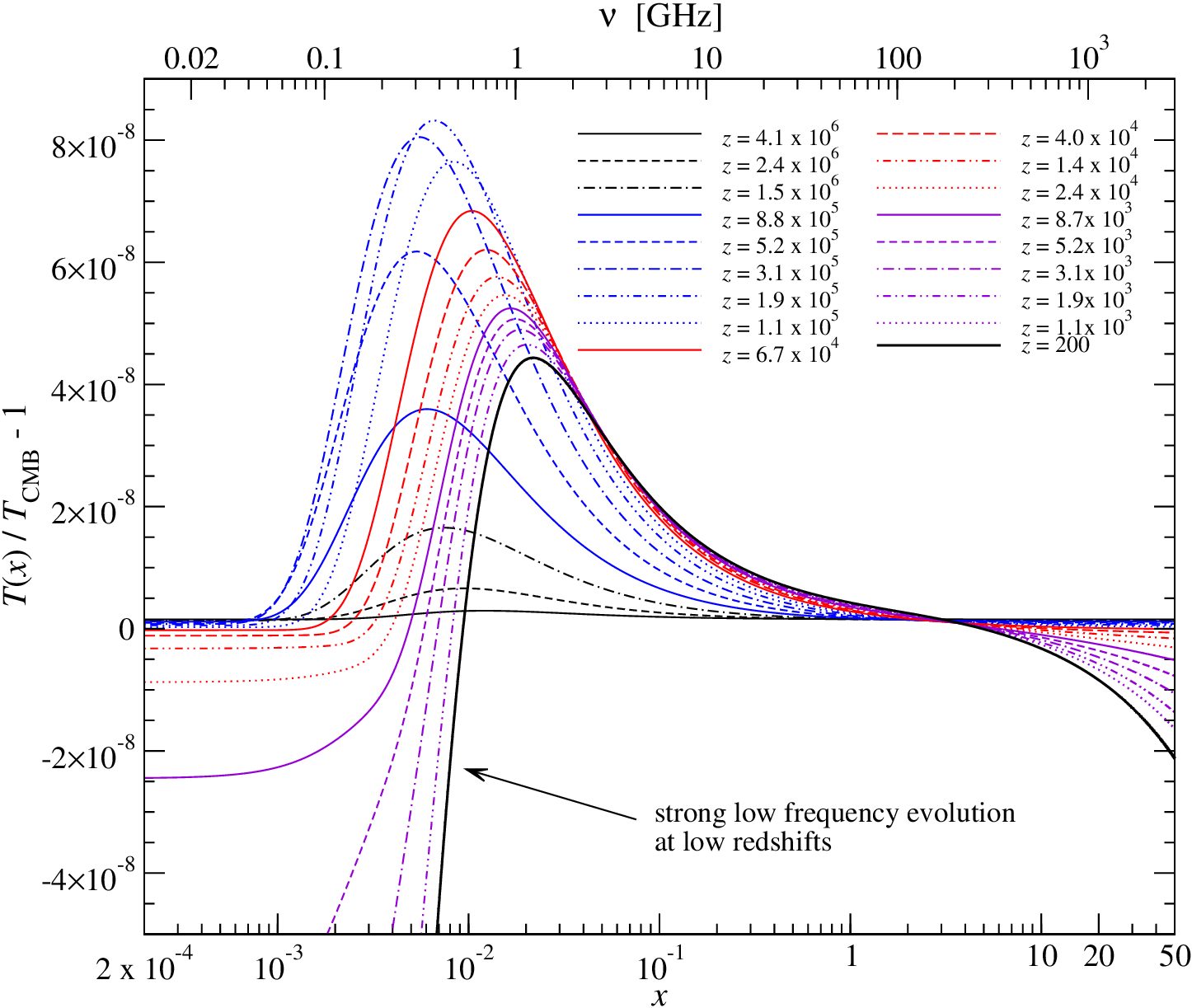}
\caption{Evolution of the CMB spectral distortion caused by the continuous cooling from electrons. At low redshifts one can see the effect of electrons starting to cool significantly below the temperature of the photons, which leads to strong free-free absorption at very low frequencies.}
\label{fig:Te_x_cool_evol}
\end{figure}
In Fig.~\ref{fig:Te_x_cool} we show the corresponding CMB spectral distortion in the two cases {discussed above}. 
Here two aspects are very important: firstly, the amplitude of the distortion is strongly underestimated when one assumes that the Compton equilibrium temperature is just $\Tz$, i.e., enforces $\rho_{\rm eq}=1$.
In this case the distortions do not build up in the full way, as the difference of the electron temperature is artificially reduced. Since the electron temperature appears in the exponential factor of the DC and BR emission and absorption term, this leads to a crucial difference at low frequencies.

Secondly, the distortions at both very low and at very high frequencies are rather large.
This is connected mainly with the low redshift evolution of the distortion. Once the Universe enters the recombination epochs, the temperature of the electrons can drop significantly below the temperature of the photon field (cf. Fig.~\ref{fig:Te_cooling}). This implies significant absorption by BR at low frequencies, and also a sizeable down-scattering of CMB photons at high frequencies, in an attempt to reheat the electrons.
Interestingly, the high and low frequency distortion is very similar in the two considered cases. This also suggests that this part of the distortion is introduced at low redshifts, where the electron temperature in both cases is practically the same (cf. Fig.~\ref{fig:Te_cooling}).

To illustrate this aspect of the problem, in Fig.~\ref{fig:Te_x_cool_evol} we present a sequence of spectra starting at redshifts during which distortions are quickly thermalized ($z\sim 10^6$), passing through the epoch of $\mu$-type distortions ($z\sim 10^5$), followed by the $y$-type era ($z\sim 10^4$), and ending well after recombination. 
Close to the initial time one can observe the slightly higher temperature at both low and very high frequencies, which is the result of the consistent initial condition. 
At the final redshift the distortion is neither a pure $\mu$- nor a pure $y$-type distortion. At high frequencies it has some characteristics of a {\it negative} $y$-type distortion, while around $\sim 1\,$GHz it looks like a {\it negative}  $\mu$-type distortion.
At very low frequencies the free-free distortion dominates, as explained above.
One can see from Fig.~\ref{fig:Te_x_cool_evol} that the free-free distortion indeed starts to appear at rather late times, when the electron temperature departs by more than $\Delta T/T\sim 10^{-8}$ from the photons.
We found that $n_x$ according to Eq.~\eqref{eq:n_x_app} with parameters $\mu_\infty=-\pot{2.22}{-9}$, $\xc=\pot{1.5}{-2}$, $\phi_{\rm f}-1=-\pot{8.0}{-10}$, $\ye=-\pot{4.3}{-10}$ and $\yff=-\pot{4.17}{-12}$ represents the total spectral distortion rather well (cf. Fig.~\ref{fig:Te_x_cool}).
These effective values for $\mu_\infty$ and $\ye$ are several times below the limits that might be achieved with {\sc Pixie}, implying that measuring this effect will be very hard.

{With the values of $\mu_\infty$ and $\ye$ one can estimate the amount of energy that was release during the $\mu$-era ($50000\lesssim z \lesssim \pot{2}{6}$) and $y$-era ($z\lesssim 50000$), using the simple expressions \citep{Sunyaev1970mu}
$\mu_\infty\approx 1.4 \Delta \rho_\gamma/\rho_\gamma$ and $\ye \approx \frac{1}{4} \Delta \rho_\gamma/\rho_\gamma$, resulting in $\Delta \rho_\gamma/\rho_\gamma|_\mu \approx \pot{1.6}{-9}$ and $\Delta \rho_\gamma/\rho_\gamma|_\mu \approx \pot{1.7}{-9}$. This is consistent with the simple estimates carried out in Sect.~\ref{sec:Q_cooling}, supporting the precision of the code regarding energy conservation.}

\begin{figure}
\centering
\includegraphics[width=0.99\columnwidth]{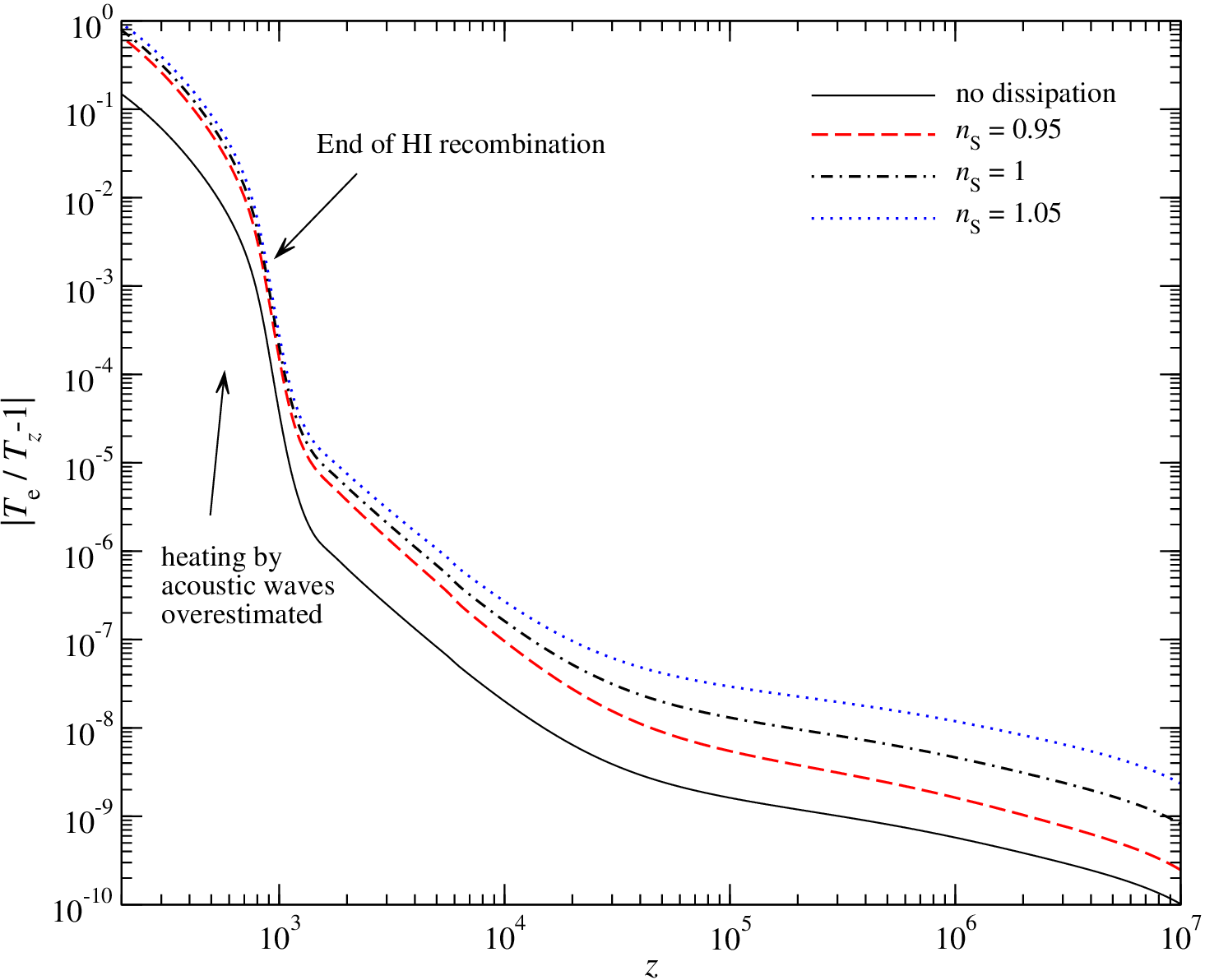}
\caption{Evolution of the electron temperature for the standard thermal history with dissipation of energy by acoustic waves. As a result of the heating the electrons are always above $\Tz$.  For comparison we show the electron temperature obtained when neglecting this effect, for which in contrast one has $\Te<\Tz$ at all times.}
\label{fig:Te_acoustic}
\end{figure}
\begin{figure}
\centering
\includegraphics[width=0.99\columnwidth]{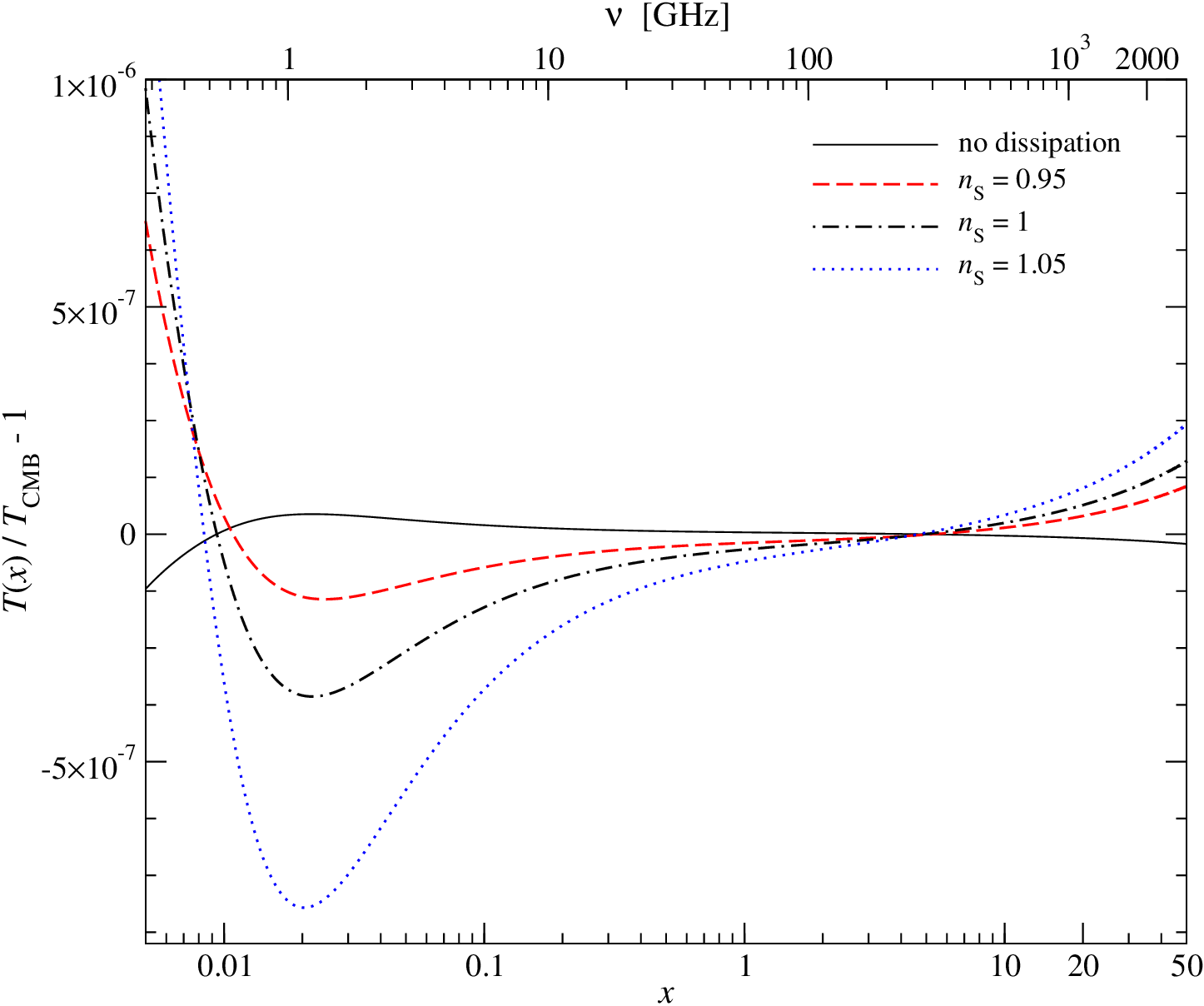}
\caption{CMB spectral distortion at $z=200$ caused by the dissipation of energy from acoustic waves for different values of $\nS$. For comparison we show the distortion obtained when neglecting this effect.}
\label{fig:Te_x_acoustic}
\end{figure}
\subsection{Dissipation of energy from acoustic waves}
\label{sec:waves}
As next example we computed the distortions arising from the dissipation of energy in acoustic waves, again starting at $\zs=\pot{2}{7}$ and solving the problem down to $\ze=200$.
In Fig.~\ref{fig:Te_acoustic} we show the evolution of the matter temperature and in Fig.~\ref{fig:Te_x_acoustic} we present the corresponding spectral distortions in the CMB. In both cases we varied the value of the spectral index, $\nS$.

As one can see from Fig.~\ref{fig:Te_acoustic}, for all cases with dissipation the electrons are heated significantly above $\Tz$.
For $\nS < 1$ slightly more energy is dissipated at low $z$ than in the case $\nS\gtrsim 1$ \citep[cf.][]{Hu1994}. Also the total energy release is smaller in cases with $\nS<1$. 
This implies that the amplitude of the distortions is larger for larger values of $\nS$, and also that the ratio of $\mu$- to $y$-type contribution increases with $\nS$ (cf. Table~\ref{tab:acoustic}).

As mentioned above, that at late times, during recombination, the electrons are significantly hotter when the heating by damping acoustic waves is included. 
We expect that our computation significantly overestimates the effect in this epoch, and for a precise inclusion of this process during recombination a more detailed treatment will be required.
In particular the $y$-type contribution to the final spectral distortion could require revision.
However, for the purpose of this paper the used approximations will suffice.

\begin{table}
\centering
\caption{Spectral distortion caused by the dissipation of acoustic waves. We give the parameters for the simple fitting formula, Eq.~\eqref{eq:n_x_app}. For convenience we defined $\Delta\phi_{\rm f}=\phi_{\rm f}-1$.}
\begin{tabular}{cccccc}
\hline
$\nS$ &  $\mu_\infty\,[10^{-8}]$ & $\xc\,[10^{-2}]$ & $\Delta\phi_{\rm f}\,[10^{-8}]$ & $\ye\,[10^{-9}]$ & $\yff\,[10^{-11}]$ \\
\hline 
$0.95$  & $0.735$ & $1.5$ & $0.75$ & $2.2$ & $1.97$ \\
$1.0$ & $1.75$ & $1.5$ & $0.95$ & $3.5$ & $2.98$ \\
$1.05$ & $4.05$ & $1.6$ & $1.0$ & $5.9$ & $4.47$ \\
\hline
\end{tabular}
\label{tab:acoustic}
\end{table}
Looking at Fig.~\ref{fig:Te_x_acoustic}, one can clearly see that in all cases with dissipation the shape of the distortion is well represented by a $\mu$-type distortion around $\nu\sim 1\,$GHz. At high and intermediate frequencies one can again see an admixture of $y$-type distortion, and at very low frequencies, additional free-free emission from electrons with $\Te>\Tg$ becomes visible.
We checked that rescaling all the distortions to the same amplitude at $\nu\sim 1\,$GHz reveals small differences in the shape of the distortion in the {\sc Pixie} bands. However, these differences will be more difficult to distinguish, in spite of the total amplitude of the distortion being rather large.

%
In Table~\ref{tab:acoustic} we gave a summary for the parameters needed to represent the final distortion in the different cases using the simple fitting formula, Eq.~\eqref{eq:n_x_app}.
The contribution from $\mu$ is dominating in all cases and is of order $\mu_\infty\sim 10^{-8}$.
By measuring this distortion it would in principle be possible to constrain the value of $\nS$ in addition to the limits that will be obtained with \Planck \, data. 
Since very different systematic effects are involved, this would further increase our confidence in the correctness of the inflationary model.

\begin{figure}
\centering
\includegraphics[width=0.99\columnwidth]{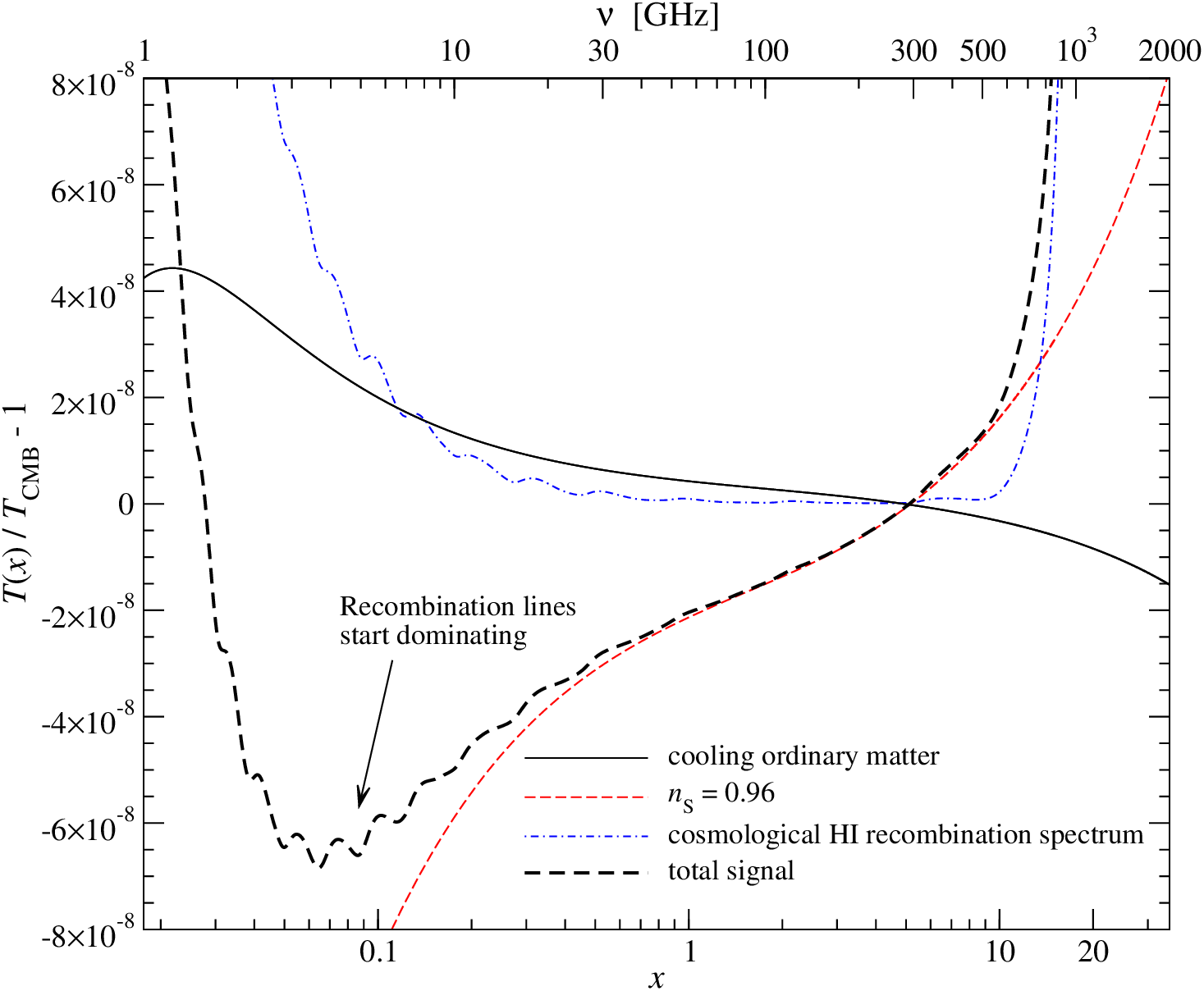}
\caption{Different cosmological signals from the early Universe.}
\label{fig:Te_x_rec_PIXIE}
\end{figure}

{Here} it is important to mention that also the cosmological recombination spectrum {\citep{Dubrovich1975, Chluba2006b, Sunyaev2009}} contributes to the signal in the {\sc Pixie} bands.
The result for the cosmological recombination spectrum computed by \citet{Chluba2006b} for a 100-shell hydrogen atom model is shown in Fig.~\ref{fig:Te_x_rec_PIXIE} together with the distortion from the damping of acoustic waves. 
{The smaller contributions of helium \citep{Dubrovich1997, Jose2008, Chluba2009c} were neglected here, however, the previously omitted free-bound contribution \citep{Chluba2006b} is included.}
For comparison we also show the distortion arising from the interaction of CMB photons with adiabatically cooling electrons alone. 
Furthermore, we present the total sum of the cosmological signals. Here we emphasize again, that even in the standard cosmological model {\it all} these signals are expected to be present in the CMB spectrum.

At high frequencies ($\nu\gtrsim 700\,$GHz) the distortion from the Lyman-$\alpha$ line emitted during hydrogen recombination at $z\sim 1400$ dominates.
However, in this frequency band the cosmic infrared background is many orders of magnitude larger than this signal, so that a measurement will be very difficult there.
At lower frequencies, on the other hand, one can see that the cosmological recombination radiation leads to a significant contribution to the distortion from acoustic damping. It slightly modifies the slope of the distortion as well as introduces small frequency-dependent variations.
These {\it wiggles} depend on the dynamics of recombination and therefore in particular should allow to constrain the baryon density and helium abundance of our Universe \citep{Chluba2008T0}.

From Fig.~\ref{fig:Te_x_rec_PIXIE} we can also conclude that when neglecting the effect caused by adiabatically cooling electrons the spectral distortion arising from the dissipation of acoustic waves is overestimated by about $\sim 20\%$ at low frequencies.
This would lead to a value for $\nS$ which is biased high. As the amplitude of the distortion is very sensitive to the exact value of $\nS$ this also demonstrates that refined computations of the spectral distortions as well as the heating rates because of acoustic damping are required to obtain accurate predictions in the light of {\sc Pixie}.
Currently, it appears that for $\nS=0.96$ {\sc Pixie} {could already} allow a $1\sigma$ detection of the effect caused by the dissipation of acoustic waves.

Finally, we would like to compare the results of our computation with the simple analytic predictions given by \citet{Hu1994}. 
There only cases with $\nS>1$ were discussed, and for $\nS=1.05$ they find $\mu_\infty\sim \pot{5}{-8}$ for a slightly different cosmology, which is in very good agreement with the result given here.
However, with {\sc CosmoTherm} were also able to compute the admixture of $y$-type distortions, which at high frequencies dominates.

\subsection{Distortions caused by annihilating particles}
\label{sec:annihil}
We now focus on the distortions that are created by annihilating particles. Recently, in particular the possibility of annihilating dark matter with Sommerfeld enhancement was actively discussed in the literature \citep[see e.g.,][and references therein]{Galli2009, Slatyer2009, Cirelli2009, Huetsi2009, Huetsi2011, Zavala2011}, resulting in rather tight constraints on this possibility from the CMB temperature and polarization anisotropies.
As mentioned above, these constraints are many orders of magnitude stronger than those deduced from {\sc Cobe/Firas}. The reason for this is that energy release can directly affect the recombination dynamics, depositing energy into the matter, while this amount of energy is tiny compared with the huge heat bath of CMB photons.

Here we therefore take two perspectives: first we compute cases that are close to the current upper bound on the annihilation efficiency, $f_{\rm ann}\sim \text{few}\times10^{-23}\,\rm eV\,s^{-1}$.
Uncertainty in the modelling of the effect of dark matter annihilation on cosmological recombination can accommodate factors of a few, so that still some range is allowed.
Second, we compute the distortions for large annihilation rates to check the consistency of our numerical treatment, and illustrate the differences with cases for decaying particles.

\begin{figure}
\centering
\includegraphics[width=0.99\columnwidth]{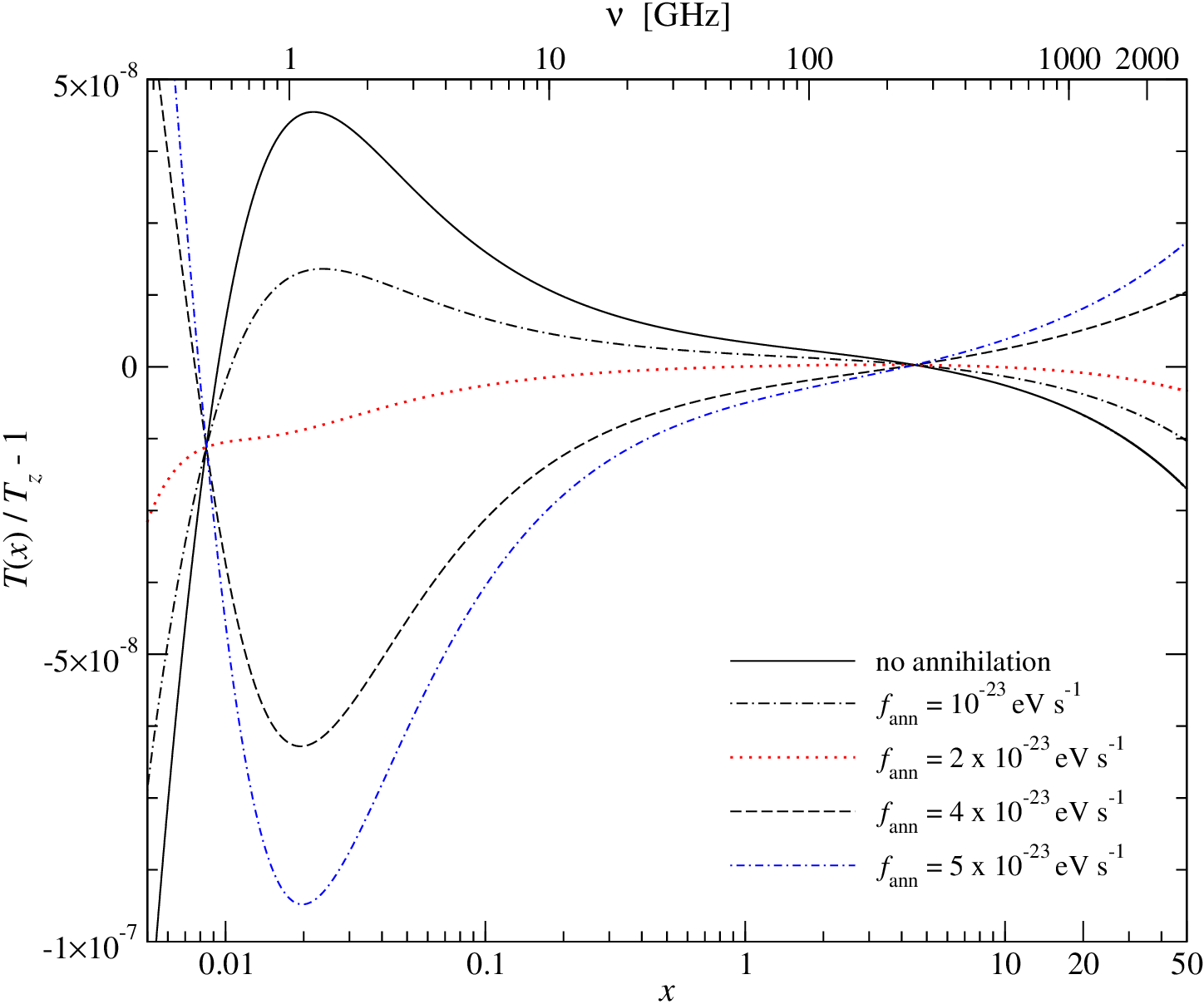}
\caption{CMB spectral distortion at $z=200$ after continuous energy injection from annihilating particles.}
\label{fig:Te_x_ann}
\end{figure}
\begin{figure}
\centering
\includegraphics[width=0.99\columnwidth]{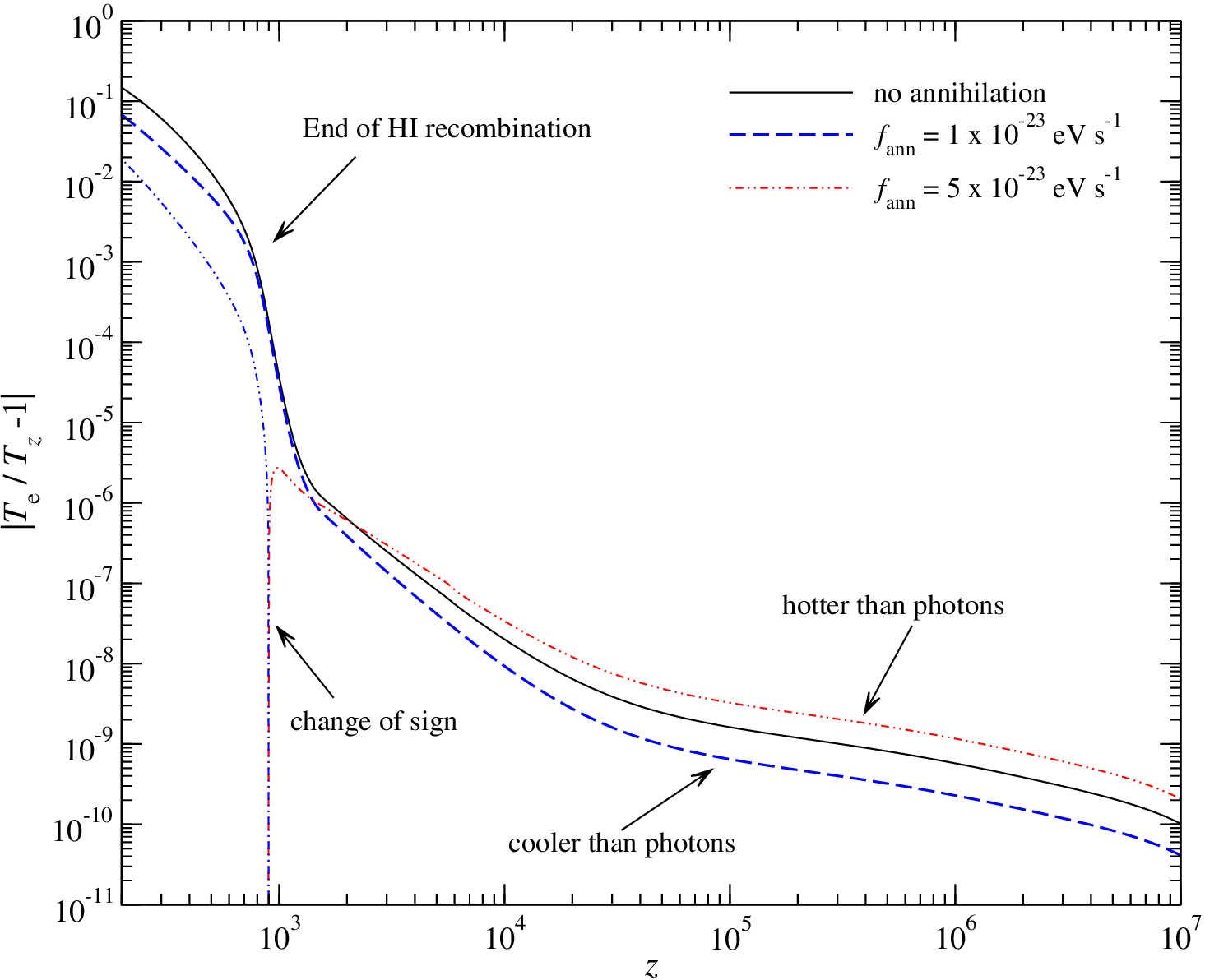}
\caption{Evolution of the electron temperature for the thermal history with annihilating particles. Red lines indicate that the electrons are hotter than the effective CMB temperature, while blue indicates cooler than this. The black/solid line gives the case without annihilation for comparison.}
\label{fig:Te_ann_z}
\end{figure}
In Fig.~\ref{fig:Te_x_ann} we present the spectral distortion after continuous energy injection by annihilating particles with different annihilation efficiencies, $f_{\rm ann}$.
For comparison we also show the result for the case without any energy release (see previous section).
To understand aspects of the solution in more detail in Fig.~\ref{fig:Te_ann_z} we furthermore present the evolution of the electron temperature for two cases with annihilation.
As Fig.~\ref{fig:Te_x_ann} suggests, for $f_{\rm ann}\lesssim \pot{5}{-24}\,\rm eV \, s^{-1}$ we expect the distortion caused by adiabatically cooling electrons to dominate. 
In this case it will be difficult to improve limits on this type of heating process by studying the CMB spectrum.

For $f_{\rm ann} =10^{-23}\,\rm eV \, s^{-1}$, one can start seeing a reduction of the positive distortion at $\nu \sim 1\,$GHz.
Also, as discussed in Sect.~\ref{sec:cooling}, for $f_{\rm ann} =\pot{2}{-23}\,\rm eV \, s^{-1}$ the net distortion introduced because of heating by annihilating particles and the cooling of electrons becomes very small distortion.
As the annihilation efficiency increases, the distortion becomes more like a $\mu$-type distortion with positive chemical potential at frequencies around $\nu\sim 1\,$GHz. However, at high frequencies we can observe a contribution from a $y$-type distortion with positive $y$-parameter. The heating by annihilating particles at early times pushes the temperature of the electrons slightly above the temperature of the photons (cf. Fig.~\ref{fig:Te_ann_z}), so that these get up-scattered.
Depending on the annihilation efficiency, this period is more or less extended. 
For small annihilation efficiency this never happens (cf. Fig.~\ref{fig:Te_ann_z}), so that the distortions is dominated by negative $\mu$ and $y$-contributions.

At low frequencies we can again see the effect of the free-free process.
Like in the previous section, both the contribution from $y$-type distortion and free-free process are mainly introduced in the redshift range corresponding to recombination $z\sim 10^3 - 10^4$.
In particular we also found a characteristic change in the extremely low frequency spectrum, when the difference of the electron and photon temperature changes sign, however, this occurred at far too low frequencies to be worth discussing any further.


\begin{figure}
\centering
\includegraphics[width=0.99\columnwidth]{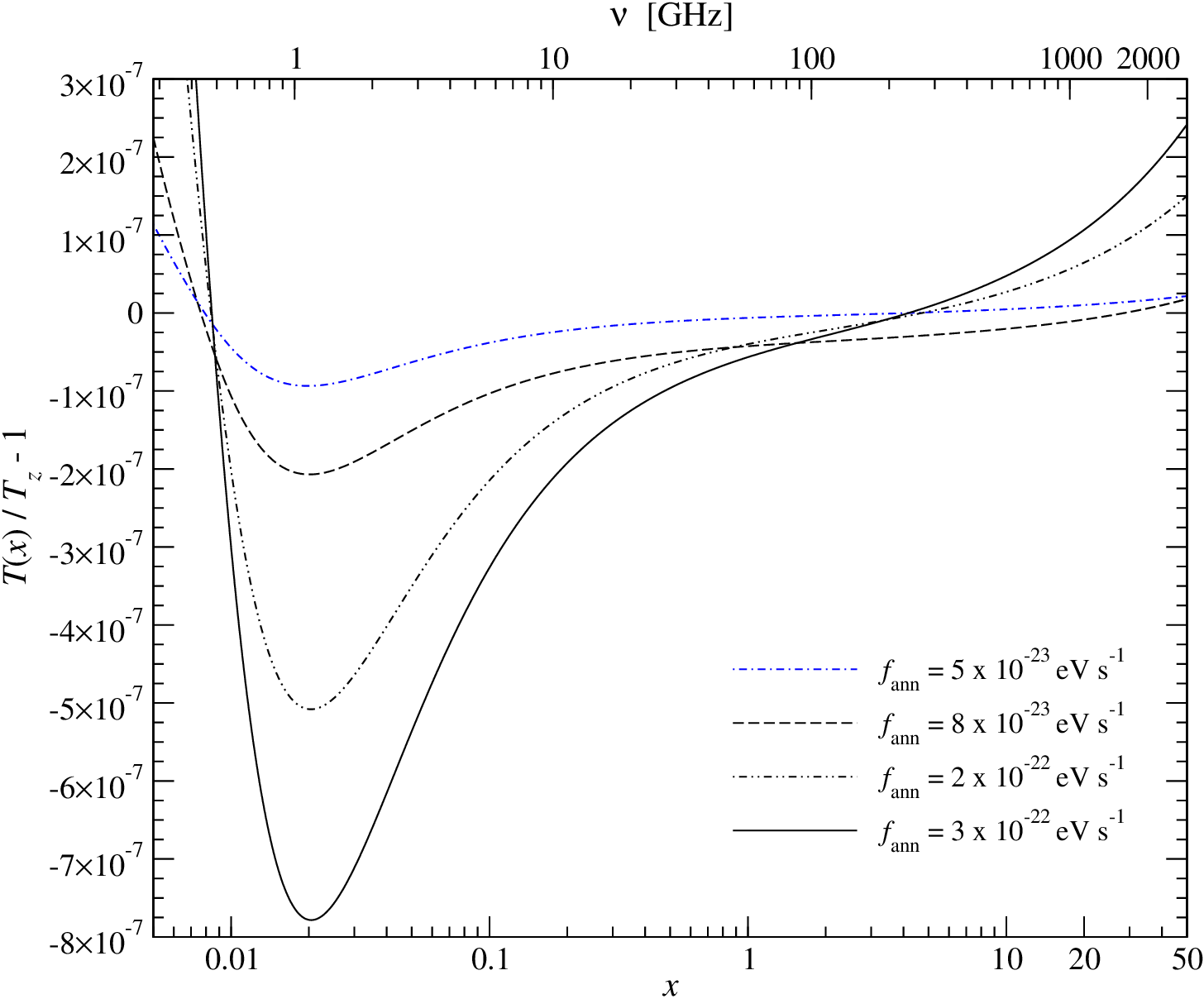}
\caption{CMB spectral distortion at $z=200$ after continuous energy injection from annihilating particles with large annihilation cross section.}
\label{fig:Te_x_ann.high}
\end{figure}
For illustration, in Fig.~\ref{fig:Te_x_ann.high} we also give some examples with very high annihilation efficiencies. As expected the amplitude of the distortion increases with $f_{\rm ann}$. Furthermore, one can observe a shift in the position of the maximal distortion close to $\nu\sim 1\,$GHz as $f_{\rm ann}$ increases.
For annihilation efficiency $f_{\rm ann}=\pot{3}{-22}\,\rm eV \, s^{-1}$ the distortion reaches $\mu_\infty \sim \pot{4}{-8}$. This is very similar to the case of acoustic damping for $\nS=1.05$, which would be well within reach of {\sc Pixie}.
We also found that the distortion in both cases looks very similar, however, current limits on $f_{\rm ann}$ from the CMB anisotropies disfavour so high annihilation efficiencies.

\begin{figure}
\centering
\includegraphics[width=0.99\columnwidth]{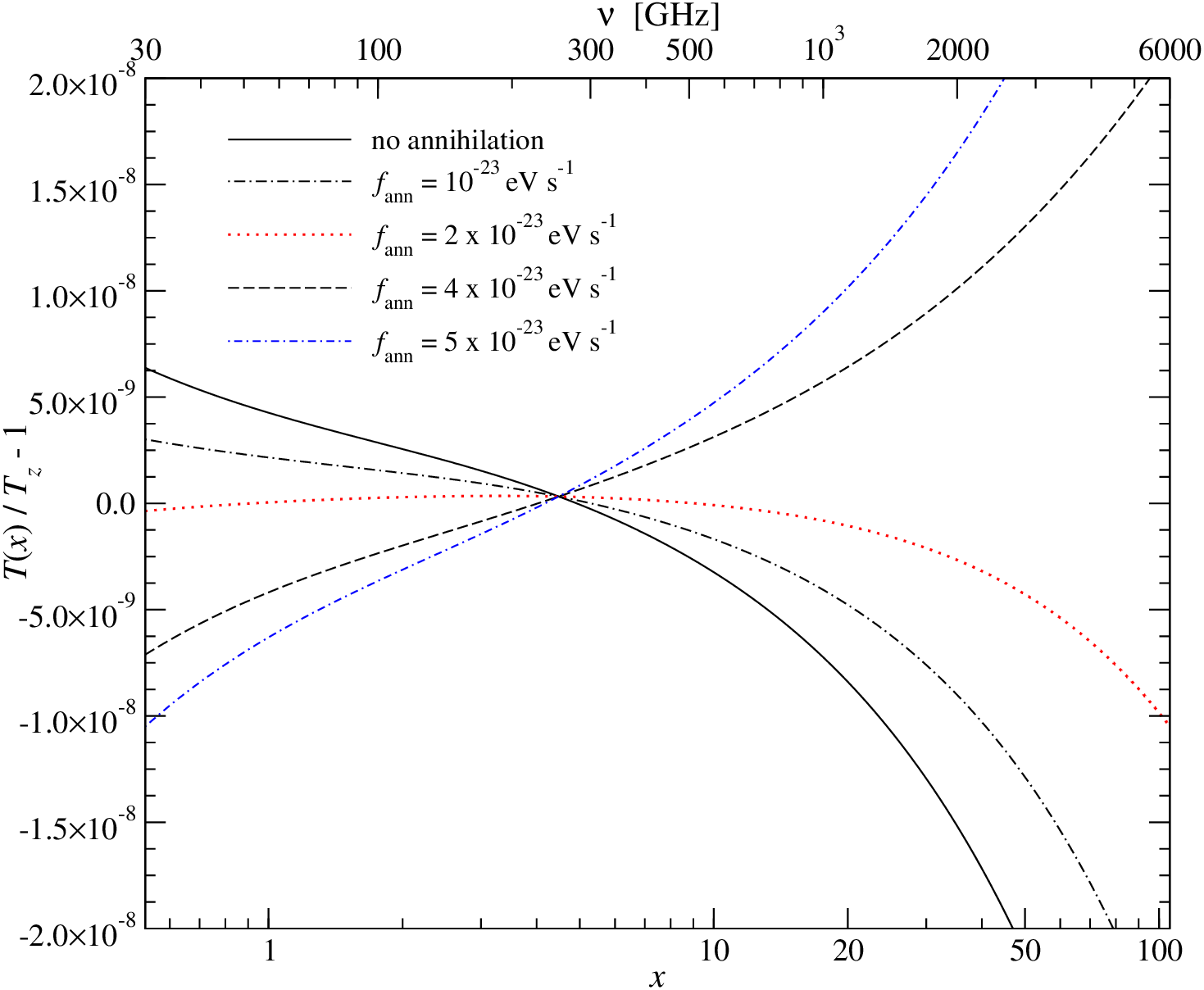}
\caption{CMB spectral distortion at $z=200$ after continuous energy injection from annihilating particles in the {\sc Pixie} bands.}
\label{fig:Te_x_ann_PIXIE}
\end{figure}
Finally, in Fig.~\ref{fig:Te_x_ann_PIXIE} we show the distortions for some of the previous cases, but focused on the spectral bands of {\sc Pixie}.
One can see that in all shown cases the distortions to lowest order have a very similar shape, while only the amplitude is varying from case to case.
Furthermore, the typical amplitude of the distortions is $\Delta T/T \sim 10^{-9}-10^{-8}$, rendering a measurement in terms of the sensitivity of {\sc Pixie} rather difficult.
At frequencies below $30\,$GHz the situation would be slightly better, as the distortion increases to the level of $\Delta T/T \gtrsim 10^{-8}$. In this frequency band also the distortions from the recombination epoch \citep{Chluba2006b, Sunyaev2009} would become visible (see Fig.~\ref{fig:Te_x_rec_PIXIE}).

Presently, we conclude that improving the limits on energy injection by annihilating particles using {\sc Pixie} will be very hard unless $f_{\rm ann}$ exceeds $\sim 10^{-22}\,\rm eV \, s^{-1}$.
With the current upper limits on the effective annihilation rate is also seems that in comparison to the distortion introduced by the dissipation of acoustic waves, those arising from annihilation would only contribute at the level of $10\%-20\%$. 
If annihilating particles are present in the Universe this might lead to problems in interpreting future spectral measurements that could be carried out by {\sc Pixie}.
This is especially severe, since the shape of the distortion in both cases is very similar, implying that more detailed forecasts will be required to assess the observational possibilities with {\sc Pixie}, {also including possible obstacles introduced by foregrounds (see comments below)}.

\begin{figure}
\centering
\includegraphics[width=0.99\columnwidth]{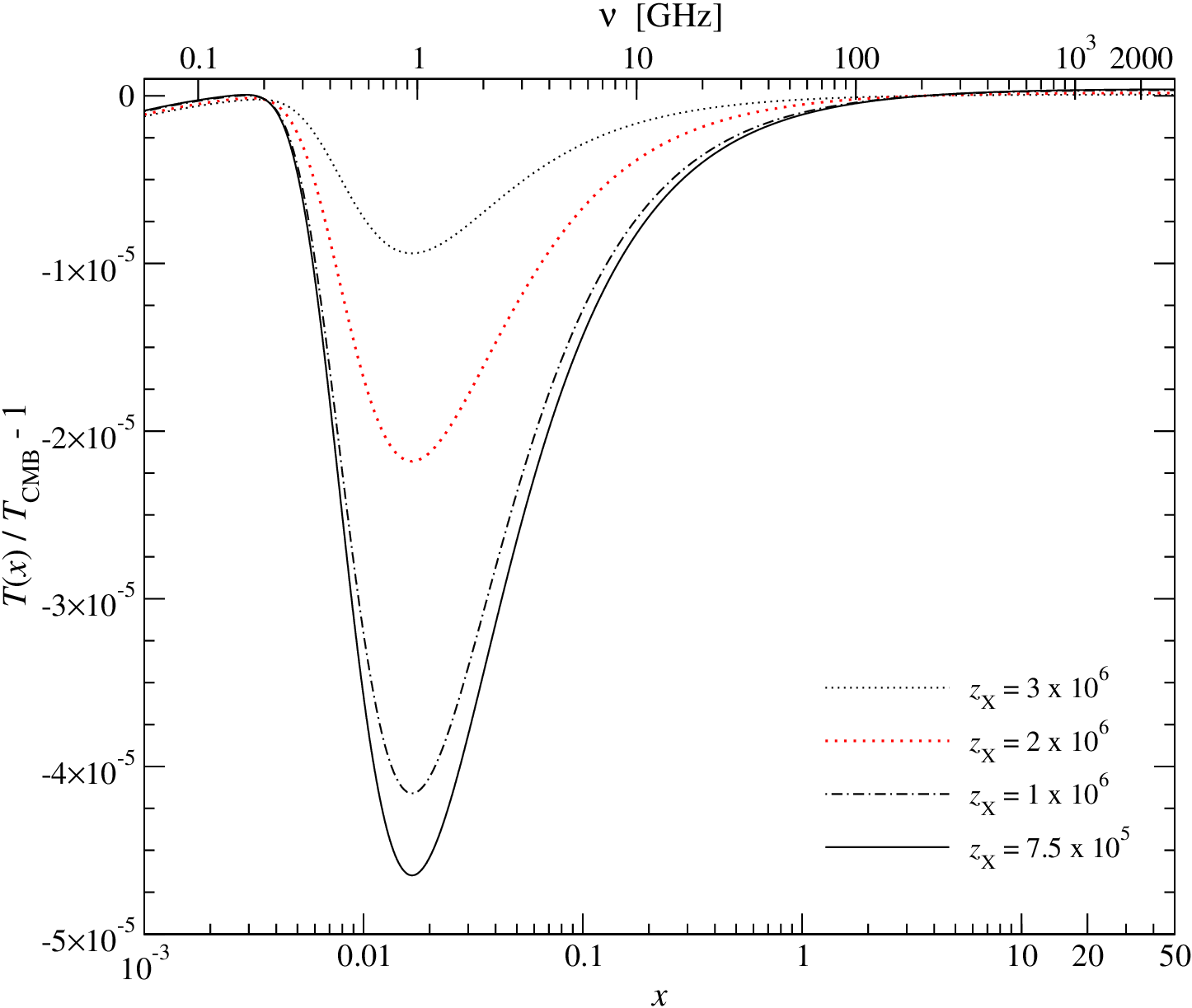}
\\
\includegraphics[width=0.99\columnwidth]{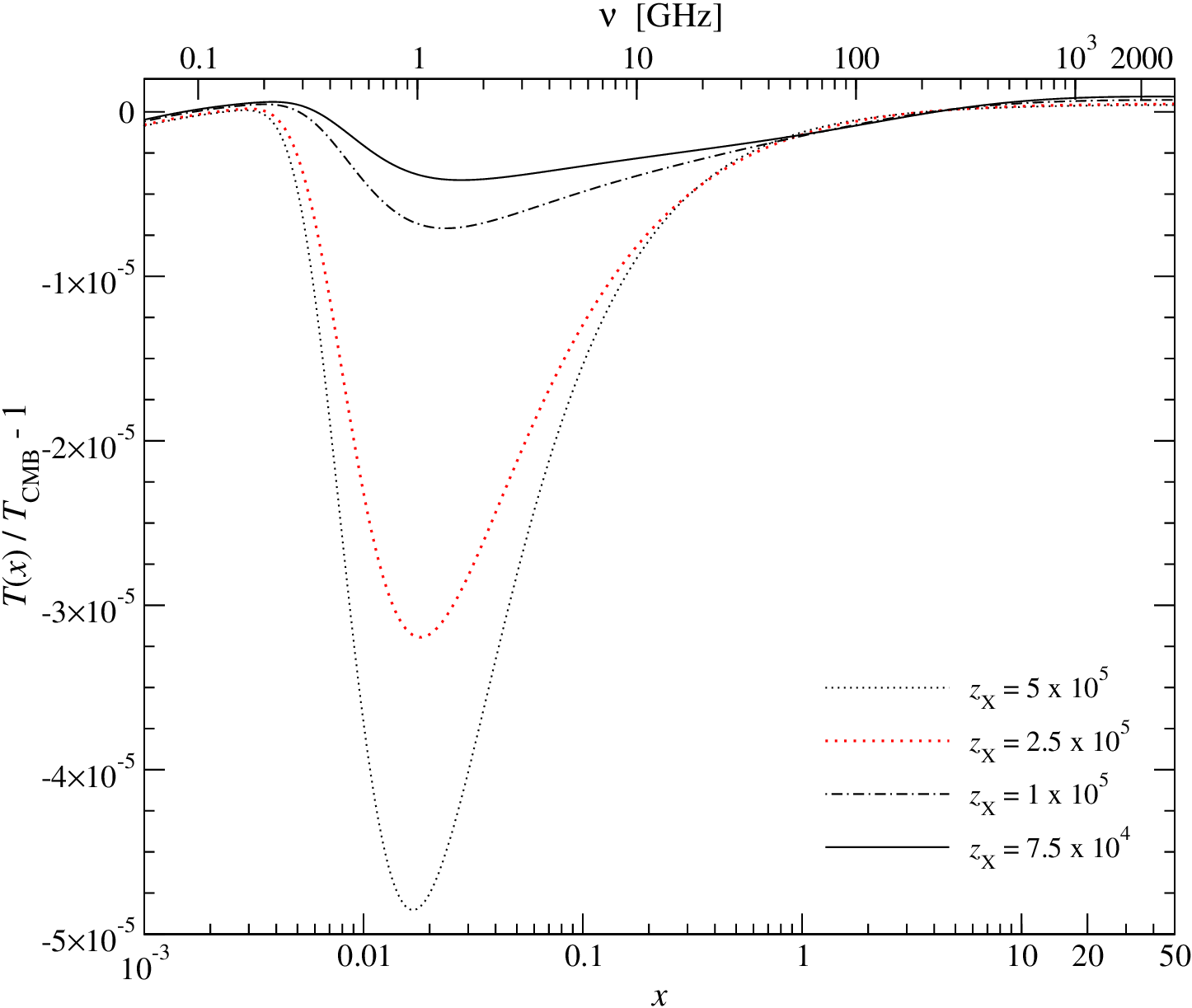}
\\
\includegraphics[width=0.99\columnwidth]{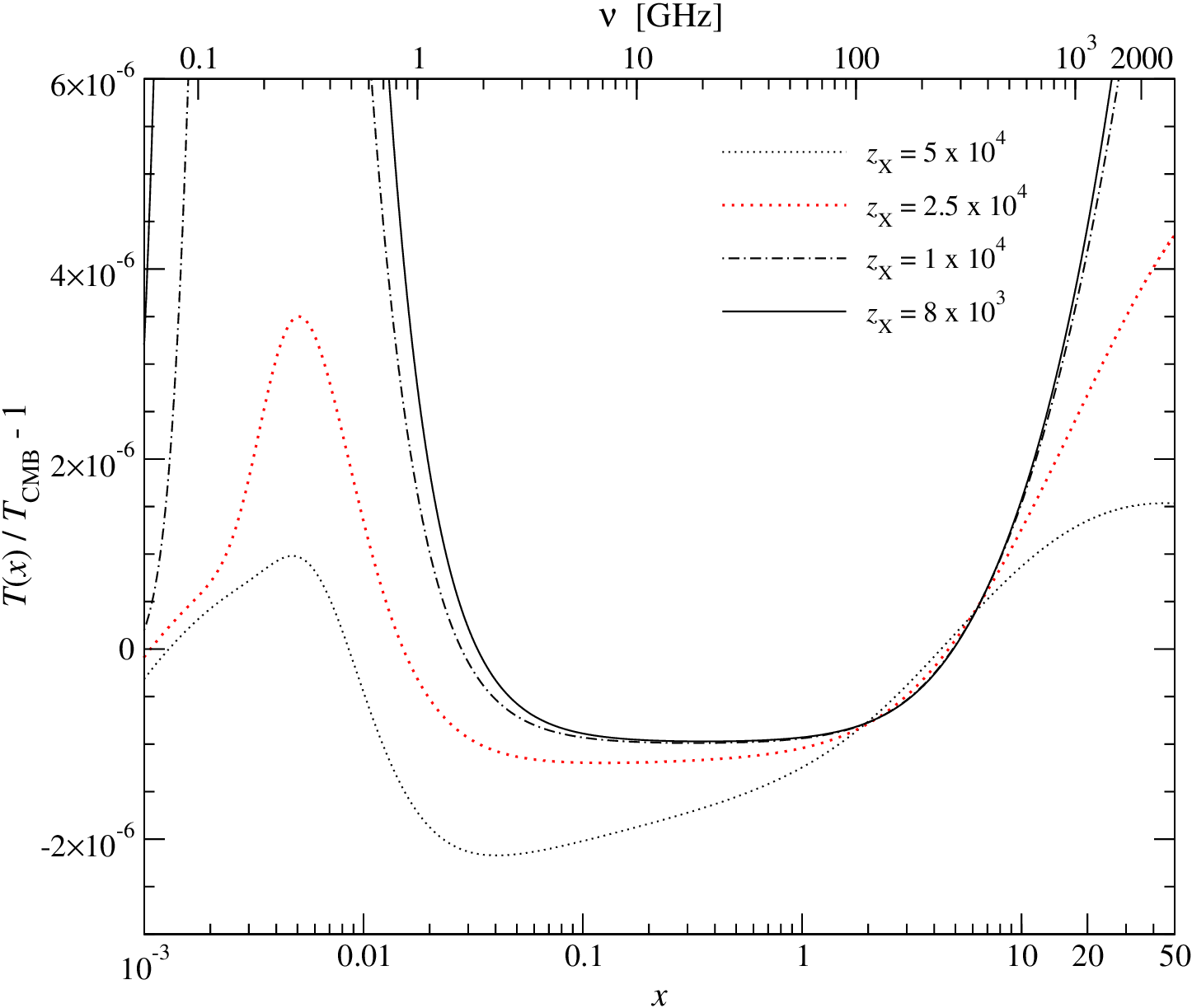}
\caption{CMB spectral distortion at $z=200$ after energy injection from decaying relic particles. In all cases we fixed $f_{\rm dec}=2\,z_{\rm X}\,$eV, which corresponds to a total energy release of $\left.\Delta \rho_\gamma/\rho_\gamma\right|_{\rm dec} \sim \pot{1.3}{-6}$. For the effective temperature of the CMB this implies $\Delta T^\ast_\gamma/\TCMB\sim -\pot{3.2}{-7}$ at $\zs=\pot{2}{7}$ and at $\ze=200$ in all cases we found $|\Delta T^\ast_\gamma/\TCMB|\sim 10^{-10}$.}
\label{fig:Te_x_decay}
\end{figure}
\subsection{Distortions caused by decaying particles}
\label{sec:decay}
In the case of decaying particles, observational constraints are less tight. Therefore one can in principle allow large amounts of energy injection well before the recombination epoch.
%
However, here we are not so much concerned with deriving detailed constraints on this process using our computations. We rather wish to demonstrate how the shape of the distortions depends in the particle lifetimes. To this end we therefore chose parameters which are not in tension with the limits deduced from {\sc Cobe/Firas} \citep{Fixsen2002}.

In Fig.~\ref{fig:Te_x_decay} we compiled the results obtained for fixed total energy release by particles of different lifetimes.
One can clearly see that for particles with short lifetimes, corresponding to $\zX\gtrsim \pot{2}{6}$, only a small part of the released energy remains visible as spectral distortion. This is because the processes of thermalization is very efficient, and does not allow much of the distortion to survive \citep[cf. also]{Burigana1991, Hu1993}.
As $\zX$ decreases to $\zX\sim \pot{5}{4}$ the low frequency feature at $\nu\sim 1\,$GHz becomes more visible. The overall shape of the distortion is well characterized by a $\mu$-type distortion, and only at very low frequencies it deviates slightly, because of the free-free distortion introduced at late times.
Since in the chosen example the effective energy release rate of the decaying particles is rather large, the free-free distortion is not as visible as in the cases discussed in the previous two sections.

As we decrease $\zX$ down to $\zX\sim \pot{5}{4}$ we can see a change in the characteristic spectral distortion. The low frequency $\mu$-type feature starts to disappear, while at intermediate and high frequencies a significant $y$-type distortion starts to mix in.
Decreasing $\zX$ even further the distortion clearly starts to be dominated by a $y$-type distortion, with a flat $\Delta T/T$ at intermediate frequencies and a characteristic rise of $\Delta T/T$ at high frequencies.
However, at low frequencies the interplay between $y$-type and free-free distortion becomes important, leading to another positive feature at $\nu\sim 500\,$MHz.

\begin{figure}
\centering
\includegraphics[width=0.99\columnwidth]{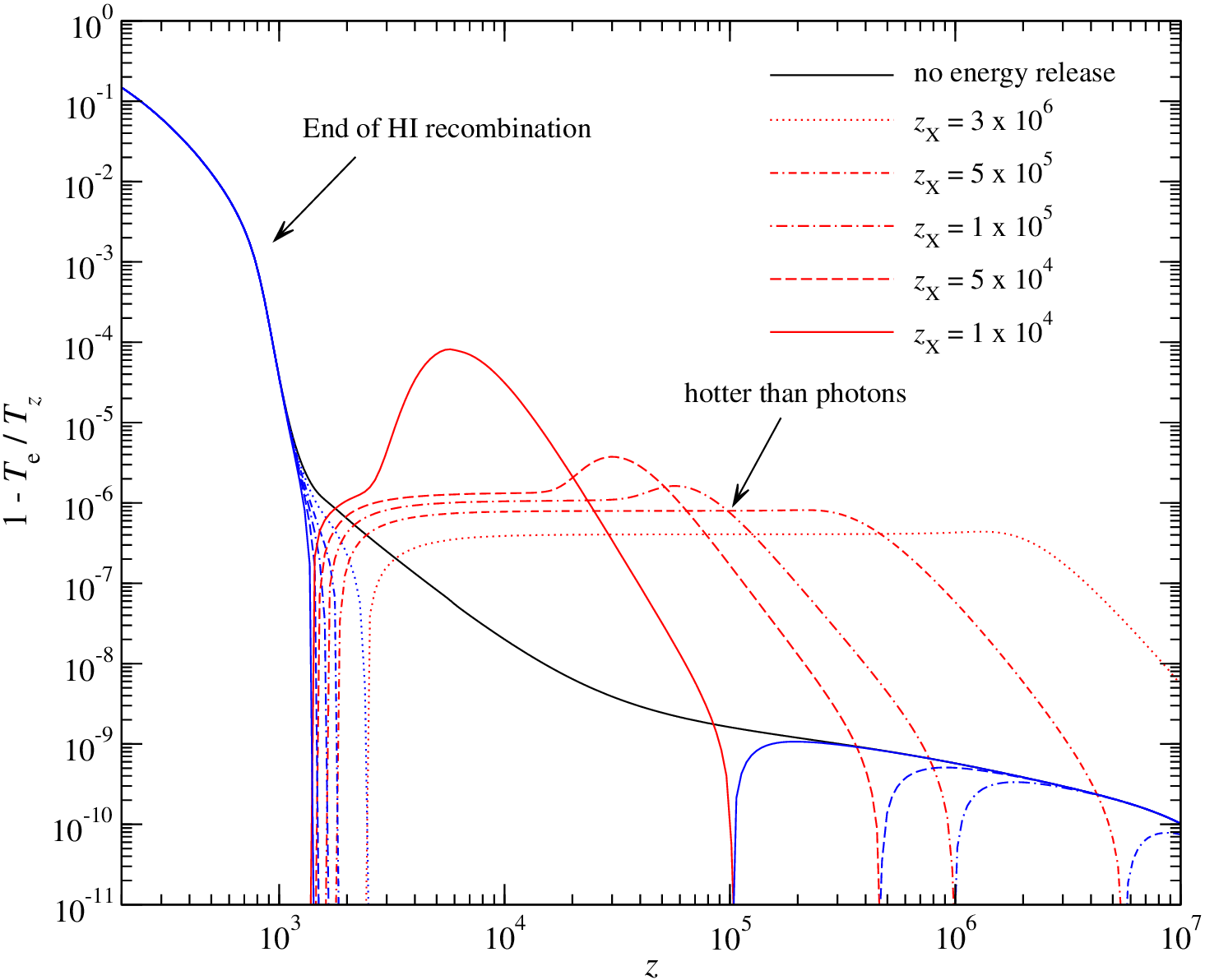}
\caption{Evolution of the electron temperature for the thermal history with decaying particles. Parameters were chosen like in Fig.~\ref{fig:Te_x_decay}. Red lines indicate that the electrons are hotter than the effective CMB temperature, while blue indicates cooler than this. The black/solid line gives the case without annihilation for comparison.}
\label{fig:Te_decay_z}
\end{figure}
\begin{figure}
\centering
\includegraphics[width=0.99\columnwidth]{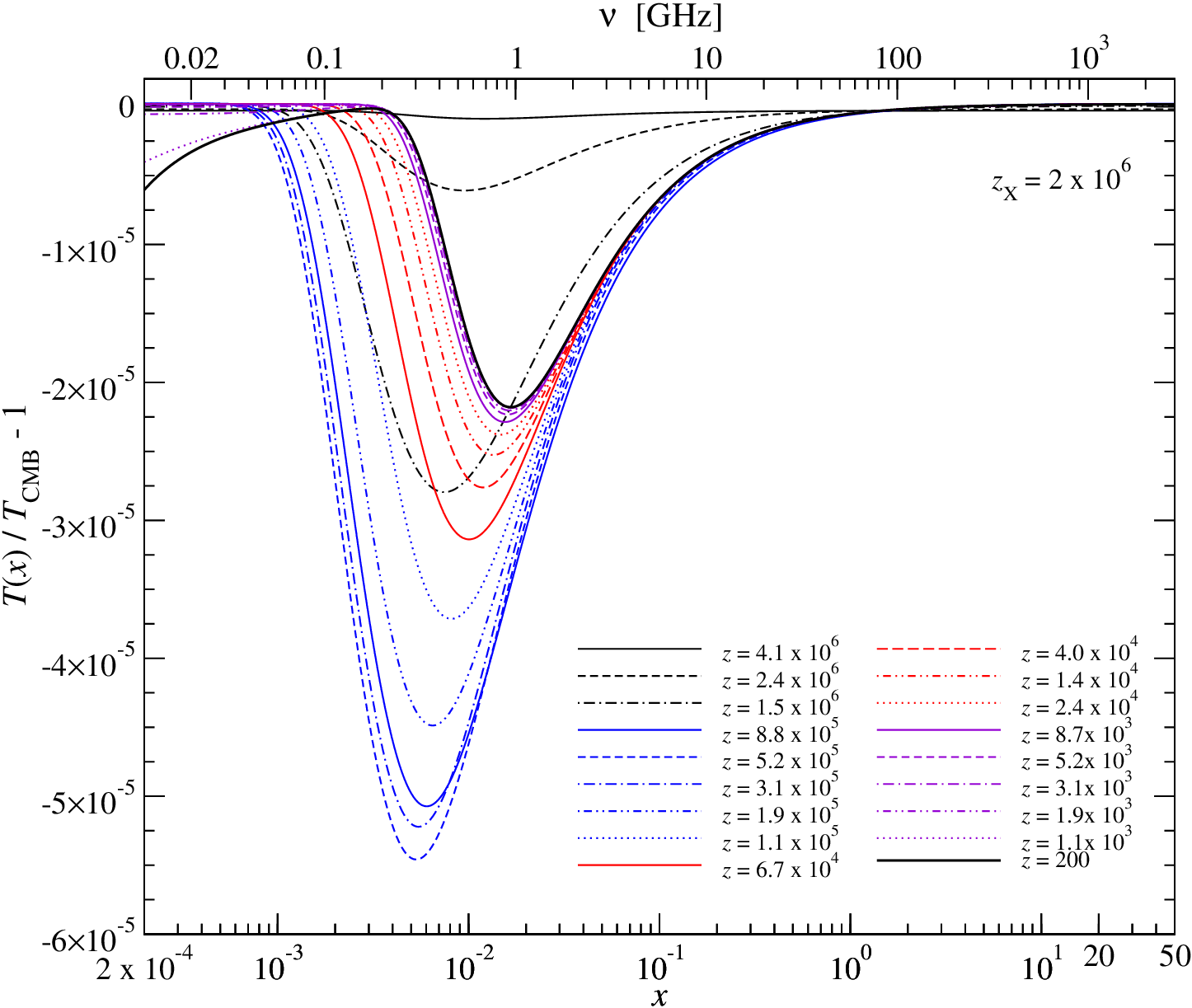}
\\
\includegraphics[width=0.99\columnwidth]{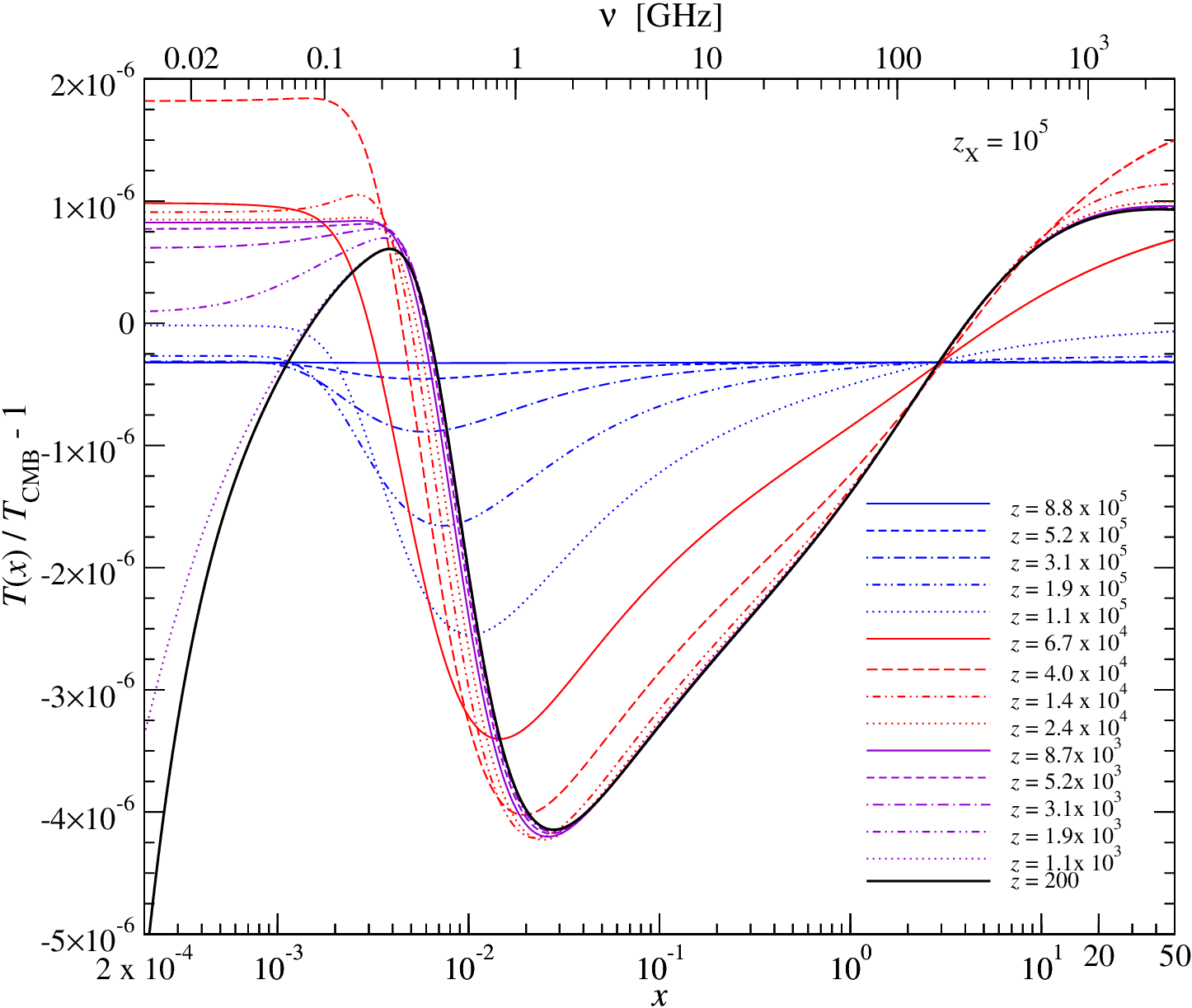}
\\
\includegraphics[width=0.99\columnwidth]{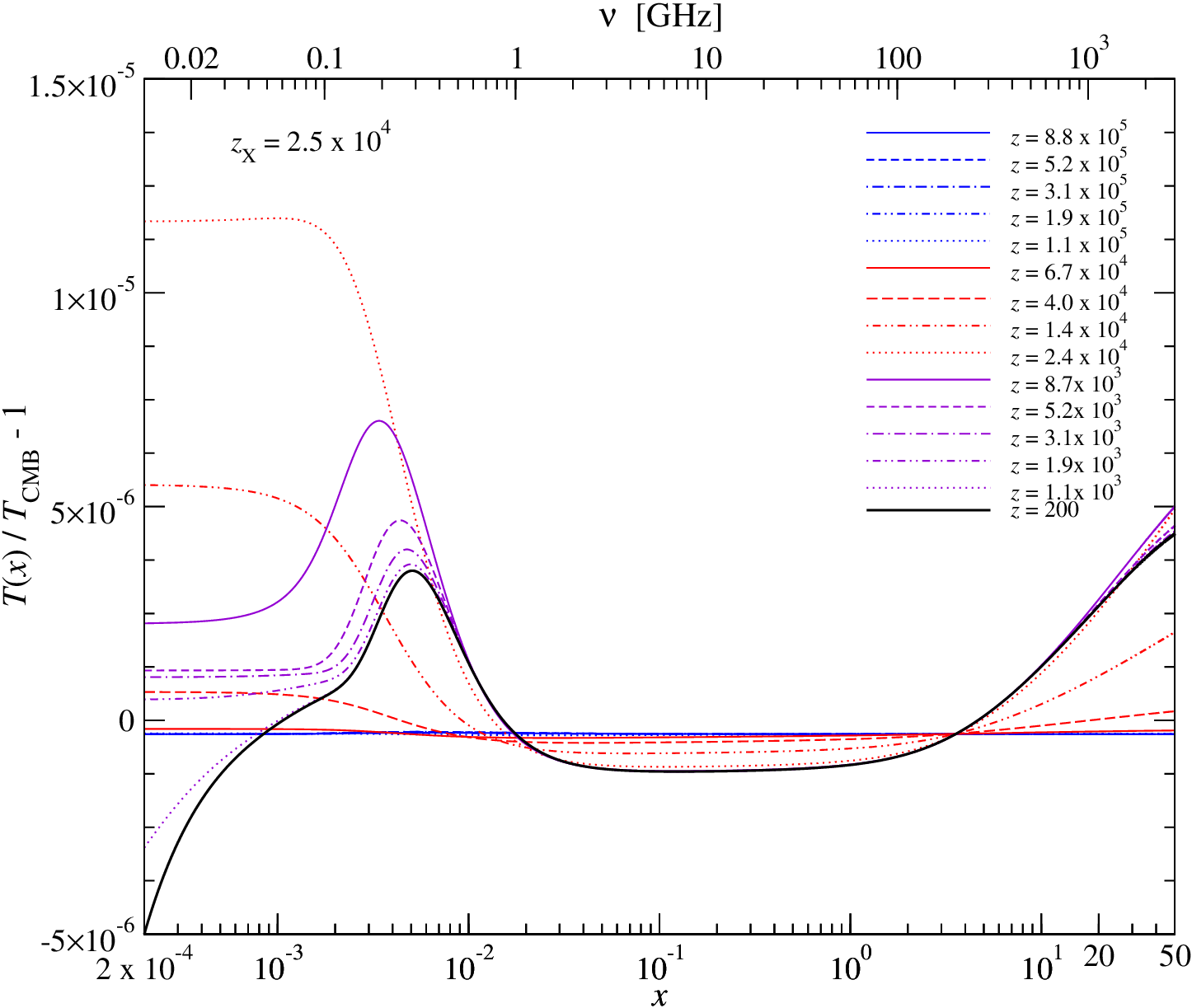}
\caption{Evolution of the CMB spectral distortion caused by the heating from decaying particles with different lifetimes. At $z\lesssim 10^4$ one can see the effect of electrons starting to cool significantly below the temperature of the photons, which leads to free-free absorption at very low frequencies.}
\label{fig:Te_x_decay_evol}
\end{figure}
\begin{figure}
\centering
\includegraphics[width=0.99\columnwidth]{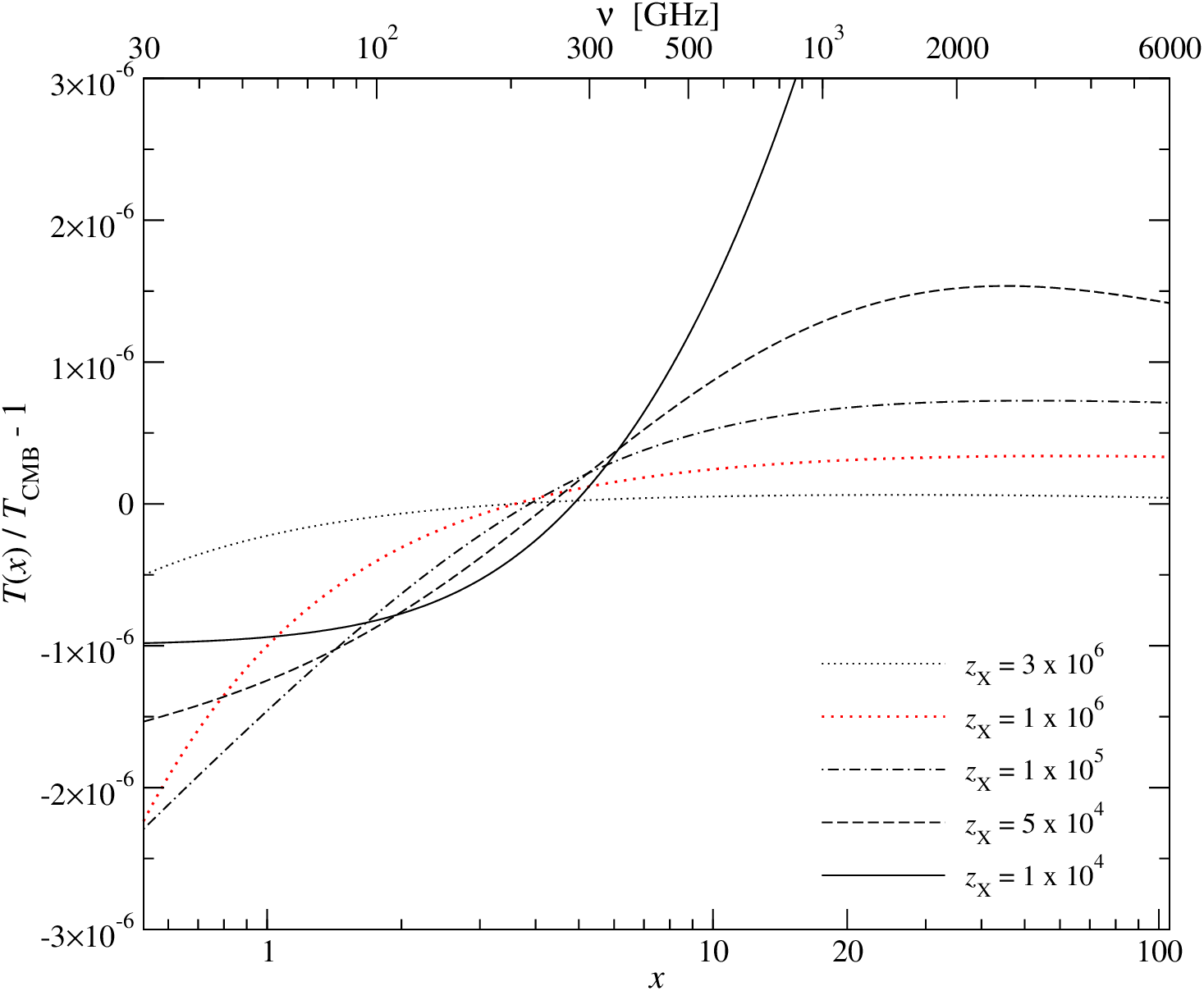}
\caption{CMB spectral distortion at $z=200$ after energy injection from decaying relic particles with different lifetimes in the {\sc Pixie} bands. In all cases we fixed $f_{\rm dec}=2\,z_{\rm X}\,$eV, which corresponds to a total energy release of $\left.\Delta \rho_\gamma/\rho_\gamma\right|_{\rm dec} \sim \pot{1.3}{-6}$. For the effective temperature of the CMB this implies $\Delta T^\ast_\gamma/\TCMB\sim -\pot{3.2}{-7}$ at $\zs=\pot{2}{7}$ and at $\ze=200$ in all cases we found $|\Delta T^\ast_\gamma/\TCMB|\sim 10^{-10}$.}
\label{fig:Te_x_decay_PIXIE}
\end{figure}
To understand the effect of decaying particles on the CMB spectrum a little better in Fig.~\ref{fig:Te_decay_z} we present the evolution of the electron temperature for some cases of Fig.~\ref{fig:Te_x_decay}.
One can see that for decreasing values of $\zX$ at high redshifts the electron temperature follows the case without energy injection for a longer period.
Then, once the heating by decaying particles becomes significant, the electron temperature becomes larger than $\Tz$. 
%
%
After the heating stops for cases with $\zX\gtrsim 10^5$ the relative difference in the electron temperature remains rather constant, with only slow evolution.
Caused by the heating the effective temperature of the CMB also increased and after it ceased the electrons simply keep the temperature dictated by the distorted CMB photon field. 

In cases with $\zX\lesssim 10^5$, however, one can observe an extended period after the maximal heating at which the electrons loose some of their heat again. Having a closer look at the cases with $\zX\gtrsim 10^5$ one can find the same there, but much less pronounced.
At high redshifts the Compton interaction is extremely fast and allows the temperatures of electrons and photons to depart only slight, even with significant energy release.
At low redshifts, Compton scattering becomes much less efficient, so that during energy release larger differences between electrons and photons are possible.
During these periods the electrons are notably hotter than the CMB, so that photons get up-scattered and a $y$-type signature can arise.

In Fig.~\ref{fig:Te_x_decay_evol} we illustrate the evolution of the CMB spectral distortion caused by the heating from decaying particles with different lifetimes.
The upper panel gives an example for a particle with short lifetime. The distortion clearly is close to a $\mu$-type distortion until very late times. The only difference is because of the effect of electrons cooling significantly below the CMB temperature at late times, introducing a small modification because of free-free absorption in the 100\,MHz frequency band.
In the central panel we give a case which at the end has the character of both $\mu$ and $y$-type distortion.
Initially, it starts like a $\mu$-type distortion, but heating continues to be significant down to $z\sim 10^5$, when electrons obtain temperatures larger than the CMB, such that photons are partially up-scattered.
At the end of the evolution the spectrum remains in a state that is a mixture.
Finally, in the lower panel of Fig.~\ref{fig:Te_x_decay_evol} we give an example for a case that looks like a pure $y$-distortion at high and intermediate frequencies.
In this case, energy is mainly released at times when Compton scattering is unable to re-establish full kinetic equilibrium with the electrons.
However, at low frequencies one can again observe the effect of cooling electrons during at after the epoch of recombination.

In Fig.~\ref{fig:Te_x_decay_PIXIE} we show the distortions for some of the previous cases, but focused on the spectral bands of {\sc Pixie}.
In contrast to the case of annihilating particles, where the shape of the distortion was rather insensitive to the effective annihilation rate, for decaying particles the shape of the distortion varies strongly with its lifetime.
This should make it possible to distinguish the effect of decaying particles from the other sources of energy release discussed so far.
For the chosen energy injection rate the typical amplitude of the distortions is $\Delta T/T \sim 10^{-7}-10^{-6}$, which is well within reach of the {\sc Pixie} sensitivities.
However, to forecast the possible constraints from {\sc Pixie} requires consideration of more cases {and realistic foreground models}.

\subsubsection{Upper limits from analytic estimates}
\label{sec:decay_estimate}
\citet{Hu1993b} provided simple analytic expressions that allow estimating the final spectral distortion after some energy release caused by decaying relic particles. 
These expressions were widely used in the literature to place limits on the possible amount of decaying particles with different lifetimes, and here we wish to compare them with the results of our computations.

To obtain the analytic estimates one can start with the simple approximations for single energy release at $\zh$.
At high redshifts, during the era of $\mu$-type distortions one has\footnote{{In a baryon dominated Universe BR is more important than DC emission. In this case the one has $\mu_\infty\approx \mu_{\rm h}\,e^{-([1+\zh]/[1+z_{\rm br}])^{5/4}}$ with $z_{\rm br}\sim \pot{6.2}{6}$ \citep{Sunyaev1970mu}.}}  \citep{Danese1982, Hu1993, Chluba2005}
\beal
\label{eq:mu_therm}
\mu_\infty&\approx \mu_{\rm h}\,e^{-([1+\zh]/[1+\zmu])^{5/2}},
\end{align}
where the thermalization redshift 
\beal
\label{eq:z_mu}
\zmu&=
\pot{1.98}{6}
\left[\frac{1-\Yp/2}{0.88}\right]^{-2/5}
\left[\frac{\Omega_{\rm b}h^2}{0.022}\right]^{-2/5}
\end{align}
%
was already used several times above.
For $\mu_{\rm h}$ \citet{Sunyaev1970mu} gave the well known approximation $\mu_{\rm h}\approx 1.4\frac{\Delta \rho_\gamma}{\rho_\gamma}$, where it is assumed that a negligible amount of photons is injected, but bulk of the energy comes out as heat.

In Eq.~\eqref{eq:mu_therm}, the exponential factor acts as a {\it visibility function for spectral distortions}. At redshifts $z\lesssim \zmu$ practically all energy ends up as CMB spectral distortion, while at $z\gtrsim\zmu$ thermalization exponentially suppresses the residual distortion with double Compton emission being the main source of photons.

\begin{figure}
\centering
\includegraphics[width=0.99\columnwidth]{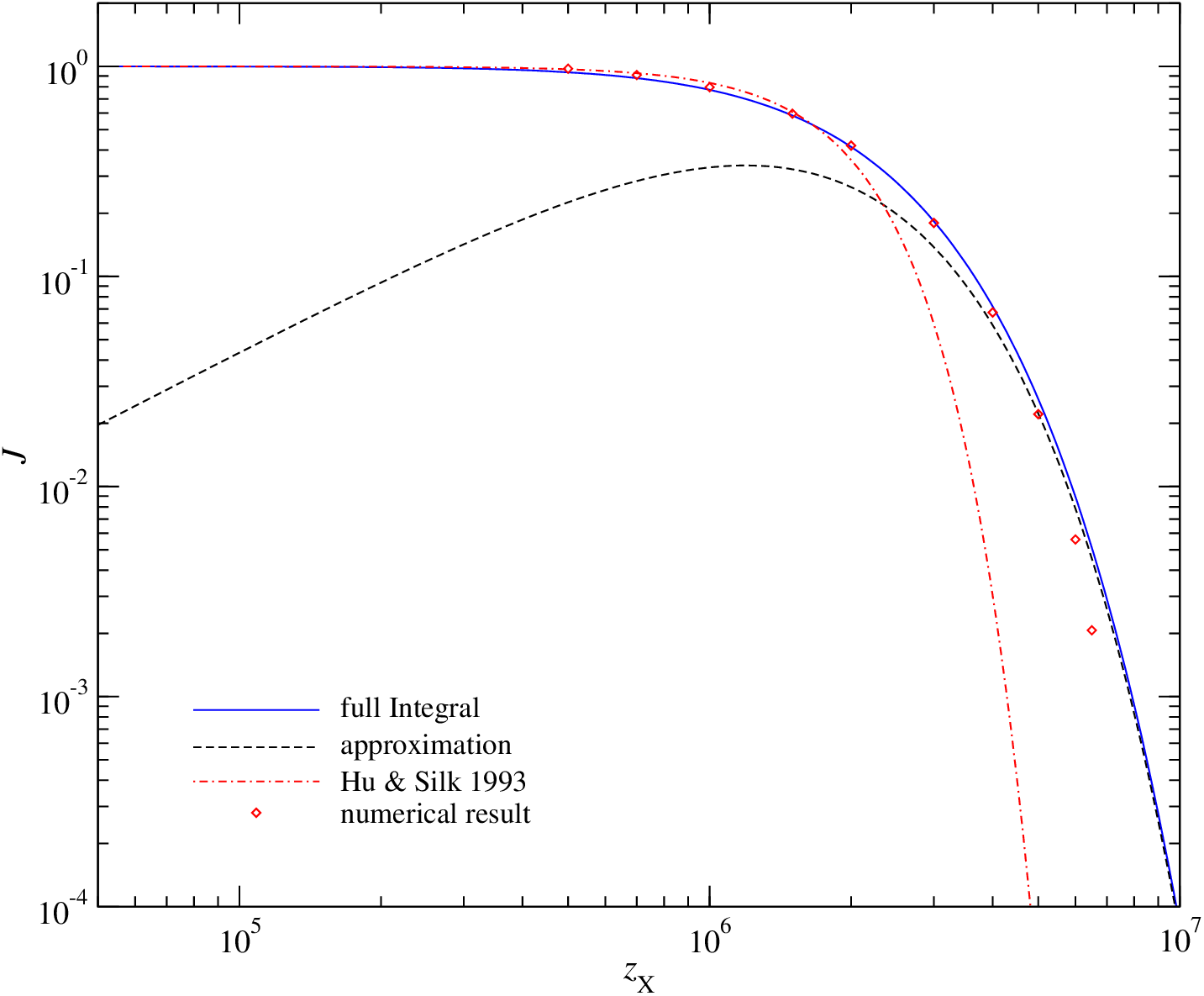}
\caption{Resulting efficiency integral $\mathcal{\bar{J}}$ for decaying particles with lifetimes corresponding to redshift $\zX$.}
\label{fig:JInt}
\end{figure}
To compute the total distortion arising in the $\mu$-era from decaying particles one simply has to calculate the spectral visibility weighted energy release rate:
\beal
\label{eq:mu_therm}
\left.\bar{\frac{\Delta \rho_\gamma}{\rho_\gamma}} \right|_{\rm dec}
&\approx \int 
\frac{1}{\rho_\gamma}
\left.\frac{\id E}{\id t} \right|_{\rm dec}\,e^{-( [1+z]/[1+\zmu])^{5/2}}\id t,
\end{align}
which, assuming radiation domination, in our parametrization, Eq.~\eqref{eq:DE_Dt_dec} reads
\beal
\label{eq:Drho_tot_dec_weighted}
\left.\bar{\frac{\Delta \rho_\gamma}{\rho_\gamma}} \right|_{\rm dec}\!\!
&\!\approx\!
10^{-5}\!\left[\frac{f^\ast_{\rm X}}{\pot{8}{5}\,{\rm eV}}\right]
\left[\frac{1-\Yp}{0.75}\right]
\left[\frac{\Omega_{\rm b}h^2}{0.022}\right]
\left[\frac{1+z_{\rm X}}{\pot{5}{4}}\right]^{-1} \!\!\mathcal{\bar{J}},
\end{align}
where we defined the integral
\bsub
\label{eq:J_weighted}
\beal
\mathcal{\bar{J}}
&=\frac{2}{\sqrt{\pi}}\,\int^{z_{\rm X}^2}_0 \id \xi \sqrt{\xi}\,e^{-[\xi+\lambda^{5/2}_{\rm X}\xi^{-5/4}]}
\\
\label{eq:JInt_appr}
&\stackrel{\lambda_{\rm X}\gtrsim 1}{\approx} \frac{2}{3} \, 2^{11/18}\,5^{4/9}\,\lambda_{\rm X}^{10/9}\,
\exp\left(-\frac{9\,\lambda_{\rm X}^{10/9}}{2^{8/9}\,5^{5/9}} \right),
\end{align}
\esub
with $\lambda_{\rm X}=[1+\zX]/[1+\zmu]$.
The integral $\mathcal{\bar{J}}$ can be easily solves numerically and is shown in Fig.~\ref{fig:JInt} together with the result from the approximation, Eq.~\eqref{eq:JInt_appr}, which works very well for $\zX\gtrsim \zmu$.

In the work of \citet{Hu1993b}, this estimate was performed in a slightly different way. There it was assumed that all the energy released by the decaying particles effectively is injected at time $t_{\rm eff}\sim \tX$.
This can be concluded from Eq.~(8) in their paper, where the exponential factor reads $e^{-(t_{\rm dC}/\tX)^{5/4}}\equiv e^{-\lambda_{\rm X}^{5/2}}$, which implies $\mathcal{\bar{J}_{\rm Hu}}\approx e^{-\lambda_{\rm X}^{5/2}}$. 
In Fig.~\ref{fig:JInt} we also plotted this version for $\mathcal{\bar{J}}$ and find that for $\zX>\zmu$ it strongly underestimates the actual value of $\mathcal{\bar{J}}$, as already pointed out by \citet{Chluba2005}.
This implies that the limits derived from {\sc Cobe/Firas} for particles with lifetimes $\tX\lesssim \pot{6}{6}\,{\rm s}$ are significantly stronger.

\begin{figure}
\centering
\includegraphics[width=0.99\columnwidth]{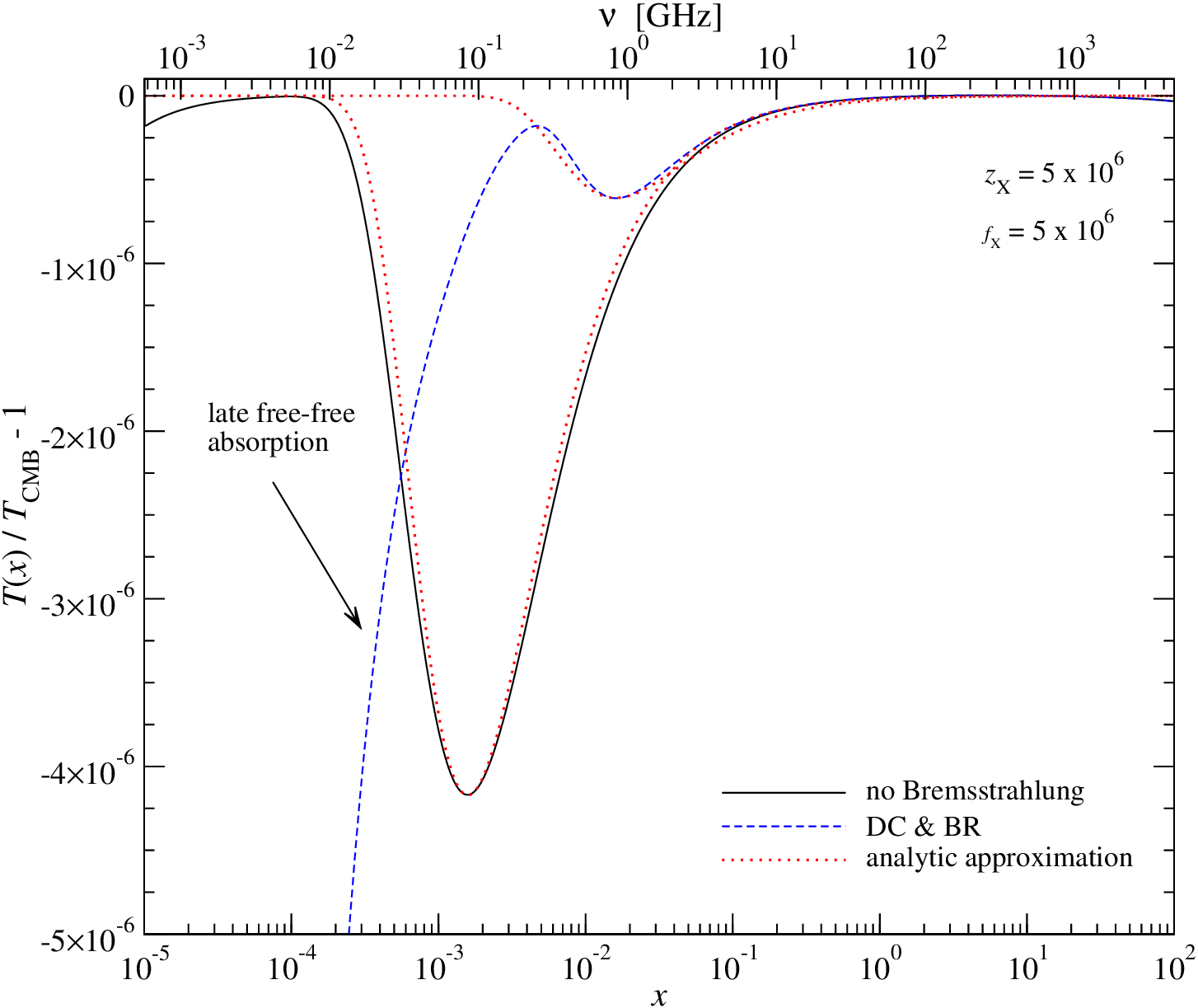}
\caption{CMB spectral distortion at $z=200$ after energy release by decaying particles with redshift $\zX=f_{\rm X}=\pot{5}{6}$ and $\Delta \rho/\rho \sim \pot{6.3}{-7}$. In this figure the importance of BR is illustrated. Also we show simple analytic approximations according to Eq.~\ref{eq:T_mu_x}, with $\mu_{\infty}=\pot{2.65}{-8}$ and $x=\pot{1.6}{-2}$ in the case with BR, and $\mu_{\infty}=\pot{1.8}{-8}$ and $x=\pot{1.6}{-3}$ in the other.}
\label{fig:no_BR}
\end{figure}
Numerically we were able to compute the efficiency function $\mathcal{\bar{J}}$ using {\sc CosmoTherm}. 
In practice $\mathcal{\bar{J}}$ just defines how much of the energy that was released remains visible as spectral distortion today.
Assuming a constant total energy release, one can therefore compute $\mathcal{\bar{J}}$ simply varying the lifetime of the particle and comparing the effective value for $\mu_\infty$ with the total amount of injected energy.
To make the results more comparable we switched off BR, since for the estimate above this was not included consistently.
This also makes it easier to define $\mu_\infty$ as the late changes in the distortion at low frequencies are not arising (see Fig.~\ref{fig:no_BR}).
From Fig.~\ref{fig:no_BR} we can also see how much the low frequency spectrum is affected by BR. The position of the maximal temperature dip in the case without BR is close to $x\sim \pot{1.6}{-3}$, while with BR it is at $x\sim\pot{1.6}{-2}$. This demonstrates the well known fact that DC becomes inefficient at low redshifts \citep[see][]{Danese1982}.
%

%
The result of this exercise is also shown in Fig.~\ref{fig:JInt} for $\Delta \rho_\gamma/\rho_\gamma\sim \pot{6.4}{-6}$.
As one can see, the agreement with the analytic estimate is excellent for this amount of energy injection.
However, for larger energy injection we found that $\mathcal{\bar{J}}_{\rm num}<\mathcal{\bar{J}}$ at $z>\zmu$.
Also, when switching on BR the simple analytic formula cannot be directly applied, as in those cases the distortion no longer is just a simple $\mu$-type distortion, but the low redshift and frequencies other 
contributions can be noticed.
In these cases, full numerical computations for each case should be carried out.

{
We also comment here that, using the simple expression, Eq.~\eqref{eq:JInt_appr}, it is clear that the distortions introduced during the epoch of electron-positron annihilation ($z\sim\pot{6}{7}$) are completely unobservable.
Although the total energy release from this epoch is $\Delta \rho/\rho\sim 1$, the distortions are suppressed by at least $\mathcal{\bar{J}}\sim \pot{5}{-30}$.
For this simple estimate the very efficient electron-electron and electron-positron Bremsstrahlung process have not been included but would lead to even faster thermalization.
Furthermore, even only with normal BR one would reach the same conclusion \citep{Sunyaev1970mu}.
Anything happening at $z\gtrsim 10^7$ with $\Delta\rho_\gamma/\rho_\gamma \lesssim 10^{-5}$ will lead to distortions that are no larger than $\mu_\infty\sim 10^{-9}$.
}

\begin{figure}
\centering
\includegraphics[width=0.99\columnwidth]{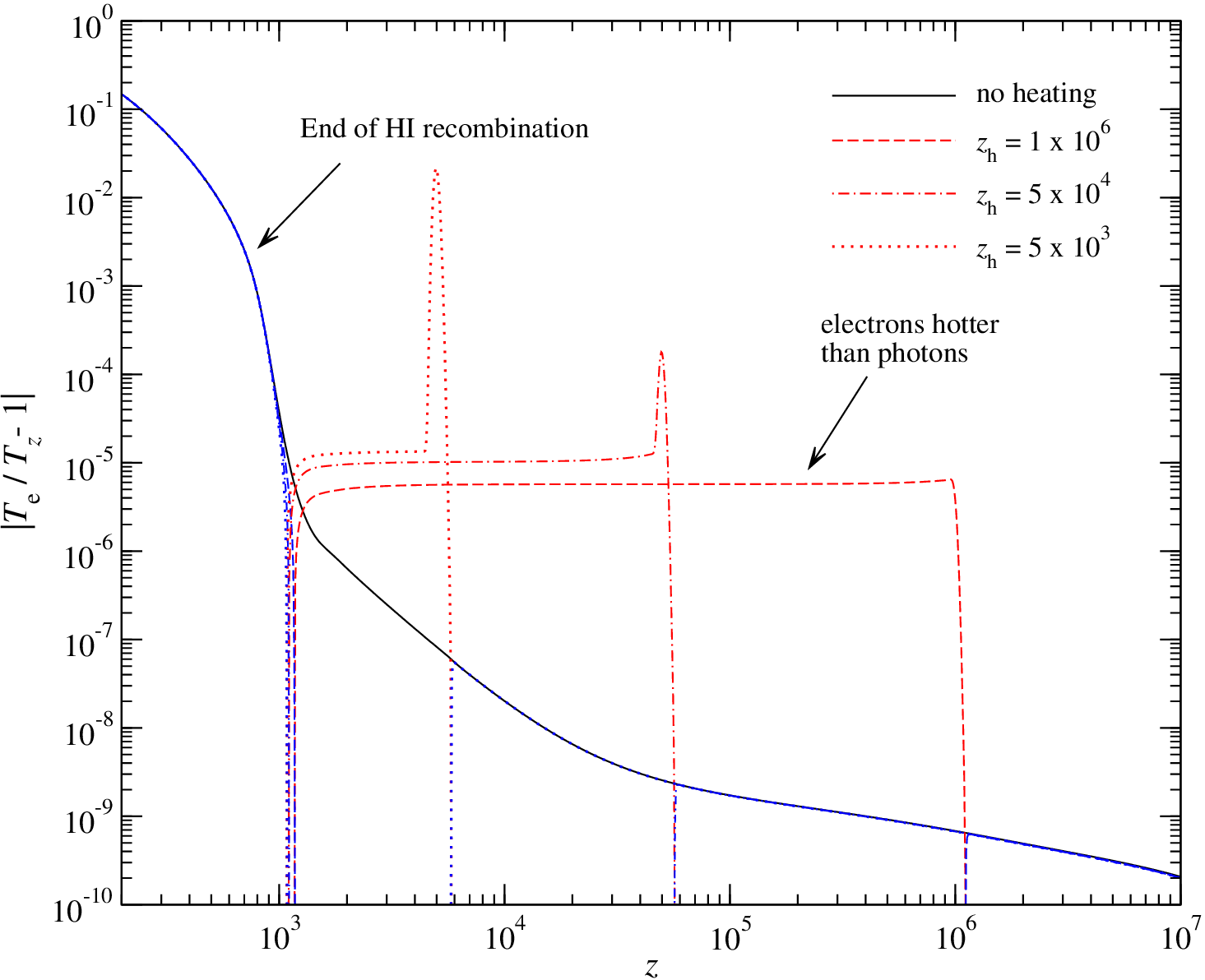}
\caption{Evolution of the electron temperature for thermal histories with quasi-instantaneous energy release. 
The total energy release was $\Delta \rho/\rho \sim 10^{-5}$ in all cases.
The black/solid line gives the case without energy injection for comparison. Red colour indicates that $\Te>\Tz$ while blue colour means $\Te<\Tz$.}
\label{fig:Te_delta_z}
\end{figure}

\begin{figure}
\centering
\includegraphics[width=0.99\columnwidth]{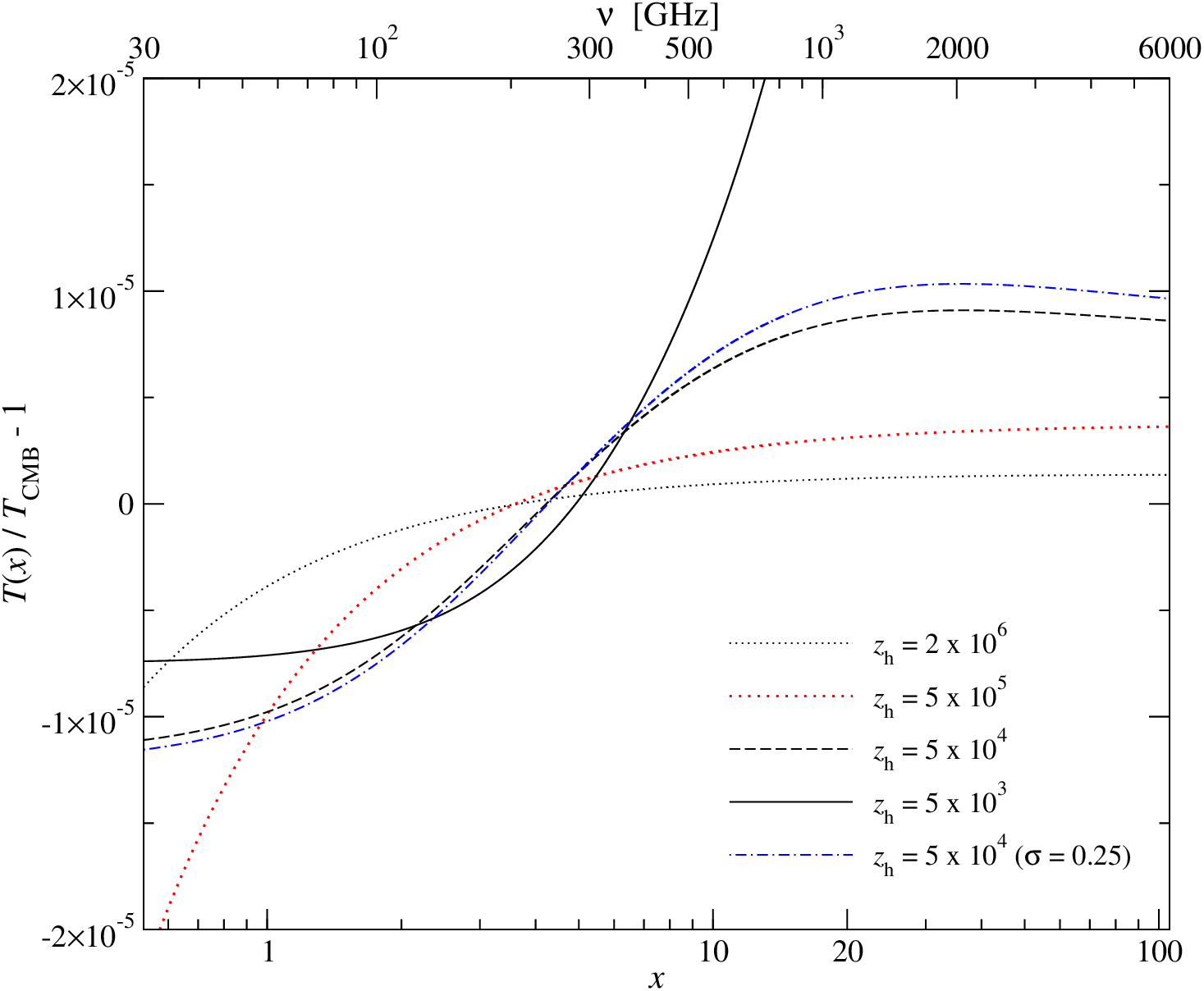}
\caption{CMB spectral distortion at $z=200$ after quasi-instantaneous energy injection at different redshifts in the {\sc Pixie} bands. The total energy release was $\Delta \rho/\rho \sim 10^{-5}$ in all cases. 
The first four curves were computed for $\sigmah=0.05\,t_{\rm h}$, while for the last one we used 
$\sigmah=0.25\,t_{\rm h}$.
}
\label{fig:Te_x_delta_PIXIE}
\end{figure}
\subsection{Distortions caused by quasi-instantaneous energy release}
\label{sec:delta}
To close our computations of spectral distortion after energy release we considered some examples with quasi-instantaneous energy injection. 
The evolution of the electron temperature is reported in Fig.~\ref{fig:Te_delta_z} and a few cases for the final spectral distortions in the {\sc Pixie} bands are shown in Fig.~\ref{fig:Te_x_delta_PIXIE}.
We started your computation at $z=\pot{4}{7}$ with the initial condition defined as explained in Sect.~\ref{sec:initial_cond}.

The electron temperature rises steeply close to $\zh$ and falls off very fast once the energy release ceases.
For the spectral distortion one can again observe the difference in the characteristic spectrum, which is dominated by $\mu$-type contributions for cases with large, and is clearly of $y$-type for very small values of $\zh$. For $\zh\sim \pot{5}{4}$ the distortion is a mixture of both.

The differences in the shape of the distortions are very visible, however, comparing to the case with decaying particles shows that quasi-instantaneous energy release gives rise to very similar distortions for injection at equivalent epochs. 
This suggests that distinguishing these two cases could be rather demanding.

To make this point even clearer, for $\zh=\pot{5}{4}$  we also ran a case with $\sigmah=0.25\,t_{\rm h}$ and the same total energy injection. The result is also shown in Fig.~\ref{fig:Te_x_delta_PIXIE}.
{Clearly, it is rather hard to distinguish the two} lines corresponding to $\zh = \pot{5}{5}$, however, the differences are still at the level of a few percent.
For sufficiently large energy release this could possibly be measured with {\sc Pixie}, however, more detailed forecasts will be necessary.

\label{sec:end_senarios}

\section{Discussion and conclusion}
\label{sec:conc}
We presented detailed computations of the CMB spectral distortions introduced by different physical processes.
We focused in particular on small distortions that could be within reach of {\sc Pixie}.
For this purpose {{\sc CosmoTherm}, a thermalization code which solves the couple Boltzmann equation for photons and electrons in the expanding isotropic Universe, was developed}.
Improved approximations for the double Compton and Bremsstrahlung emissivities, as well as the latest treatment of the cosmological recombination process were taken into account.

We demonstrated that the interaction of CMB photons with adiabatically cooling electrons and baryons results in a mixture of a {\it negative} $\mu$- and $y$-type distortion with effective $\mu\sim -\pot{2.2}{-9}$ and $y\sim -\pot{4.3}{-10}$ (see Fig.~\ref{fig:Te_x_cool}).
{For the currently quoted} sensitivities of {\sc Pixie} \citep{Kogut2011PIXIE} this effect is unobservable, {even in the most optimistic case of no foregrounds.}
{However, the rapid progress in detector technology and the possibility of extending the frequency bands of {\sc Pixie} below 30 GHz, might render this process interesting in the future.}

Furthermore, we computed {the shape} of the distortion arising from the dissipation of acoustic waves in the expanding Universe (see Fig.~\ref{fig:Te_x_acoustic}) using estimates for the energy release rates provided by \citet{Hu1994}.
We find an effective $\mu\sim \pot{8.0}{-9}$ with an admixture of $y\sim \pot{2.5}{-9}$ for $\nS=0.96$.
%
%
{It therefore} appears that for $\nS=0.96$ {\sc Pixie} {could already} allow a $1\sigma$ detection of the effect caused by the dissipation of acoustic waves. 

Nevertheless, the distortion created by the damping of acoustic waves might {be precisely measured} in the future. 
This would in principle allow us to place additional constraints on different inflationary models, however, as we discuss here for precise predictions of the expected distortion one should {also} include the contribution from the cosmological recombination radiation {\citep{Dubrovich1975, Chluba2006b, Sunyaev2009}}, which was emitted at $z\sim 1400$, {and contributes significantly at very high and low frequencies} (see Fig.~\ref{fig:Te_x_rec_PIXIE}).
In addition, our computations indicate that the energy injection rate associated with the dissipative heating process during the recombination epoch is very likely overestimated and should be refined for reliable forecasts.
This could in particular affect the admixture of $y$-type distortions arising from this heating mechanism, {and we plan to look at this problem in a future publication}.
We also emphasize that {\it all} the aforementioned distortions should arise in the standard cosmological model.
Observing the cosmological recombination spectrum together with these distortions might therefore open a way to place additional constraints on the Universe we live in.

We also computed the possible distortions arising from annihilating and decaying relic particles. In the case of annihilating particles the distortion is dominated by a $\mu$-type contribution (see Fig.~\ref{fig:Te_x_ann}), however, at very low frequencies the free-free processes leads to significant modifications. Similarly, at high frequencies a late $y$-type contribution arises.
The shape of the distortion is very similar for different annihilation efficiencies, but the amplitude and sign depend significantly on this.
For very small annihilation rates the distortion arising from cooling electrons dominate and make the annihilation signal very hard to observe.
Furthermore, if annihilating particles are present in the Universe, their effect on the CMB spectrum could be confused with the one from dissipation of acoustic waves, however, for current upper limits on the effective annihilation rate, this is likely restricted to the level of $10\%-20\%$.
We furthermore conclude that presently improving the limits on energy injection by annihilating particles using {\sc Pixie} will be very hard unless $f_{\rm ann}$ exceed  $\sim 10^{-22}\,\rm eV \, s^{-1}$.

In the case of decaying particles, the shape of the distortion strongly depends on the lifetime of the particle (see Fig.~\ref{fig:Te_x_decay_PIXIE}). This should allow distinguishing between different particle models rather easily, unless the energy deposition rates become too small.
We also revisited the simple estimates for the amplitude of the spectral distortion in the $\mu$-era, showing that the residual distortion caused by decaying particles with lifetimes shorter than $\tX\sim \pot{6}{6}\,{\rm sec}$ was underestimated.
This implies that the CMB limits on decaying particles are possibly slightly stronger than previously anticipated (see Fig.~\ref{fig:JInt}).
However, since the distortions for particles with lifetimes $\tX\sim 10^8-10^9\,{\rm sec}$ are neither pure $\mu$- nor pure $y$-type deduced upper limits are more difficult to interpret.

Finally, we also computed the distortions arising in cases with quasi-instantaneous energy release (see Fig.~\ref{fig:Te_x_delta_PIXIE}). 
These distortions are very similar to those obtained in cases with decaying particles, so that distinguishing the two mechanisms will be challenging.
However, a detailed forecast of the observational possibilities with {\sc Pixie} regarding {all these different cases should be performed to reach a final conclusion} about this.
These forecasts should include the possibilities of simultaneous energy release mechanisms. 
%
%
In addition, the distortions introduced at lower redshifts, for example by heating because of supernovae \citep{Oh2003}, or shocks during large scale structure formation \citep{Sunyaev1972b, Cen1999, Miniati2000}, have to be considered.
{Also, the effect of unresolved SZ clusters \citep{Markevitch1991}, the kinetic SZ effect from reionization \citep[e.g., see][and references therein]{McQuinn2005}, and the integrated signals from dusty galaxies {\citep[e.g., see][]{Righi2008, Viero2009, Vieira2010, Dunkley2010, Lagache2011}} will contribute at an important level. In all these cases one should consider the possibilities to include spacial-spectral and polarization information to separate the different components. 
However, we leave a detailed investigation for some future work.}
%

%

\section*{Acknowledgments}
JC is very grateful for additional financial support from
the Beatrice~D.~Tremaine fellowship 2010.
\changeA{He also thanks Mike Seiffert for his comments.}
Furthermore, the authors acknowledge the use of the GPC supercomputer at the SciNet HPC Consortium. SciNet is funded by: the Canada Foundation for Innovation under the auspices of Compute Canada; the Government of Ontario; Ontario Research Fund - Research Excellence; and the University of Toronto.

\begin{appendix}

\end{appendix}

\bibliographystyle{mn2e}
\bibliography{Lit}
\end{document}